\documentclass{aa}
\usepackage{graphicx}
\usepackage{url}       % For breaking URLs easily trough lines
\begin{document}
%\begin{article}
%\begin{opening}

\title{A coronal wave as CME footprint}
\author{C. Delann\'ee\inst{1,2}}
%C. \surname{Marqu\'e$^{1}$}, 
%A. \surname{Zhukov$^{1}$}}
%D. \surname{Biesecker$^{3}$}
\institute{Royal Observatory of Belgium, Brussels, Belgium
\and LESIA, Meudon observatory, Meudon, France}
%$^{3}$NOAA/SEC, Boulder, US
\date{Received November 20, 2009 / Accepted ...}

\abstract
{Coronal waves are disturbances propagating over a large portion of the solar disc. 
Despite the fact that they resemble a wave originating from a single point, 
several authors argue that they are due to the restructuring of the solar magnetic 
field during a coronal mass ejection.}
{I report a coronal wave observed in Fe{\sc xii}, soft X-ray, and H$\alpha$ 
on November 6, 2006, jointly with a CME and a flare to show the spatial and 
temporal relation of the wave and the CME. I also take advantage of the spectral 
resolution of the wave to obtain an approximation of its temperature. Finally, I 
compare the magnetic field topology to the wave location.}
{I overlay all the observations. To highlight the faint structures, and especially 
stationary structures produced on the passage of the wave, I use images obtained 
by subtracting a pre-event image from the observations then I apply wavelet 
transform. The temperature is obtained by comparing theoretical emissions to the 
ratio of the emission of the wave front through two different filters in soft X-ray. 
Using the synoptic magnetogram obtain with the Mickelson Doppler Imager (MDI/SoHO) and the Potential Field Source Surface (pfss) of the solar software (SSWIDL), 
I draw the magnetic field lines at the borders of the magnetic field topological 
domains.}
{The observations show bright fronts that 
are co-spatial when observed at the same time, therefore I believe that the same 
wave is observed in the different band passes. The 
wave front is not observable in H$\alpha$, but just in Fe{\sc xii} 
and in soft X-ray, indicating that the temperature of the wave front is
higher than the chromospheric temperature. The ratio of emission through two SXI 
filters gives a temperature of $7\ 10^6$ K for the wave front. The northern-most edge of the CME footprint related to this wave corresponds to the last location of the wave 
on the limb. The image processing reveals stationary brightenings produced on the passage 
of the wave front. The potential spherical extrapolation of the sun shows that 
those stationary brightenings are lying in jumps of magnetic field lines 
connectivity.}
{Coronal waves are hot structures. They are the footprint of a 
CME and are related to the magnetic field topology.}

\maketitle

\section{Introduction}
Global coronal waves are large scale disturbances that can be observed in 
the solar atmosphere in coronal and chromospheric spectral wavelengths. 
H$\alpha$ waves were first discovered in association 
with flares and named Moreton waves (Moreton 1960, 1961, Moreton and Ramsey 1960). More recently, 
waves were observed in Fe{\sc xii}, usually named EIT waves (Dere et al. 1997, Thompson et al. 1998, Attrill et al. 2007, Long et al. 2008), and finally 
in soft X-rays (Khan and Hudson 2000, Khan and Aurass 2002; Narukage et al. 2002). These 
waves are bright structures that usually propagate in a restricted arc,  narrow in H$\alpha$,
thicker and diffuse in Fe{\sc xii}, traveling on the solar 
disc across distances that can be up to the solar radius. Their 
morphology and evolution may be compared to surface gravity waves propagating 
on the surface of a water.

Warmuth et al. (2004) show that when the waves appear in H$\alpha$, EUV 
and soft X-ray spectral band passes simultaneously, they are nearly co-spatial.
The study conducted in Warmuth et al. (2004) is in 
contradiction with other observational studies: e.g. Eto et al. (2002) who 
found the Moreton wave to precede the diffuse coronal wave, and 
Klassen et al. (2000).
The later reference analyzes in detail several coronal waves observed in 
Fe{\sc xii}, compares their velocities to the usual values given in the 
literature and concludes that the coronal waves are slower than the Moreton waves. However, Vr{\v s}nak et al. (2002) indicate several properties 
of the dynamic (deceleration of the coronal wave front) and the observation 
of the Fe{\sc xii} waves (low cadence of images and visibility of the EIT 
waves much farther than the Moreton waves) that have to be taken into 
account to resolve this discrepancy.
%Narukage et al. (2002). Narukage et al. (2002) mention that 
%the wave fronts in soft X-ray 
%and in H$\alpha$ are rougthly overlaid but produced a sketch of 
%the fronts location showing quite a shift between the two fronts
%(see Fig. 1k in Narukage et al. 2002). The derived velocities 
%from these estimates of the front location, are 629 km s$^{-1}$
%for the soft X-ray front and 491 km s$^{-1}$ for the H$\alpha$
%front. In Fig. 1 of Narukage et al. (2002), several images are
%presented of the H$\alpha$ and soft X-ray observations of the
%waves that we use to be overlaid (see Fig. 1). The two fronts
%are fully overlaid despite the 1 minute delay between the two 
%images, the shapes of the fronts are very similar, 
%which make the sketch of the fronts locations (Fig. 1k of 
%Narukage et al. 2002) unreasonnable. The time origin of the 
%waves is the same and estimated at 04:32:00 UT (Narukage 
%et al. 2002). Therefore, using the estimated velocities derived
%for each wave front, they should be separated by 41.4 Mm, i.e.
%roughtly separated by the thickness of the wave front, which is
%not observed. To correct the discrepancy between the velocity
%and the overlay of the wave fronts Narukage et al. (2002)
%introduce a number which should correspond to the location of the
%wave fronts at their origin. This origin in location is not
%discussed in the article and contradicts the fact that the model
%developed later on, implies that the origin of the two waves has 
%to be the same. Therefore, the study in Narukage et al. (2002)
%seems quite misled and not conclusive. 
%
For example, Eto et al. (2002) find that the wave fronts observed in H$\alpha$ 
and in Fe{\sc xii} have different velocities and conclude that the 
coronal wave is physically distinct from the H$\alpha$ wave.
However, going through the data set again, one can find that the flare related to the emission of the wave
begins 4 min. earlier than the first image obtained by EIT 
(Extreme ultra-violet Imaging Telescope, Delaboudin\`ere et al. 1995) 
showing the flare
and peaked 1 min. later. In H$\alpha$, the wave front is clearly
visible over 12 min., beginning 1 min. after the EIT image showing
the flare, ending just before the second EIT image of the event
showing the flare, i.e. the first one showing the EIT wave. Due to the low cadence of images of EIT, this set of data cannot give a clear spatial correspondence between the EUV and H$\alpha$ wave. Regarding the velocities of each wave fronts derived from the observations, as the edge of the wave fronts are very different (in EUV it is diffuse, and sharper in H$\alpha$), differences in estimation of location of the border of each waves can give different estimation of the velocities, that may have lead Eto et al. (2002) to find different velocities for the two wave fronts.
%Therefore, due to the low cadence of images of EIT, Eto et al. (2002)
%could study the spatial relation of the two observed waves. A filament
%starts to oscillates in H$\alpha$ one minute before the second EIT image. 
%This oscillation is taken to be the signature of the passage of the Moreton %wave by the filament.
%As Eto et al. (2002) locate the EIT wave front slightly closer to the flare 
%site than the filament, the derived velocity of the EIT wave front is 
%smaller than the Moreton wave front. However, I do not locate the EIT wave 
%front of this particular event at the same place than Eto et al. (2002) 
%does: having a closer look to the image at 06:13 UT of the Figure 5 in 
%Eto et al. (2002), I see a bright structure that links an active region 
%in the northern hemisphere and the oscillating filament, this brightening 
%could belong to the EIT wave front. The use of running 
%difference images with low contrast do not permit to locate clearly the 
%EIT wave front, but the image, produced using the process described in the 
%section 2 (see Figure 1), clearly shows the EIT wave a little ahead of the filament. However, I still have to believe that the oscillation of the filament is due to the passage of the Moreton wave that is no more visible in H$\alpha$.
Co-spatial H$\alpha$ and Fe{\sc xii} wave fronts are studied using
observations in Fe{\sc xii} and in H$\alpha$ obtained at the same time (e.g. Warmuth et al. 2004; Delann\'ee et al. 2007; Thompson et al. 2000).
Each of these studies conclude that the observed H$\alpha$ and 
Fe{\sc xii} wave fronts are co-spatial. Therefore, I believe that
the Moreton wave and the EIT wave are the two manifestations
of the same structure and I disregard any interpretation of the H$\alpha$
and Fe{\sc xii} being two structures moving at different speed
(e.g. Eto et al. 2002; Chen et al. 2002).

Balasubramaniam et al. (2007) show that, when a Moreton wave passes through a filament, it  can 
be enhanced in both the H$\alpha$ line center and in the wings 
of the line. This fact contradicts
the amplitude of the wave estimated from the thickness of the front 
and the Doppler shift of the plasma inside the front. Therefore, 
Balasubramaniam et al. (2007) suggest that the Moreton waves are 
possibly high coronal structures emitting in H$\alpha$. Gilbert et 
al. (2004) show for two events that the observations of wave fronts 
in He{\sc i} and in Fe{\sc xii} data are co-spatial. If the
chromospheric emission would have been due to a magnetosonic wave
propagating in the chromosphere then due to the much denser medium
the speed of this wave would have been much slower and the overlay
of the two emissions (chromospheric and coronal) would 
not have been possible. Therefore, Gilbert et al. (2004) conclude that
the chromospheric emission of the wave front is not due to a wave 
propagating in the chromosphere but is rather an imprint of a 
disturbance in the corona. I also believe that the chromospheric 
emission of the observed waves is due 
to a high altitude structure that can sometime produce emission in 
lower temperature. Wills-Davey et al. (2007) noted that the coronal waves are frequently observed only in Fe{\sc xii} without any 
H$\alpha$ nor soft X-ray counterpart. Therefore, what kind of physical process can make the wave sometime observable and other time invisible in H$\alpha$?
This question also rises from this study.

Biesecker et al. (2002) compare the occurrence of a coronal wave in the catalog established by Thompson and Myer (2009) with the occurrence of a flare that is defined by a significant, impulsive, transient increase in X-ray flux recorded by NOAA GOES 3 X-ray monitor. Looking at well defined and less defined coronal waves (see the catalog given in Thompson and Myer 2009, for an explanation of the confidence level of the occurrence of a wave), Biesecker et al. (2002) find that only 66 of the 173 observed waves, occurred in conjunction with a flare that is defined by a significant, impulsive, transient increase in X-ray flux recorded by NOAA GOES 3 X-ray monitor.  This relation seems very low as I do not know any detailed study of a coronal wave that appears without any flare. In that sens, Cliver et al. (2005) show that very
week (A-class) flares can be related to Fe{\sc xii} waves, indicating that the definition of a "significant increase in X-ray flux" may impact the results obtained by Biesecker et al. (2002). Cliver et al. 
(2005) note that Moreton waves were closely associated with flares during
1960-1970 - indeed they were often referred to as "flare waves". Taking the idea of coronal waves and Moreton waves as the same structure, they both appear to be produced together with a flare from which they both seem to originate. 
Therefore, many studies modeled them as magnetosonic waves propagating 
freely in the solar corona, originating from a pressure pulse that also 
produces the flare (Warmuth et al. 2007 and references therein).

Coronal waves can appear conjointly with type II radio emission 
(Klassen et al., 2000). As the type II are radio emission slowly 
decreasing in frequency, they are commonly interpreted as signature of waves 
steepening into shocks and propagating outwards through the corona (Wild and McCready, 1950; Nelson and Melrose, 1985). The association of the type II radio burst and the coronal wave reinforces their interpretation in terms of magnetosonic waves.
However, Wills-Davey et al. (2007) discussed several reasons why the 
observed coronal waves cannot be magnetosonic waves. The principal
counterargument is that the observed structures have a large range
of velocities (from 100 to 400 km s$^{-1}$), meaning that the
coronal conditions have to be quite different at different time 
which seems unreasonable. Another counterargument is that the magnetosonic waves
have to find quite high plasma beta values to persist over large distances: 
in Wu et al. (2001) for example, a ring shaped wave propagates on the solar 
surface, but the simulation is performed using $\beta>1$ everywhere in the 
corona, which is much higher than the typical value used in other articles 
(i.e. $\beta=0.01$) to study coronal phenomenon. I would like to add to 
these counterarguments that the corona is very structured. Due to 
interaction between photospheric magnetic flux concentrations, the coronal plasma $\beta$ 
may vary a lot from very low value (less than 1) close to the flux concentration, to high values (more than 1) where the interaction of the flux concentration almost annuls the coronal magnetic field (Gary et al. 2001, Warmuth et al. 2005). These variations may exist even in a small area. In 
these conditions, a magnetosonic wave mode
would be very deformed, reflected, dissipated, or converted to another mode when it encounters a magnetic structure, i.e. active regions, filaments and bright points, which are everywhere on the solar disc, hindering the propagation of a self-similar structure over large distances. I note here that, in the numerical simulations performed in Wang et al. (2000), even if the plasma $\beta$ is quite high and the conversion into another wave mode is not taken into account, the structure of the initial atmosphere (in density and in magnetic field) deforms the simulated wave making it different to the observed one during the final stages of simulation. Finally, Podladchikova \& Berghmans (2007) show that the ring shaped observed coronal waves present patches of high light intensity. Podladchikova \& Berghmans (2007) and Attrill et al. (2007), show that these patches are rotating. The rotation of the ring shaped observed coronal waves cannot be modeled by a freely propagating magnetosonic wave. Therefore, I disregard any freely magnetosonic wave model for the observed wave structures.

Even though it may be stated that every wave appears in conjunction with a flare, most flares do not produce 
any observable coronal waves (Delann\'ee and Aulanier 1999, Cliver et al. 2005, Chen et al. 2006), 
so the flares do not seem to be the origin of the waves but a counter part of 
another phenomenon as well as the wave itself. The statistical
analysis performed in Biesecker et al. (2002), shows that
there is a strong correlation between the Fe{\sc xii} wave and the CMEs. 
Even if this study is biased by errors or threshold while doing the catalogs, 
this result is confirmed by studies on smaller numbers of individually observed events.
Warmuth et al. (2004, 12 events) and Delann\'ee et al. (2000, 17 events) 
both find that 100\% of Fe{\sc xii} waves are produced in conjunction with a CME. 
I also note that I do not know a published case of a Fe{\sc xii} wave
produced without any CME. So, I believe that 
each coronal wave is related to a CME where it finds its origin.
These two facts, i.e. the lack of a coronal wave for every flare and the presence of a CME in conjunction with a coronal wave, lead several authors to 
interpret coronal waves in terms of signatures of CMEs that are 
due to magnetic field lines opening: piston driven waves/shocks (Thompson et al. 2000; Chen et al. 2002; Cliver et al. 2005; Warmuth et al. 2005), electric currents and plasma compression generated by expanding 
magnetic field lines (Delann\'ee and Aulanier 1999, Delann\'ee 
et al. 2008), magnetic field line reconnection in the quiet 
Sun induced by the CME expansion (Attrill et al. 2007). These interpretations imply that every coronal wave is the footprint on the solar disc of a CME. 

Vr{\v s}nak et al. (2006) and Patsourakos \& Vourlidas (2009) compare 
the locations of the CME leg and of the wave front, both of them find 
that the wave moves ahead of the CME leg, contradicting the models given 
in the previous paragraph. On the other hand, Chen (2009), Attrill et al. (2009) and Cohen et al. (2009) provides examples of events on the limb showing a strong spatial mapping between the CME and the coronal wave. In this article, I study a coronal wave that 
appeared on November 6, 2006 on the east limb associated with a CME to 
determine whether the coronal wave is the CME footpoint or not. 
I analyze this event using Derotated Base Difference Image (DBDI) processing (Delann\'ee \& Aulanier 1999, Thompson et al. 2000) instead of the usual 
running difference images used in Vr{\v s}nak et al. (2006) and 
Patsourakos et al. (2009). This last process reveals the 
brightness variation of structures from one image to the following 
one: if a structure is bright in an image and remains as bright in the 
following one, it appears gray in the running difference images; 
if the brightness of the structure is still high but a bit
decreased, the structure appears dark in the running difference 
image. In the case of the coronal waves, they are commonly 
described as a bright front surrounding a dimmed region (see e.g. 
Thompson et al. 1998, Warmuth et al. 2004). This description
comes from the use of the running difference images, but appears to 
be (partly) an artifact of this process because the brightness of the 
wave front slowly decreases with the time (Delann\'ee et al. 2007). 
If instead of using the previous image to be subtracted with its following 
one, I use an image obtained before the appearance of the structure, then 
the same structure with slowly decreasing brightness would appear bright on
two consecutive images obtained after the subtraction. So doing
DBDI allows to reveal structures that last for a long 
time even if its brightness decreases. The correction of the solar
rotation is very needed as the sun may rotate quite a lot
in the time interval of the two images used to obtain a DBDI.
Using the DBDI method, Delann\'ee et al. (2007) showed that a coronal 
wave appears to have a slightly different morphology and behavior
than using the running difference images.

Delann\'ee \& Aulanier (1999), Delann\'ee (2000) 
Delann\'ee et al. (2007), Attril et al. (2007a,b) Attrill et al. (2009) and Cohen et al. (2009) study 8 
coronal waves using DBDI processing of data. They find that the 
coronal waves are not fully propagating: some portion of the wave 
reach a final location that remain 
bright for 10 min. to one hour depending on the spectral band 
pass used for the observations.
Delann\'ee and Aulanier (1999), Delann\'ee et al. (2007) and Delann\'ee (2009), Attrill et al. (2009) and Cohen et al. (2009)
reveal that the stationary brightenings of the coronal waves are 
lying in jumps of connectivity of magnetic field lines. These jumps of 
connectivity are known to be sites at which electric current is easily produced (Sweet 1958; Aulanier et al. 2006). The existence of electric currents 
are related to occurrence of hot coronal structures (Magara and Longcope 2001). 
Therefore Delann\'ee and Aulanier (1999) and Delann\'ee et al. (2007) 
conjecture that the stationary brightenings of the coronal waves 
are due to Joule heating in the jumps of magnetic field line 
connectivity. Doing the image processing of instrument observations and comparing the occurrence of structures with extrapolated magnetic field topology is a very long study to lead therefore, 
until today, only five coronal waves are studied using this 
processing. On this basis, Warmuth et al. (2004) claim that 
the stationary brightenings are very rare and that this electric 
current model cannot be suited to explain the occurrence of the 
large majority of coronal waves. I am using this same processing 
of analysis to find several properties of a coronal 
wave: its link with the CME (is it lying at the footprint of the 
CME leg?), its dynamics (is it propagating or stationary?) and 
its temperature (is it a cool or hot structure?). Sec. 1 presents 
the data used in this study. Sec. 2 analyzes the data and Sec. 3 
summarizes the analysis and concludes on the relation between 
the CME and the wave.

\section{Data}

\subsection{Instruments of observations}

I investigate simultaneous multi-wavelength data. The soft 
X-ray images are obtained with the Soft X-ray Imager 
(SXI, Pizzo et al. 2005) on board the Geosynchronous Operational 
Environmental Satellite 13 (GOES-13). The image cadence 
is about one minute. Four different filters are used: "Polyimide Thin", 
"Polyimide Thick", "Beryllium Thin", "Tin"; therefore the cadence of images is about 4 minutes for each filter. All these filters are sensitive to 
light emission from hot plasma with temperatures ranging from 1 
to 10 MK. The extreme ultra-violet images at 195 \AA~are obtained 
using the Extreme ultra-violet Imaging Telescope 
(EIT, Delaboudini\`ere et al. 1995) on board the Solar and Heliospheric 
Observatory (SoHO). The cadence of these images is 12 minutes. This 
filter is mainly sensitive to the emission of Fe{\sc xii} 
present at about 1.5 MK. The H$\alpha$ images are obtained using 
the Optical Solar Patrol Network (OSPAN) at the Sacramento Peak 
Observatory. The cadence of images is about one minute. The H$\alpha$ line ($\lambda = 6365 \AA$) is sensitive to plasma at a temperature of $10^4$ K. The images of the upper corona are obtained using the orange filter of Large Angle 
Solar Coronagraph C2 (LASCO, Brueckner et al. 1995) on board 
SoHO. This filter is mainly sensitive to the plasma density as 
its band pass is very broad. 

\subsection{Extrapolated magnetic field}

The magnetogram and the spherical 
potential magnetic field extrapolations come from the potential field source 
surface package from Solarsoft in the Interactive Data Language
(pfss/SSWIDL, Schrivjer and deRosa 2003). A synoptic magnetogram
is constructed from the data obtained with the Michelson Doppler
Imager (MDI/SoHO, Scherrer et al. 1995), and updated every 4 hours to 
include the data near the central meridian
of the sun. The synoptic magnetogram on November 6, 2006 does not show 
any strong magnetic polarity in the area of the eruption. This is likely
because there was not any magnetic polarity during the previous solar 
rotation and because the flare is behind 
the limb. After November 7, 2006 the magnetogram shows a magnetic bipole 
appearing from behind the limb in the south, at the longitude of the flare 
site. This bipole is close to the central meridian on November 13, where 
it is well resolved. Unfortunately, between November 7 and 13, it evolves expanding, rotating, emerging, dispersing. This evolution might influence the magnetic field topology during this period.
The PFSS runs at the Lockheed Martin Solar and Astrophysics Laboratory 
(LMSAL) using a synoptic magnetogram to extrapolate the potential magnetic 
field over the whole solar surface. The results are just retrieved on a
local computer where a small procedure of visualization permits to draw a 
selection of magnetic field lines. Drawing the magnetic field lines that 
have one footpoint close to each other and the other one the farthest from 
each other, I locate the limits of the domains of connectivity of the 
magnetic field lines. I checked the magnetic field topology evolution 
of the region 
affected by the wave from November 7 to 13: it remains quite similar despite 
the distortion of the magnetic bipole. 

Moreover, McIntosh et al. (2007) explain the refilling of coronal loops that are opened during an eruption by the reformation of the pre-eruptive topology. Similarly, Lynch et al. (2008) show that after an eruption the magnetic field reconfigures to 
its pre-eruption topology in a numerical simulation of an eruption. These considerations allow me to use a synoptic 
magnetogram obtained at a later time than the day of the eruption to compare 
its topology to the structures appearing during the wave front propagation.

As commonly observed, the large scale magnetic field
is potential (e.g. Delann\'ee and Aulanier 1999; Wang et al. 2002), even
in active regions, the magnetic field shear is much less than 1 Mm$^{-1}$
(Mandrini et al. 2002). Therefore, potential magnetic field is suitable for comparison to the structures appearing during the event of November 6, 2006, 
as it affects a large portion of the Sun.  

I choose to use the data obtained 
on November 10, 2006 at 12:04 UT, the closest in time from the eruption in 
which the dipole seems resolved enough. The potential magnetic field lines 
extrapolated from the synoptic magnetogram obtained on November 10, 2006, 
are shifted to correct the rotation of the Sun between November 6 and 10, 
in order to allow an overlay with the observations of November 6.

\subsection{DBDIs}

The H$\alpha$, Fe{\sc xii} and soft X-ray images are Derotated 
Base Difference Images (DBDIs): each image is corrected for the 
differential rotation to the time of the first SXI image at 
17:25 UT before being subtracted with an image obtained using 
the same filter at the closest time before the beginning of the 
flare (i.e. 17:39 UT). The wave 
appears partly above the limb and partly on the solar disc. I used an old version of the drot\_map from the solarsoft to correct the solar differential rotation. This old version add, without any correction, the over-limb observation to the corrected on disc one. Therefore, the over-limb observed structures presented in the figures of this article are not affected by any correction. In an image corrected for the solar rotation for a short 
period of time, the structures on the disc and 
close to the limb are corrected by only a very small shift (2 EIT pixels,
1.4 SXI pixel and 0.13 pixels of Figs. \ref{figure1}, \ref{figure2} and \ref{figure3} at the equator and close to the 
limb during the 2 hours duration of the event) ensuring 
continuity between the structures on the disc and the ones above the limb. 
The DBDIs presented here are highly 
contrasted to show the very small brightness changes of 
the structures, which makes the DBDIs quite noisy. 

I try to highlight the structures as much as possible by highly processing them. I apply several filters during the processing, which results in the DBDIs presented in Figs. \ref{figure2} and \ref{figure3}. The images in Figs. \ref{figure2ter} to \ref{figure3h} (with the exception of those in Fig. \ref{figure3}) are obtained by summing 4 neighboring pixels then expanding the images to their original size. This first process smooths the intensity spatial variation, removing the isolated flickering pixel. Then, the intensity of each pixel is replaced by the hyperbolic tangent of its intensity, which permit to exaggerate the faint structure while minimizing the flickering pixels. The images presented in Fig. \ref{figure2ter} undergo this process solely. 

The images in Figs. \ref{figure2quater}, \ref{figure3c} undergo a wavelet filtering removing the spatial frequencies less than 20 pixels. This wavelet transform permits to obtain an image showing the large scale variations at the typical scale of the wave front observed in Fe{\sc xii}. However, the wave front is formed of both small and large structures. So, the wavelet transformed image is used as a third filter to be multiplied by the original DBDI.

The soft X-ray observations are very contaminated by the scattered light of the flare. Therefore I choose a sub-field of view to cut as much as possible the flaring site, while retaining as much as possible of the wave front motion. The first wavelet transform of SXI data is chosen to remove any signal smaller than 12 pixels and larger than 70 pixels, permitting to remove a part of the scattered light due to the flare. The intensity spatial variation is then fitted to a 5 degree polynomial surface using the {\it sfit} function of IDL. The smoothed surface is removed from the filtered image to obtain an image without the scattered light emission. Finally, this image is multiplied by the original DBDI to obtain the images presented in Figs. \ref{figure3g} and \ref{figure3h}. 

All of this processing creates artifacts. The summing of pixel removes the flickering pixels but smooths the signal to the neighboring pixels, increasing the spatial expansion of the signal of the wave. The wavelet transform removes the flickering pixels and the bright points, however some bright points are clearly observed with increasing intensity during the passage of the wave front, and maybe slightly before the passage of the wave front. Therefore, applying this process removes a part of the wave front signal. Moreover, the parameters used for the wavelet transform enhance the interference fringes, due to the grill supporting the filters, which could not be removed by the cleaning functions of SSWIDL. The fringes appear as horizontal and vertical straight lines that overlay and perturb the signal of the wave front. Finally, the 5$^{th}$ degree polynomial surface fitting the spatial variation of the intensity in SXI, oscillates close to the border of the field of view, and does not fit the signal very well. After the subtraction of the fitted polynomial surface as a background, the image shows some artificial increase or decrease of the brightness close to the border of the sub-field of view. The highly processed images are not used to obtain the plots in Fig. \ref{figlightcurve}. All the structures described in the following section are verified in the raw observations, to be sure that they are not artifacts of the processing. For comparison between the DBDI and the processed images, I present them both. All the images in Figs. \ref{figure1}, \ref{figure2} and \ref{figure3} are co-aligned and have the same field of view, showing the western half solar disc. 
The images with a field of view above the limb (see Figs. \ref{figure2ter} and \ref{figure3c}) are also co-aligned and are rotated so that extended structures in altitude appear to be vertical. In the same way, the on-disc subfield-of-view image (see Figs. \ref{figure2quater}, \ref{figure3g} and \ref{figure3h}) are all co-aligned and show the same field of view.

Several movies of the observations are provided on line to make 
their description clearer to the reader. All the images used in the movies are DBDIs, have the same field of view and are separated by spectral emission lines.

The LASCO images 
are difference images obtained by subtracting a pre-event 
image. They are overlaid with the closest in time EIT DBDI. 

\section{Analysis}
\subsection{Phenomenology}
\begin{figure*}
\includegraphics[width=8cm]{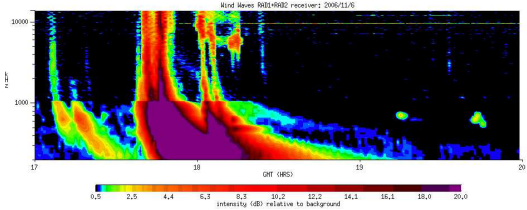}
\caption{Type II radio emission obtained at Sagamore Hill 
Observatory. The radio emission begins at 17:39 UT.}
\label{figradio}
\end{figure*}

\begin{figure*}
\includegraphics[width=2.98cm]{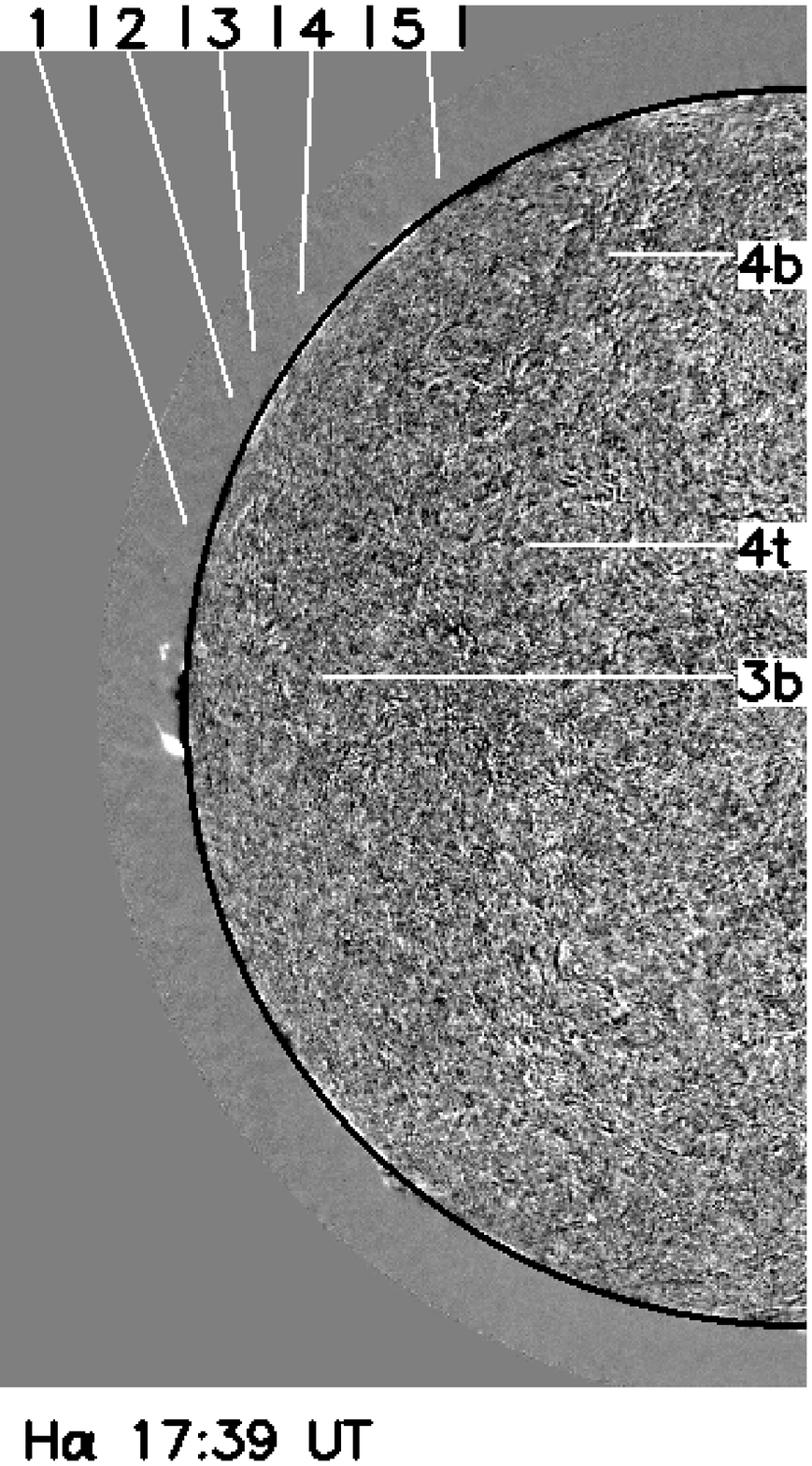}
\includegraphics[width=2.98cm]{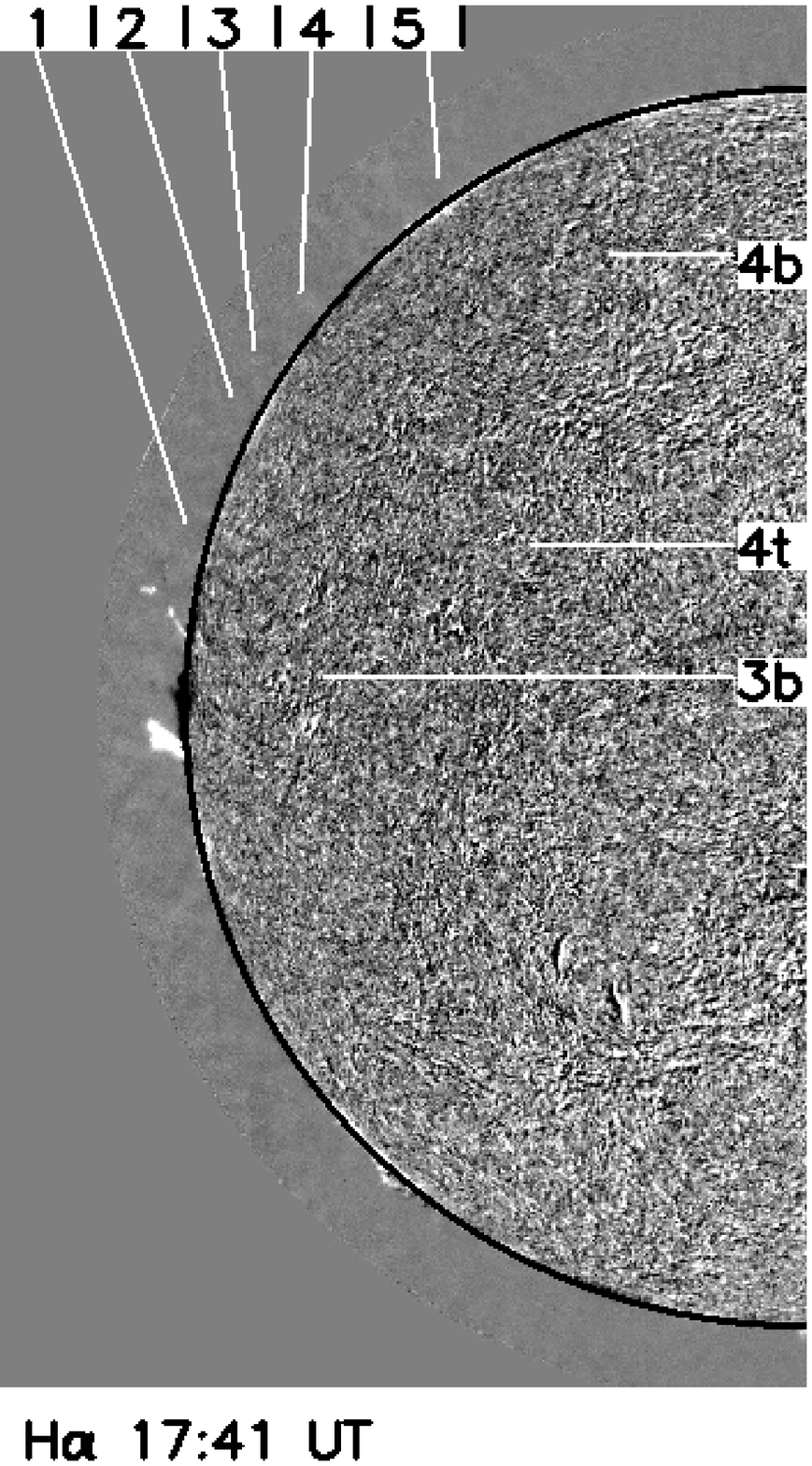}
\includegraphics[width=2.98cm]{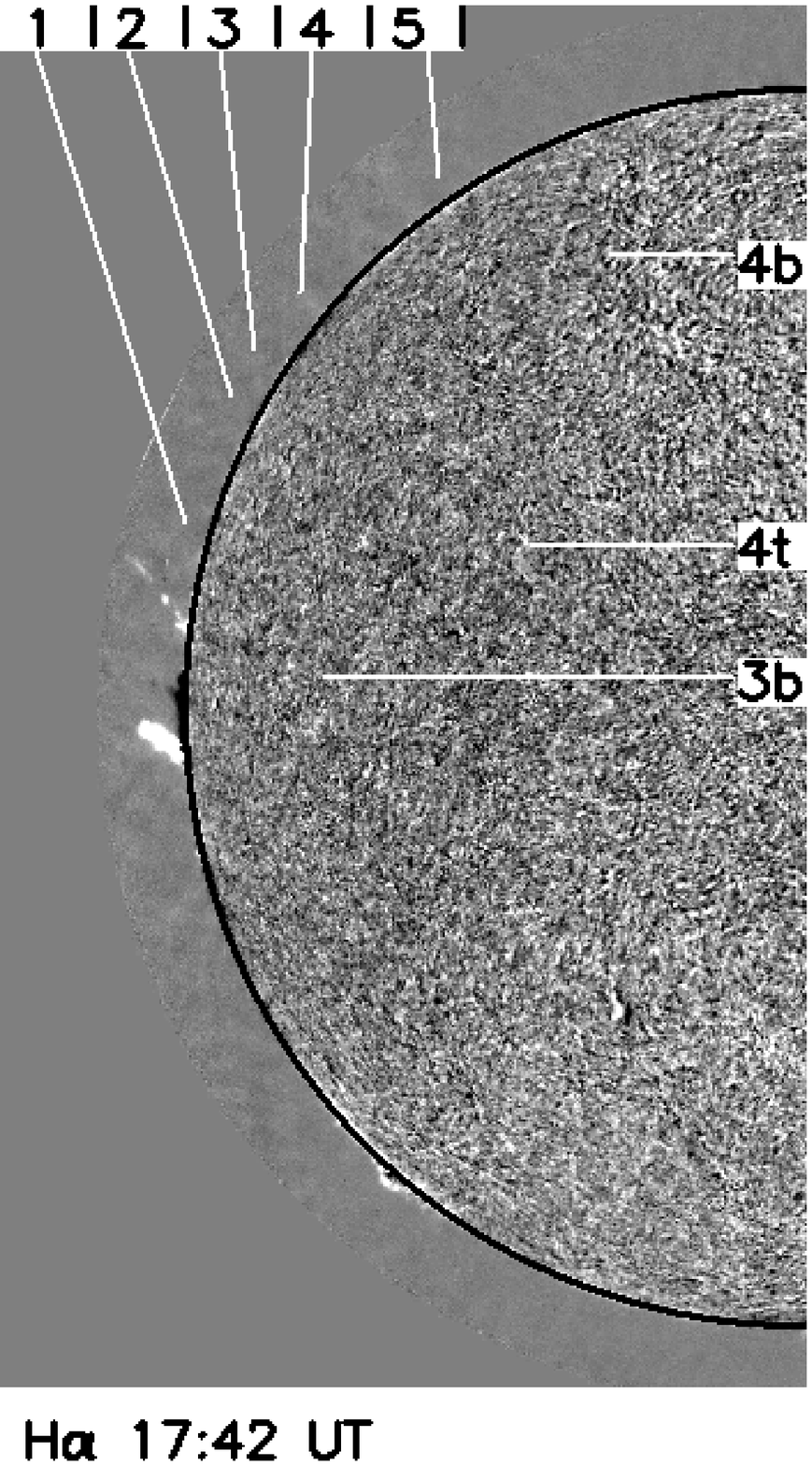}
\includegraphics[width=2.98cm]{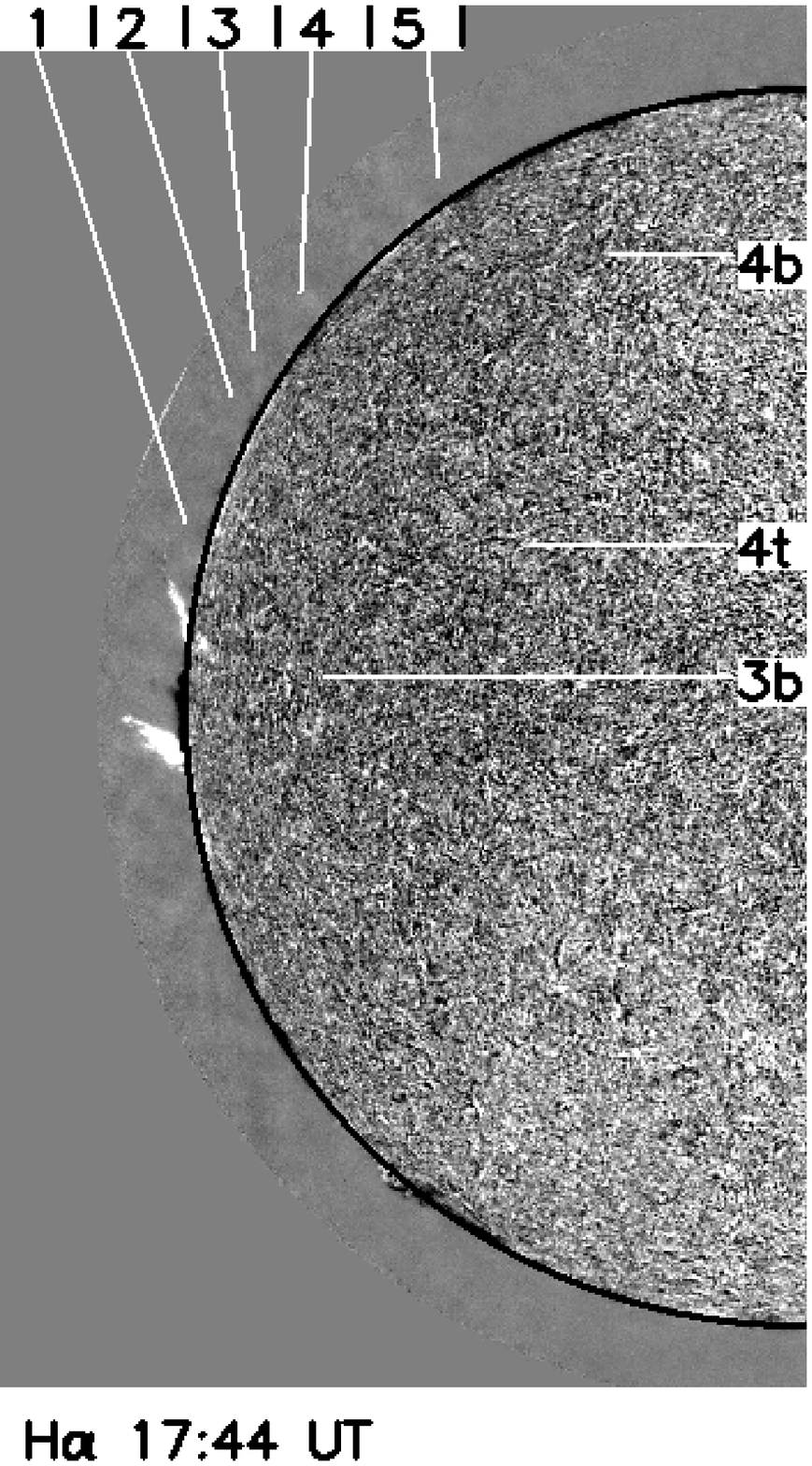}
\includegraphics[width=2.98cm]{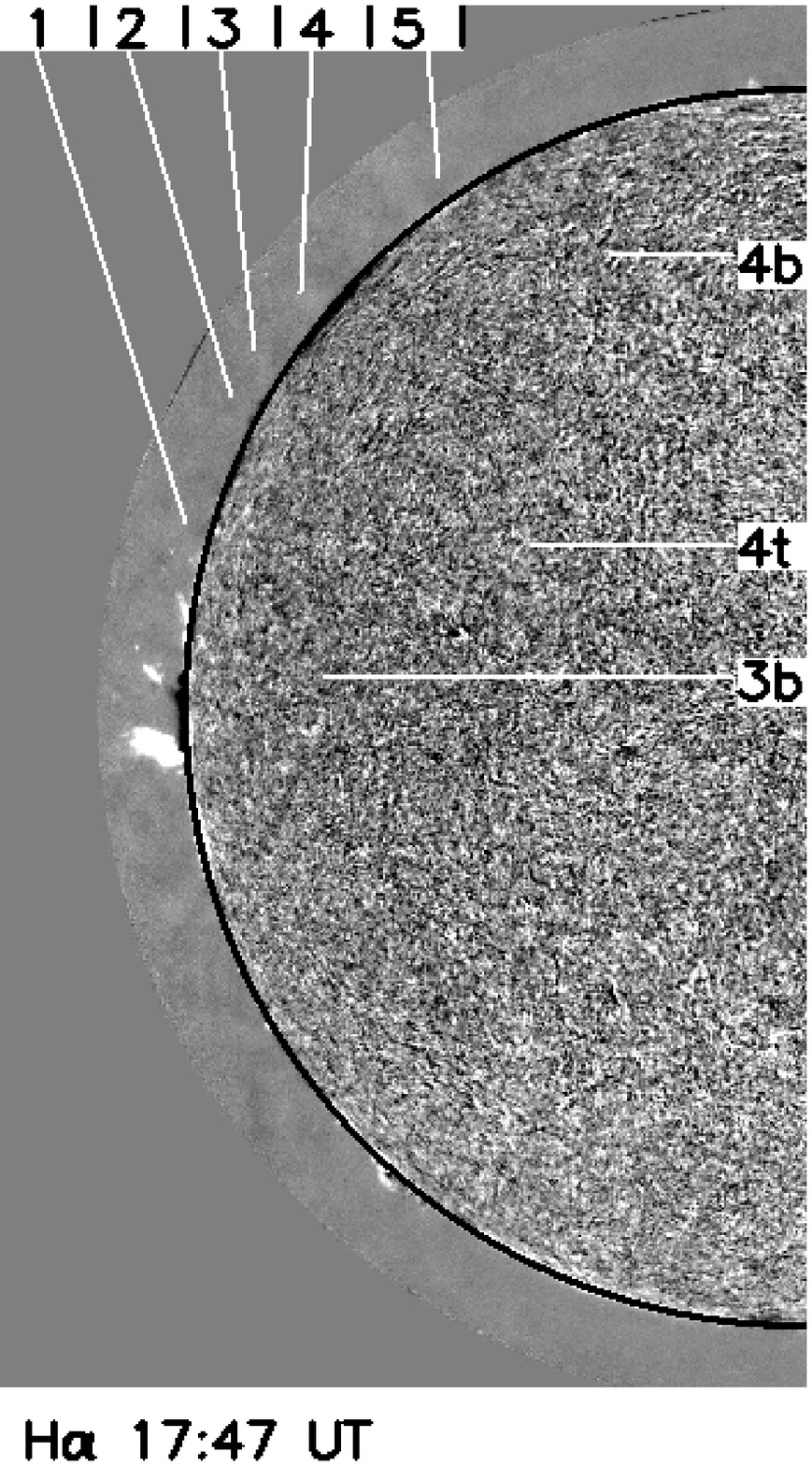}
\includegraphics[width=2.98cm]{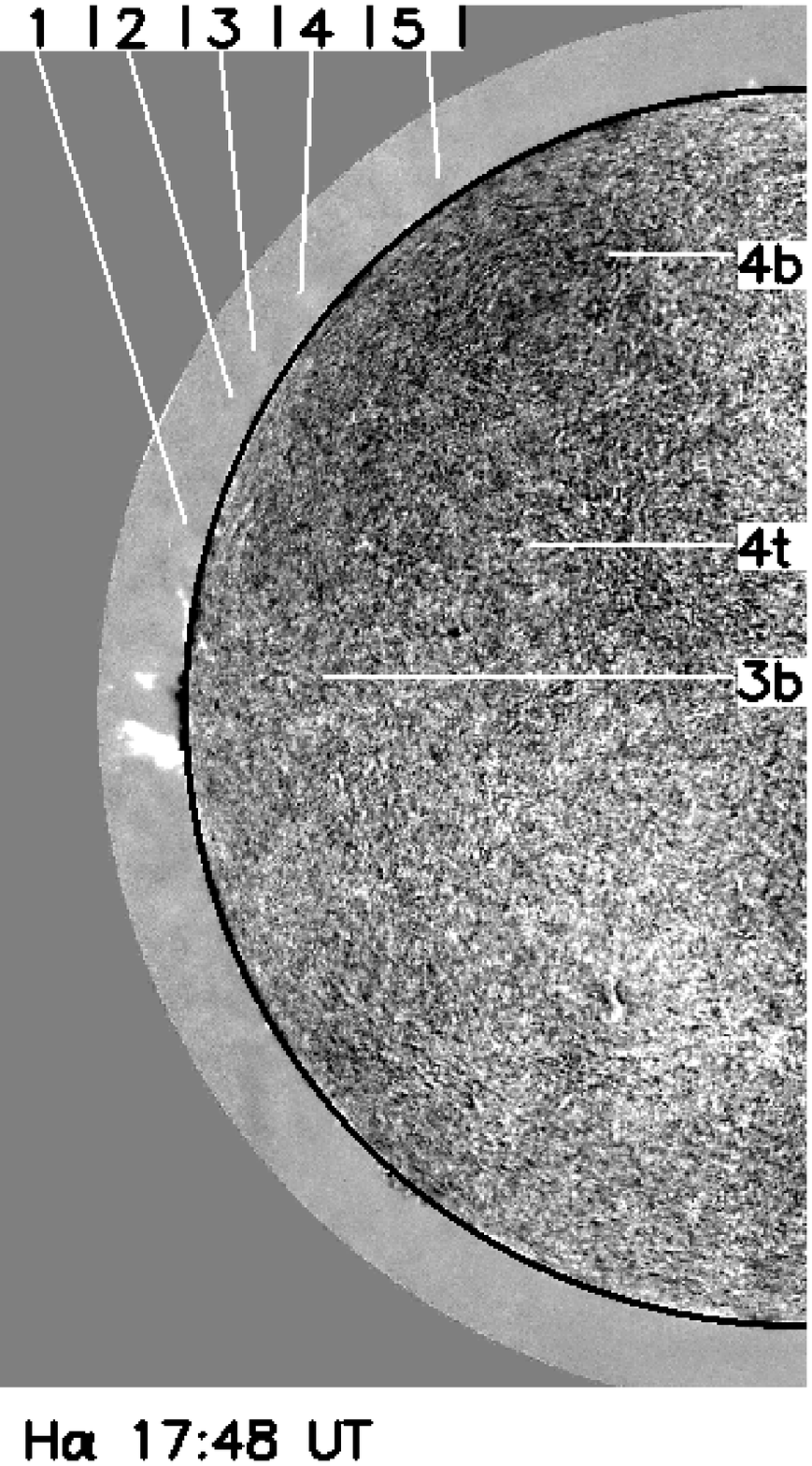}
\caption{DBDIs obtained by the OSPAN instrument using a large band 
pass filter centered on the H$\alpha$ spectral emission line. 
The numbers indicate the successive appearance of the 
wave in SXI. The field of view contains half of the solar disc and is overlaid to the field of view of images in Figs. \ref{figure2} and \ref{figure3}. A small prominence is erupting from behind the limb at 
17:40 UT. One small brightness is visible at the position 1 
from 17:44 UT to 17:47 UT. Except for this brightening, no Moreton wave 
is visible neither on the limb nor on the disc.}
\label{figure1}
\end{figure*}

On November 6, 2006, the wave is observed together with a C8.8 
flare and a CME traveling through the LASCO/C2 field of view at 
a speed of 976 km s$^{-1}$ (according to CACTUS/SIDC software). 
The flare strength determined by the soft X-ray emission intensity is likely a minimum as the source region (likely AR 0923) is located slightly behind the east limb in the southern hemisphere (S06). The GOES 13
 satellite detects the beginning of the flare at 17:43 UT, but radio emission is detected by the Sagamore Hill Observatory as early as 17:39 UT (Fig. \ref{figradio}). The H$\alpha$ images show that a small prominence is ejected 
into the corona at the location of the flare visible in SXI and in EIT (see Fig. \ref{figure1} and the H$\alpha$ movie). 
The source region of the eruption may be a magnetic bipole located in AR 0923. Correcting the solar rotation from November 10th 12:04 UT to November 6th 17:17 UT, I estimate that the active region is located approximately 15$^\circ$ behind the eastern limb at the time of the eruption. Of course, this method of estimation of the location of the active region behind the limb cannot take into account the migration, the emergence and the distortion of the active region that can shift its location in the 4 days delay between the observation and the date by which the active region is visible on the solar disc.

\begin{figure*}

\includegraphics[width=2.98cm]{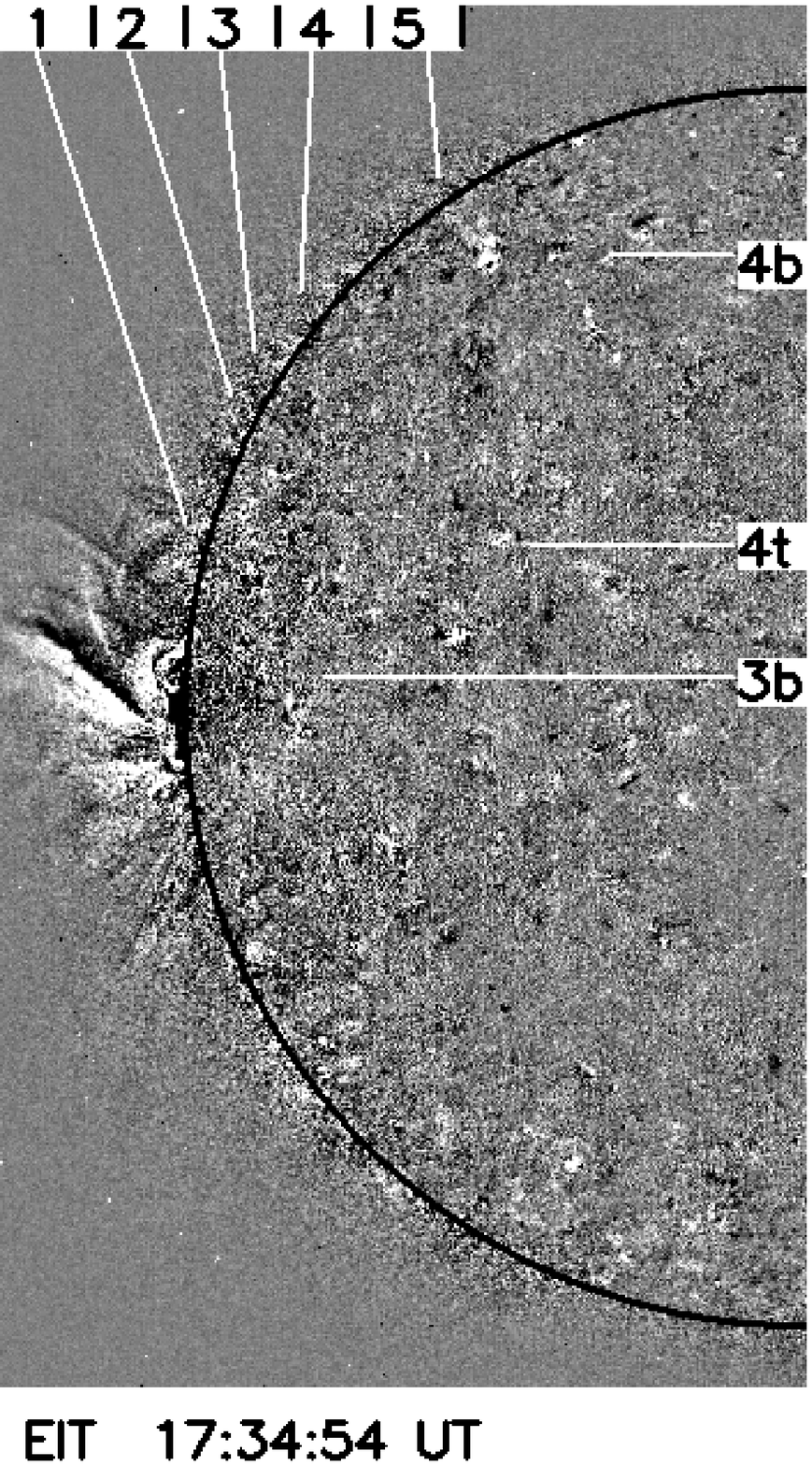}
\includegraphics[width=2.98cm]{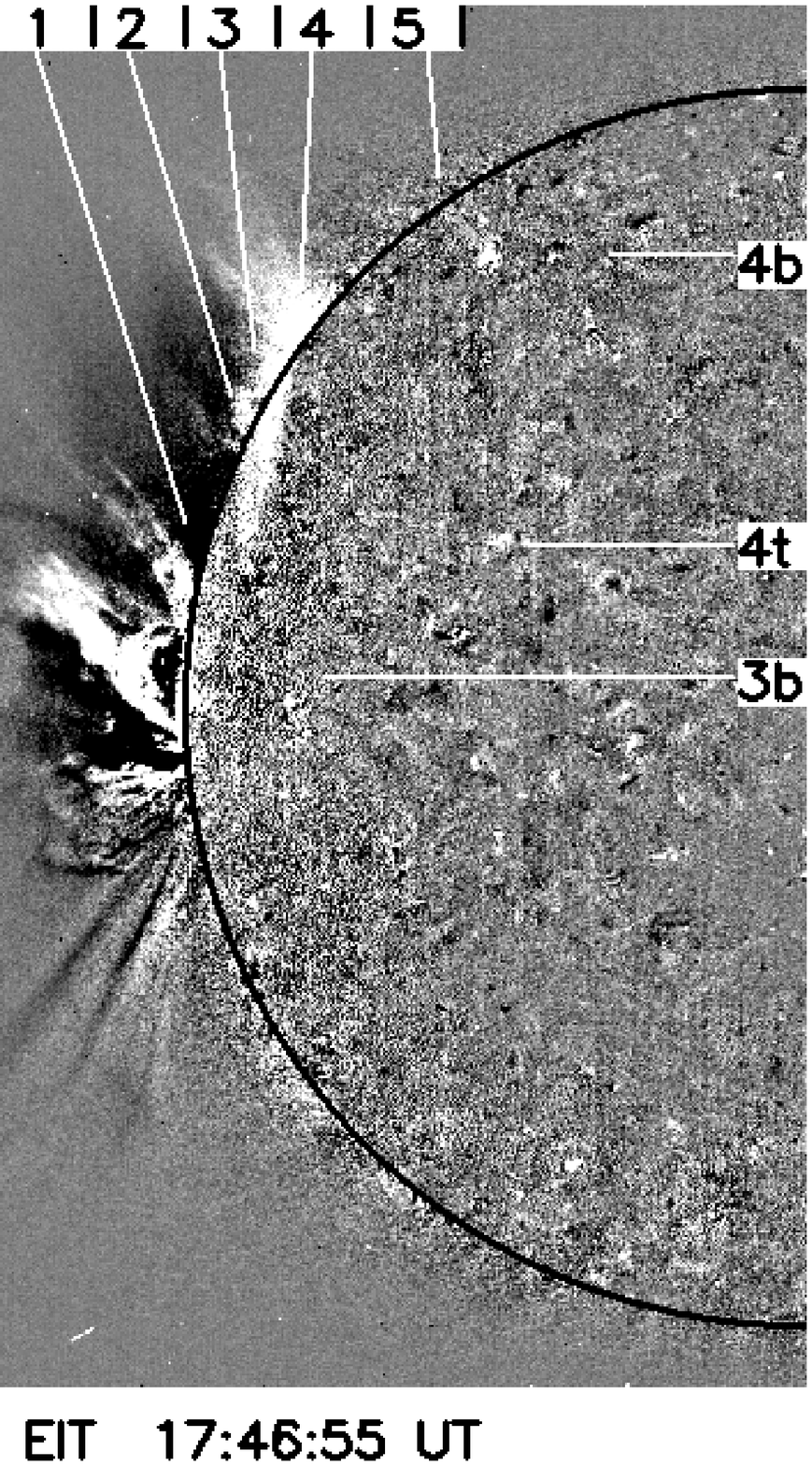}
\includegraphics[width=2.98cm]{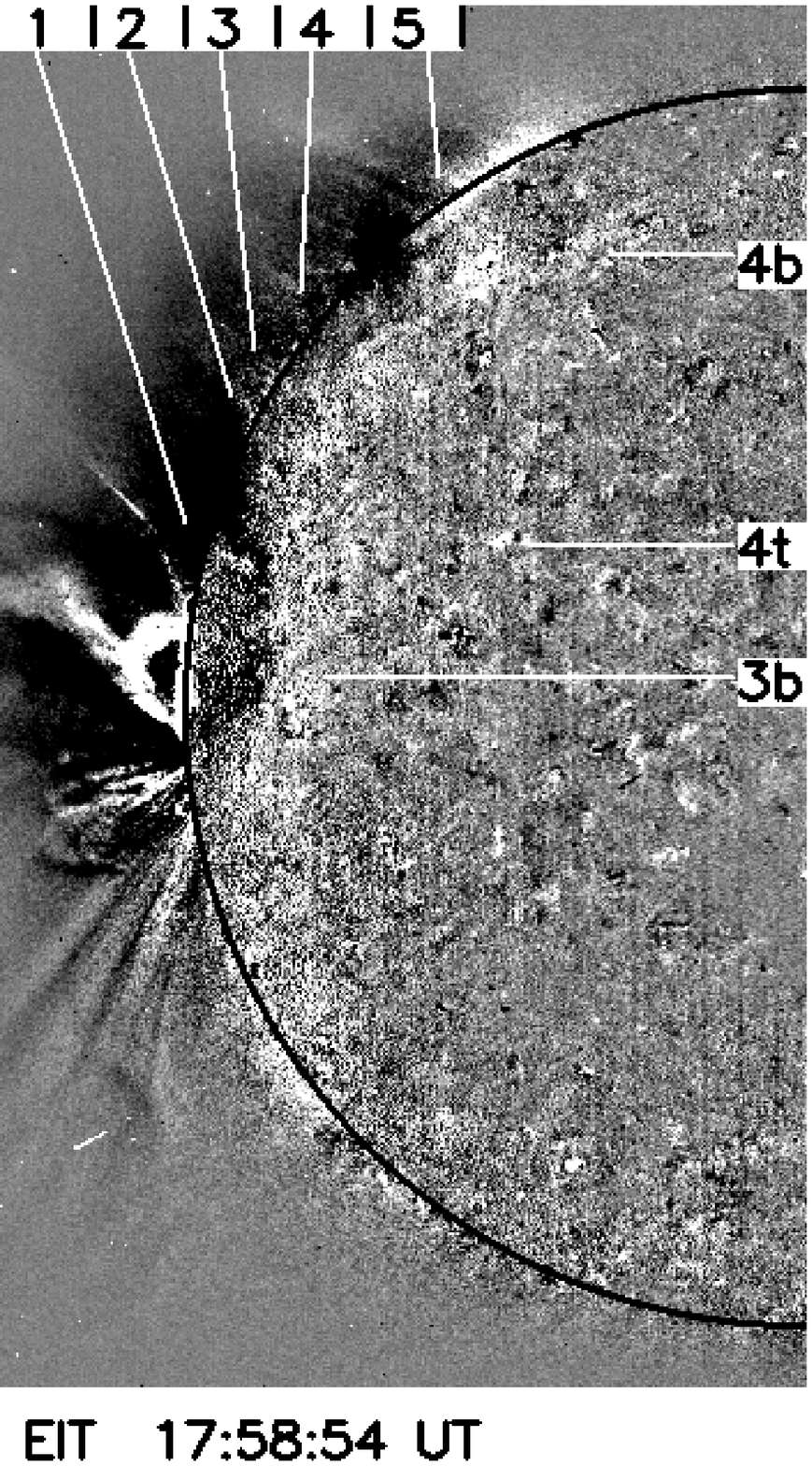}
\includegraphics[width=2.98cm]{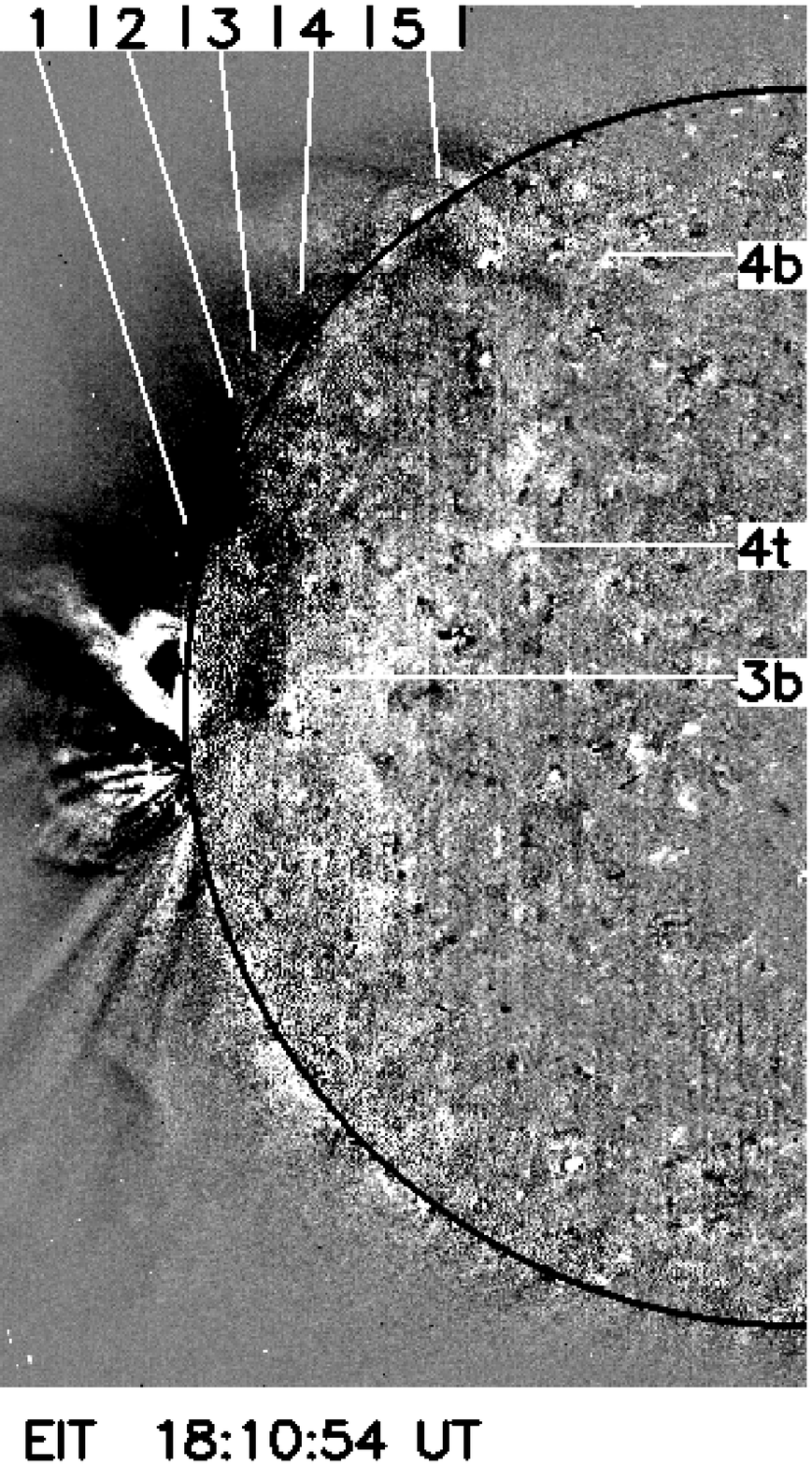}
\includegraphics[width=2.98cm]{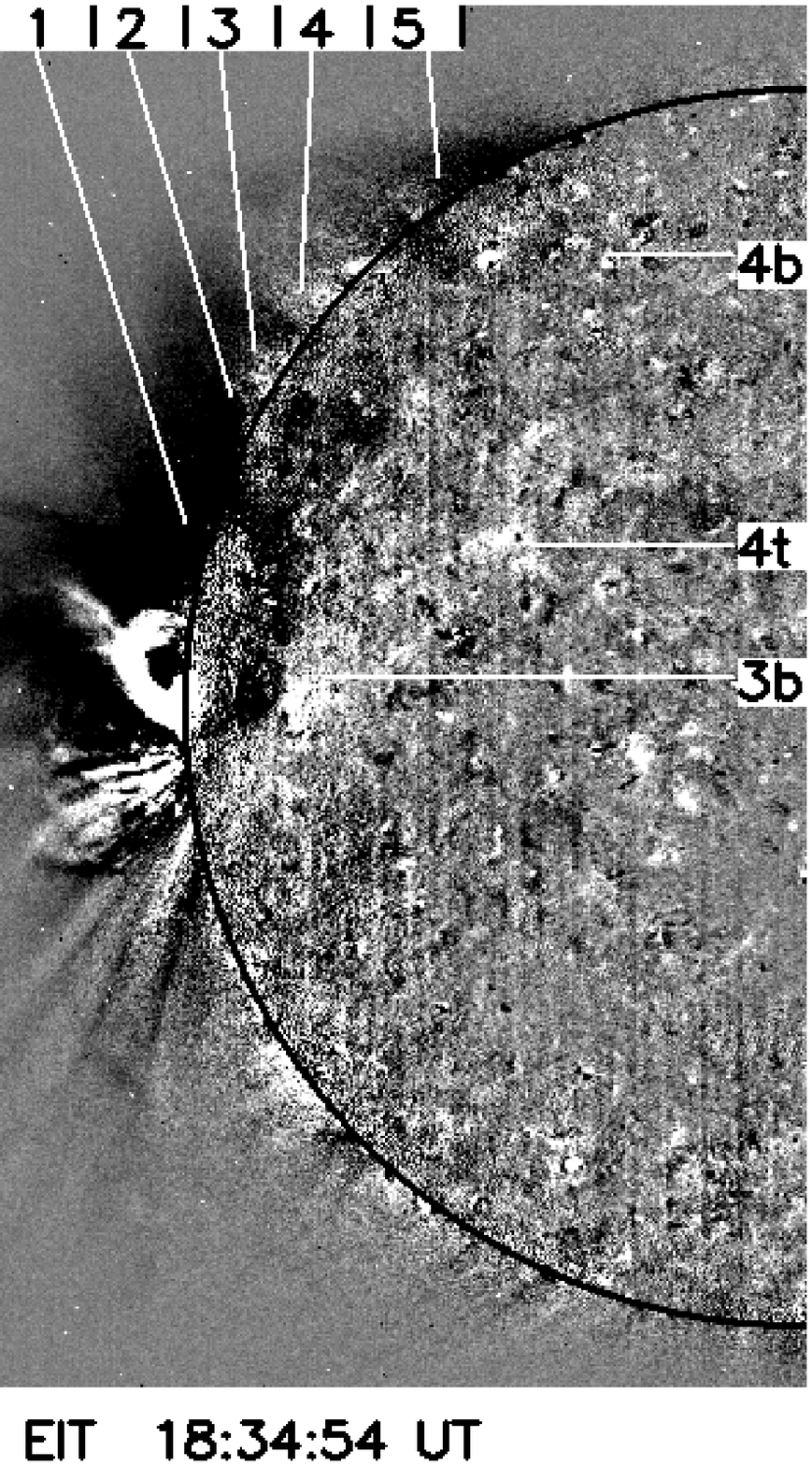}
\includegraphics[width=2.98cm]{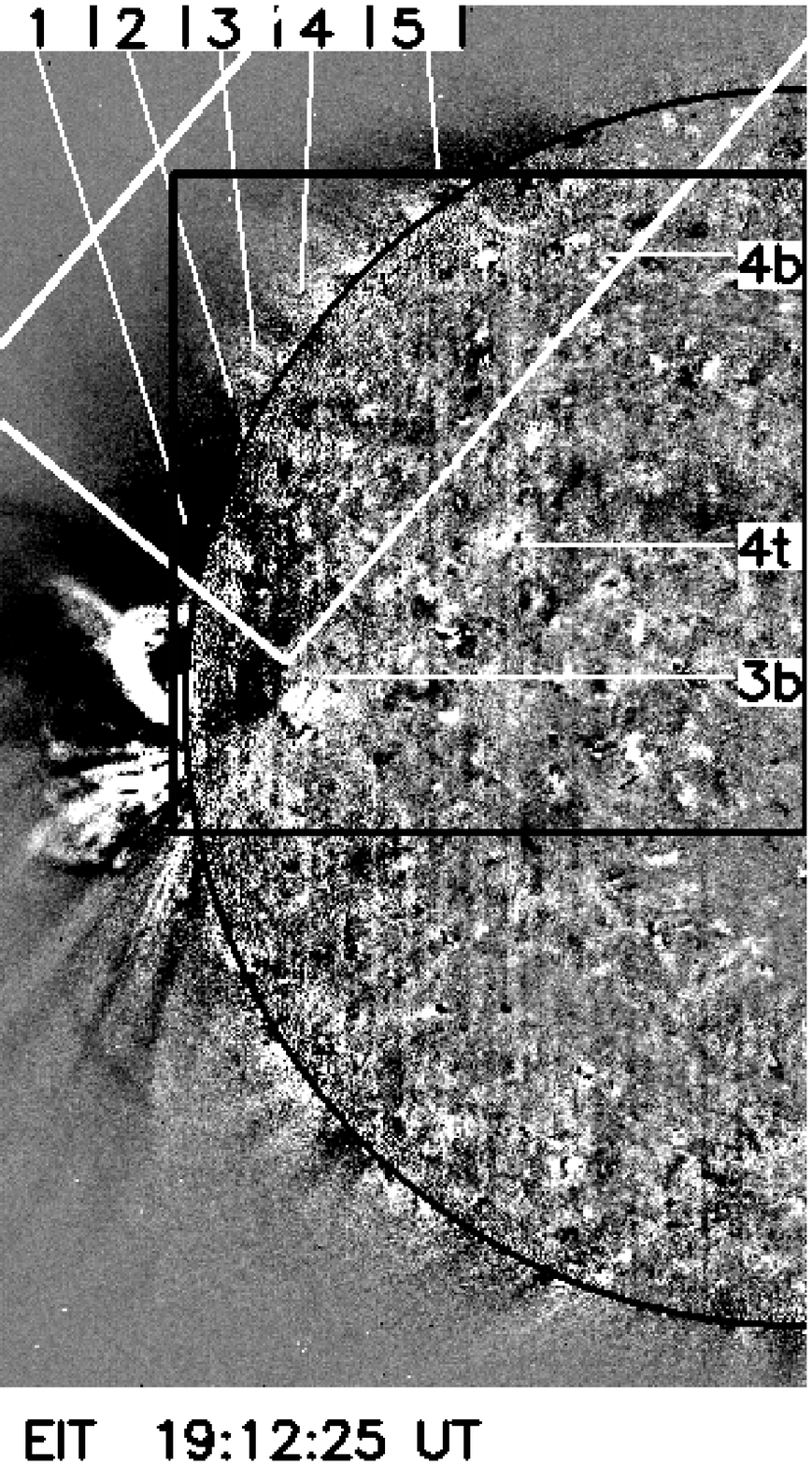}
\caption{DBDIs obtained using the 195 \AA~filter of EIT. Half 
of the solar surface is shown. The numbers indicate the successive 
appearance of the wave in SXI. The field of view contains half of the solar disc and is overlaid to the field of view of images in Figs. \ref{figure1} and \ref{figure3}. The large white rectangle on the last panel
shows partly the sub-field of view of the images displayed in Fig. \ref{figure2ter}. The large black rectangle on the last panel shows the sub-field of view of the images displayed in Fig. \ref{figure2quater}. The 
EIT wave is observed progressing from an active region located slightly 
behind the limb to the northern coronal hole where it stops. The last 
position of the EIT wave above the limb is very darkened after 18:10 UT. On the disc, the EIT wave progresses from the active region to the east, and stops at bright point pointed by labels.}
\label{figure2}
\end{figure*}

The wave is clearly visible in two band passes: in Fe {\sc xii} 
and in soft X-ray; in H$\alpha$, it is possibly observed at only one location. When it is observed at approximately the same time 
in EIT and SXI the wave front is almost 
co-spatial (see detailed description of the observations here after, and Fig. \ref{figure3g}). Therefore, I believe that this set of data 
presents multi-wavelength observations of the same structure. The wave moves from the southern hemisphere across the equator to the northern hemisphere and 
extends onto the disc toward the central meridian. The wave seems to originate from the flare site even if its propagation is not isotropic around 
this location. Between the flare and the wave front, a dimmed 
region develops (e.g. Figs. \ref{figure2}, \ref{figure3}).

I now compare the three sets of observations in each band pass in detail. The H$\alpha$ data does not show very much of the 
wave. A very small brightening appears at 17:44 UT at location 1 on Fig. \ref{figure1}. It has faded out at 17:47 UT. No motion of this 
brightness is visible. Therefore, it could be a stationary brightening of the Moreton wave. However, the differencing processing reveals the earth atmospheric turbulence effect on the image that appears as flickering bright and dark patches at the limb. So, the brightening appearing at location 2 in H$\alpha$ could also be a seeing artifact revealed by the difference of images technique. I checked in the images obtained in the red and blue wings of the H$\alpha$ spectral line if any signal of a Moreton wave could appear, and did not find any.

\begin{figure*}

\includegraphics[width=3cm]{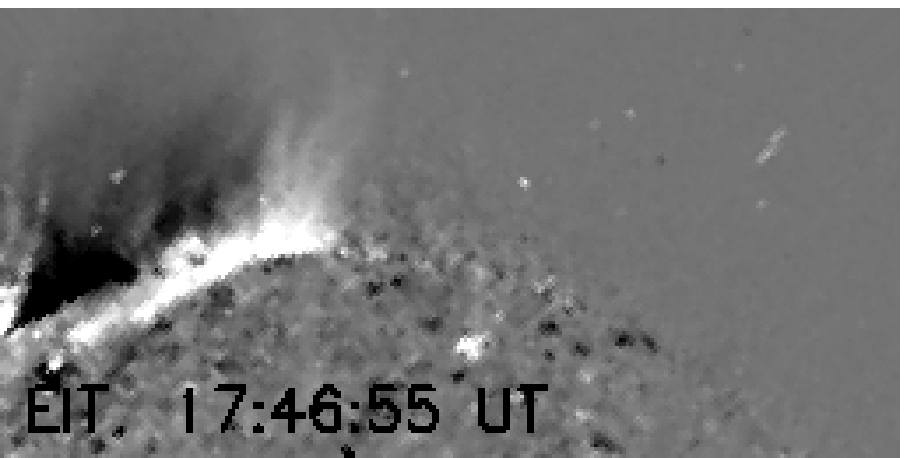}
\includegraphics[width=3cm]{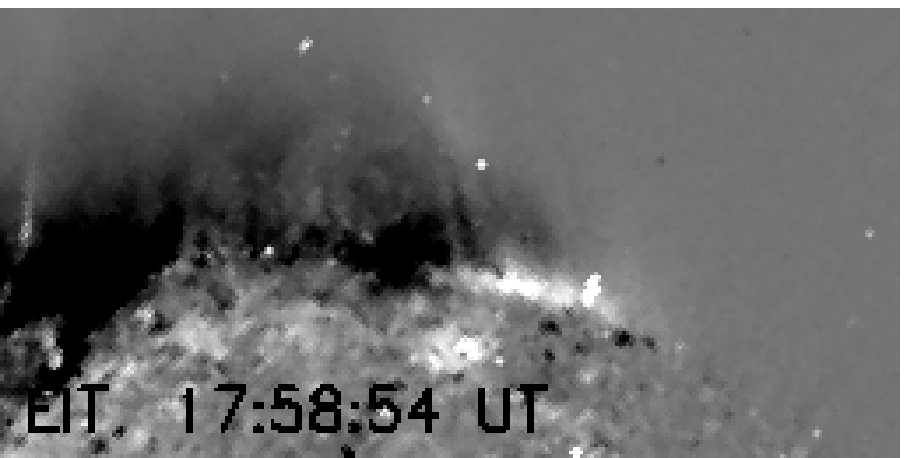}
\includegraphics[width=3cm]{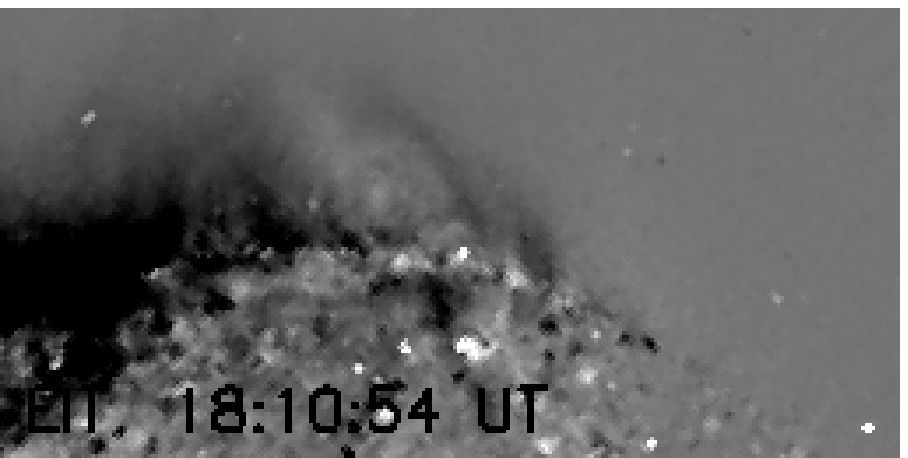}
\includegraphics[width=3cm]{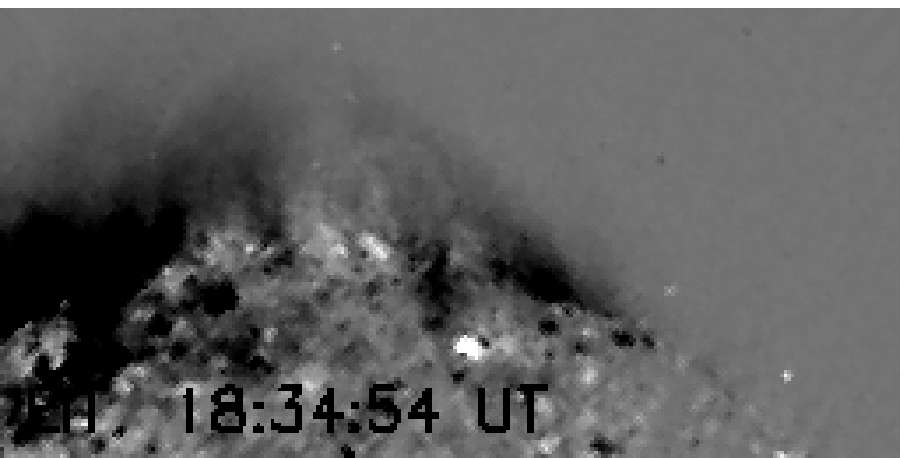}
\includegraphics[width=3cm]{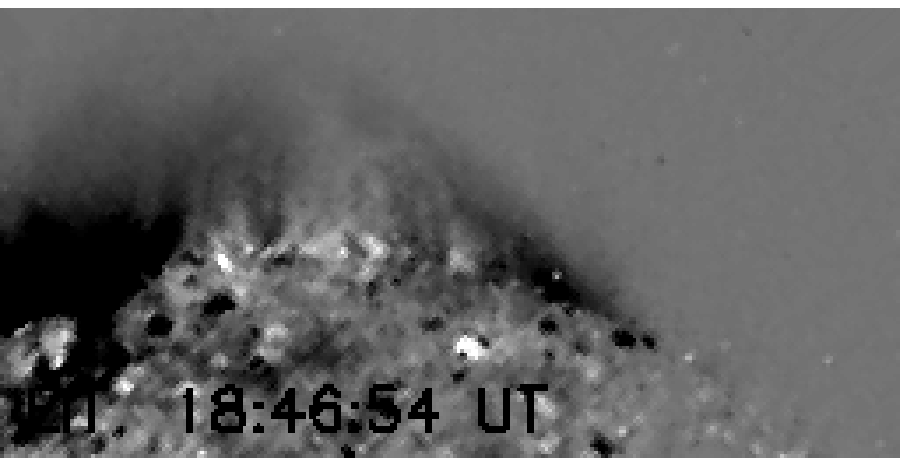}
\includegraphics[width=3cm]{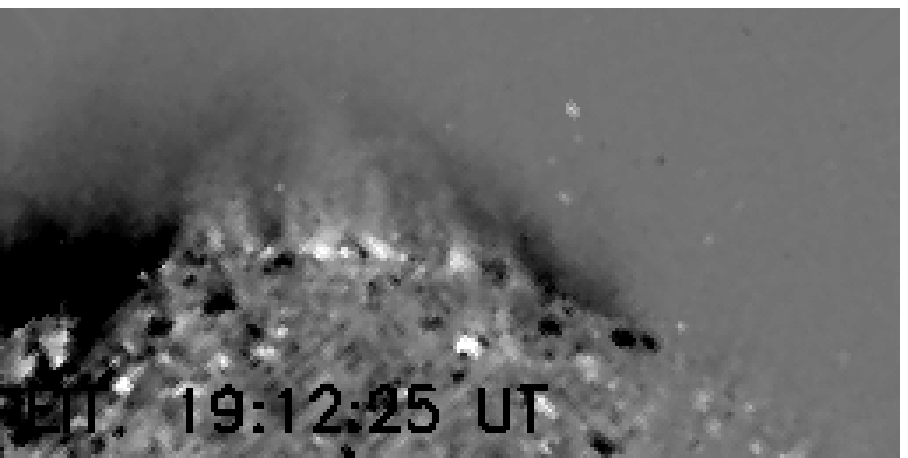}
\includegraphics[width=3cm]{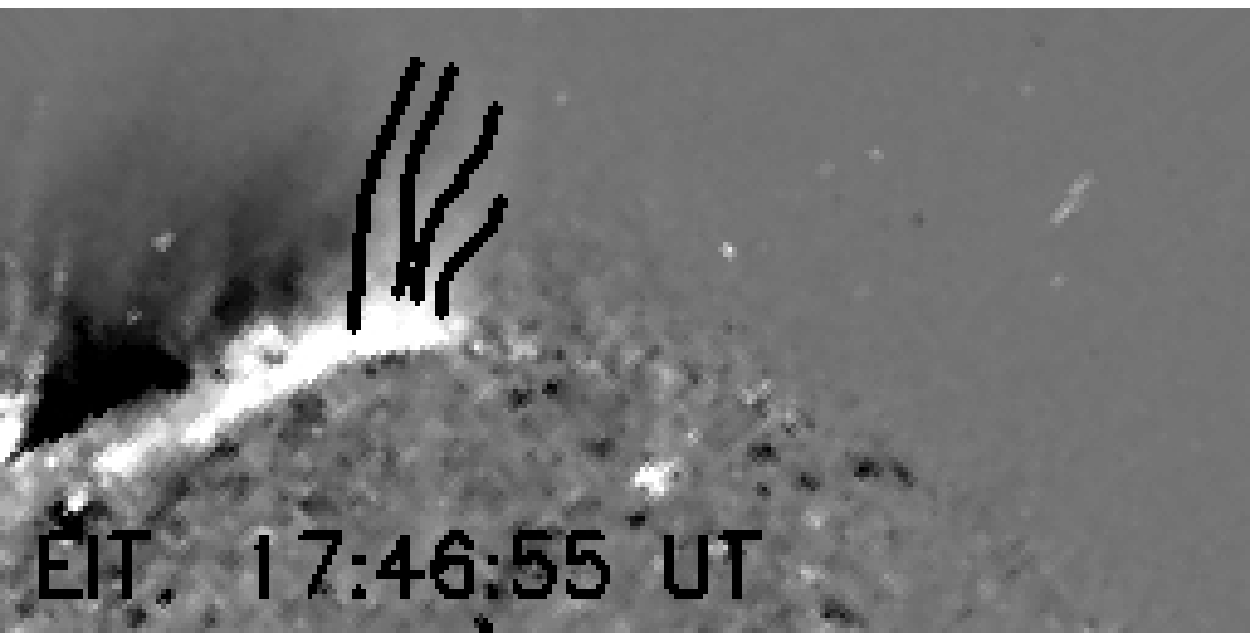}
\includegraphics[width=3cm]{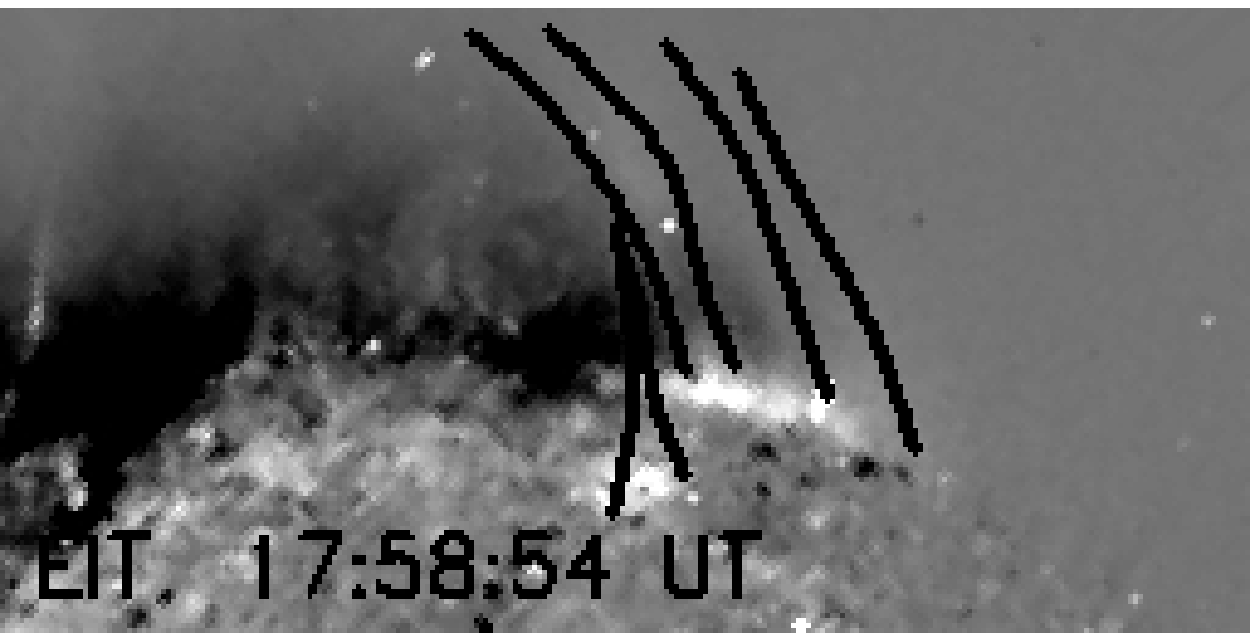}
\includegraphics[width=3cm]{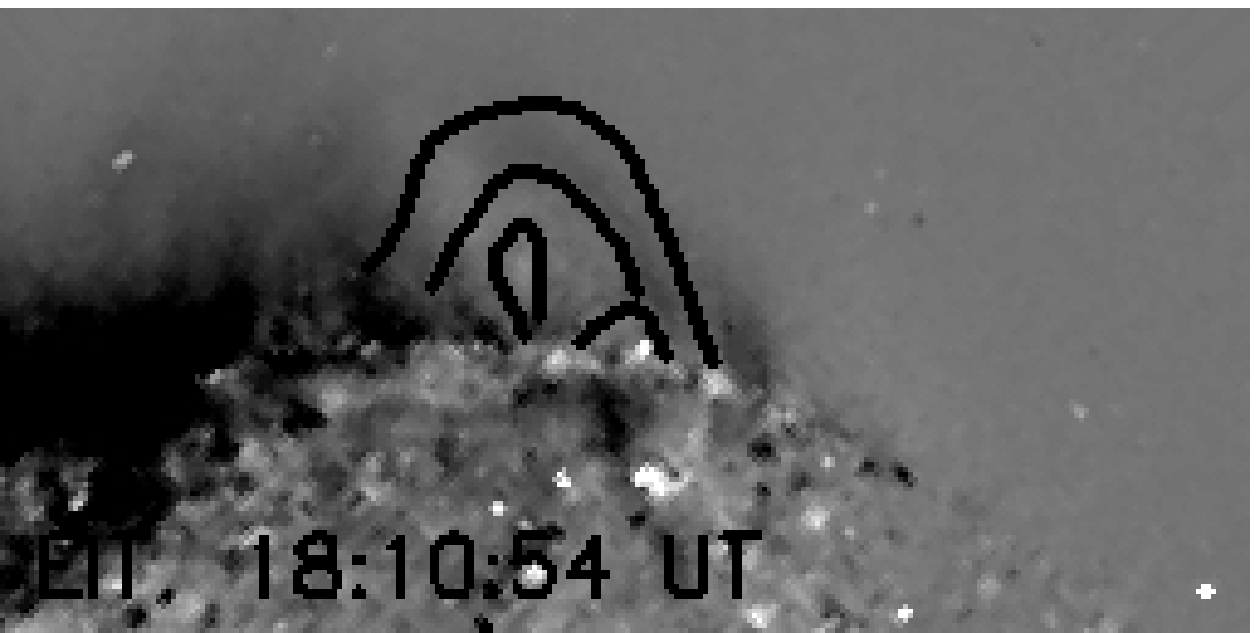}
\includegraphics[width=3cm]{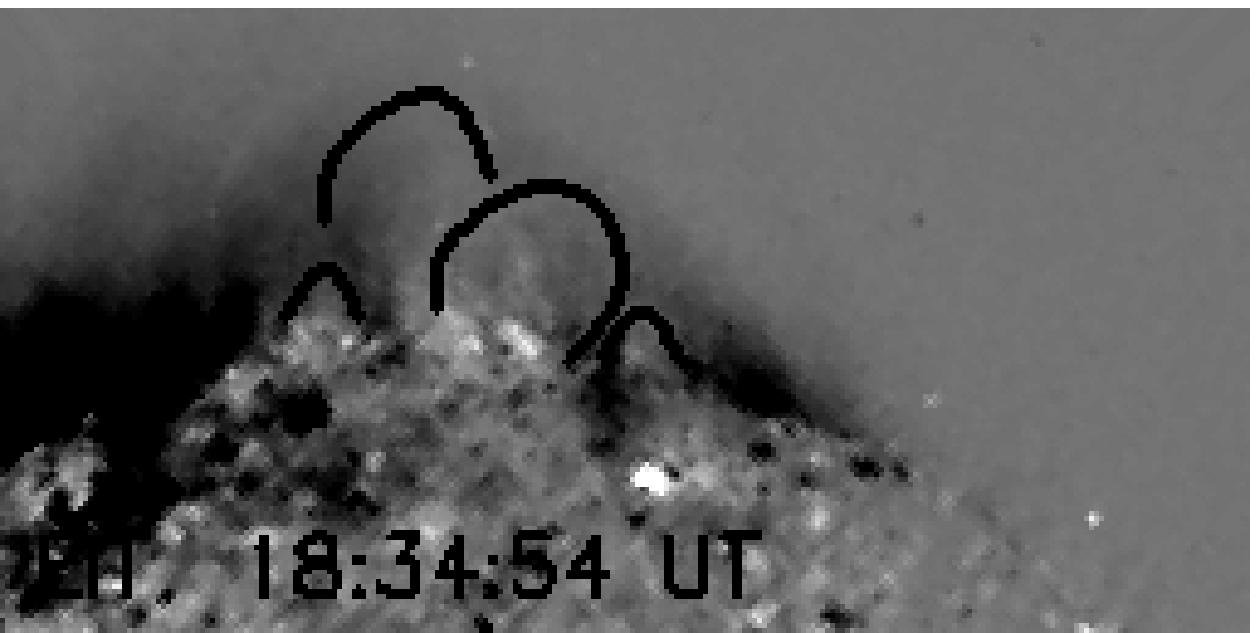}
\includegraphics[width=3cm]{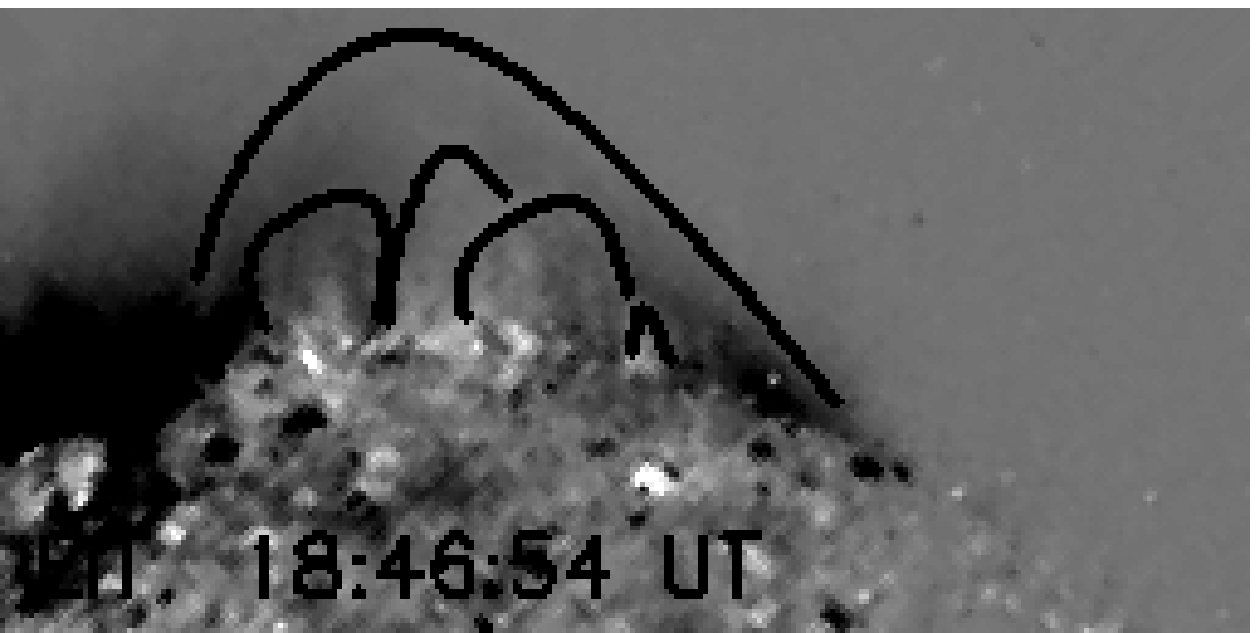}
\includegraphics[width=3cm]{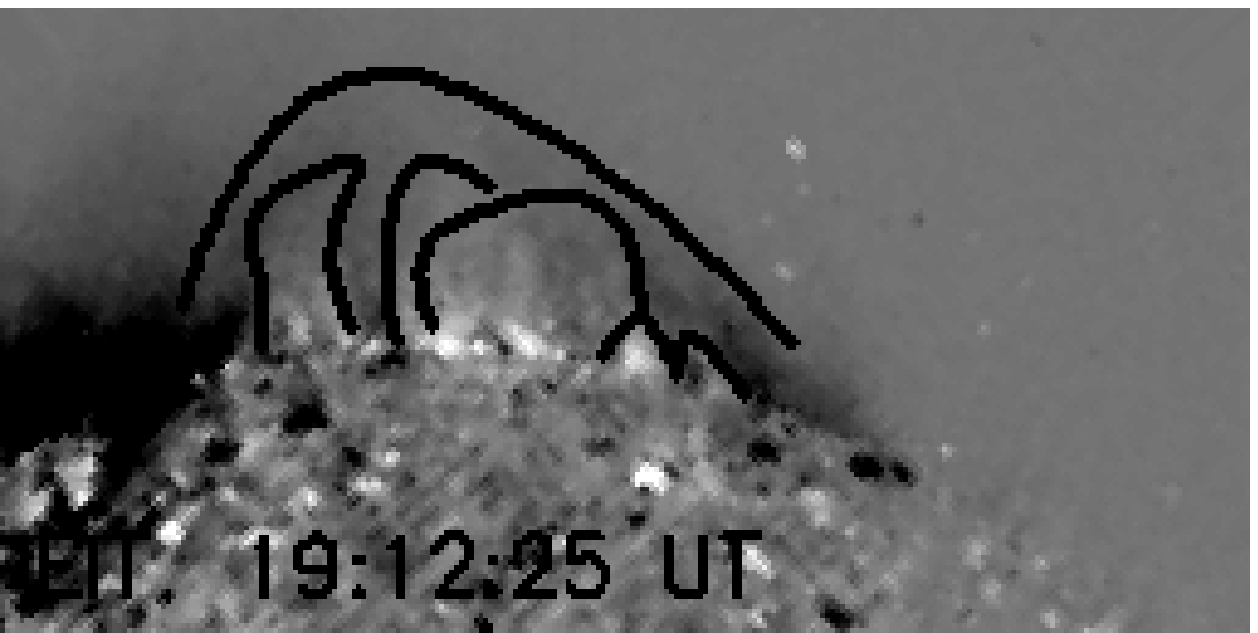}
\caption{Sub-field of view, shown in the white rectangle in the last panel of the Fig. 
\ref{figure2}, focusing on the over-limb structures observed in 195 \AA. The images are highly processed (see the text). The second line shows the same images as in the first line but with some over-limb structures highlighted with black drawings.}
\label{figure2ter}
\end{figure*}

\begin{figure*}
\includegraphics[width=3cm]{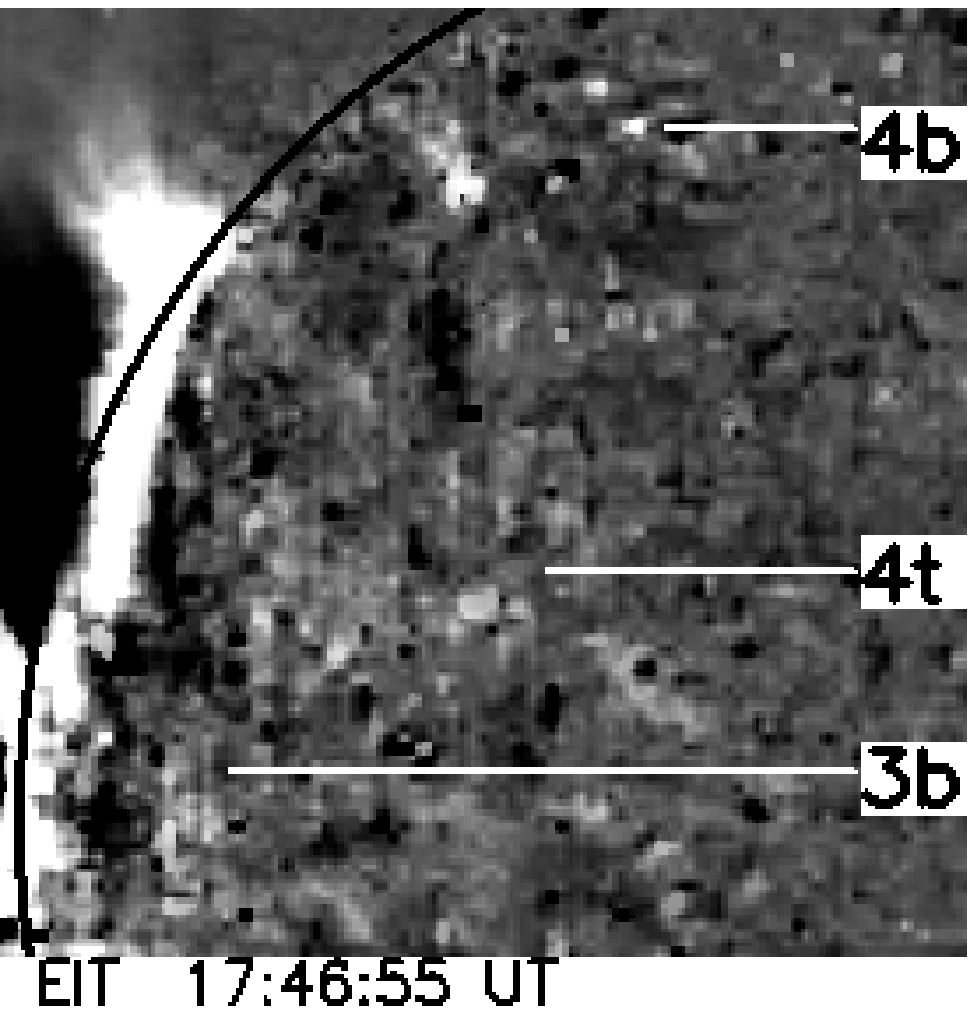}
\includegraphics[width=3cm]{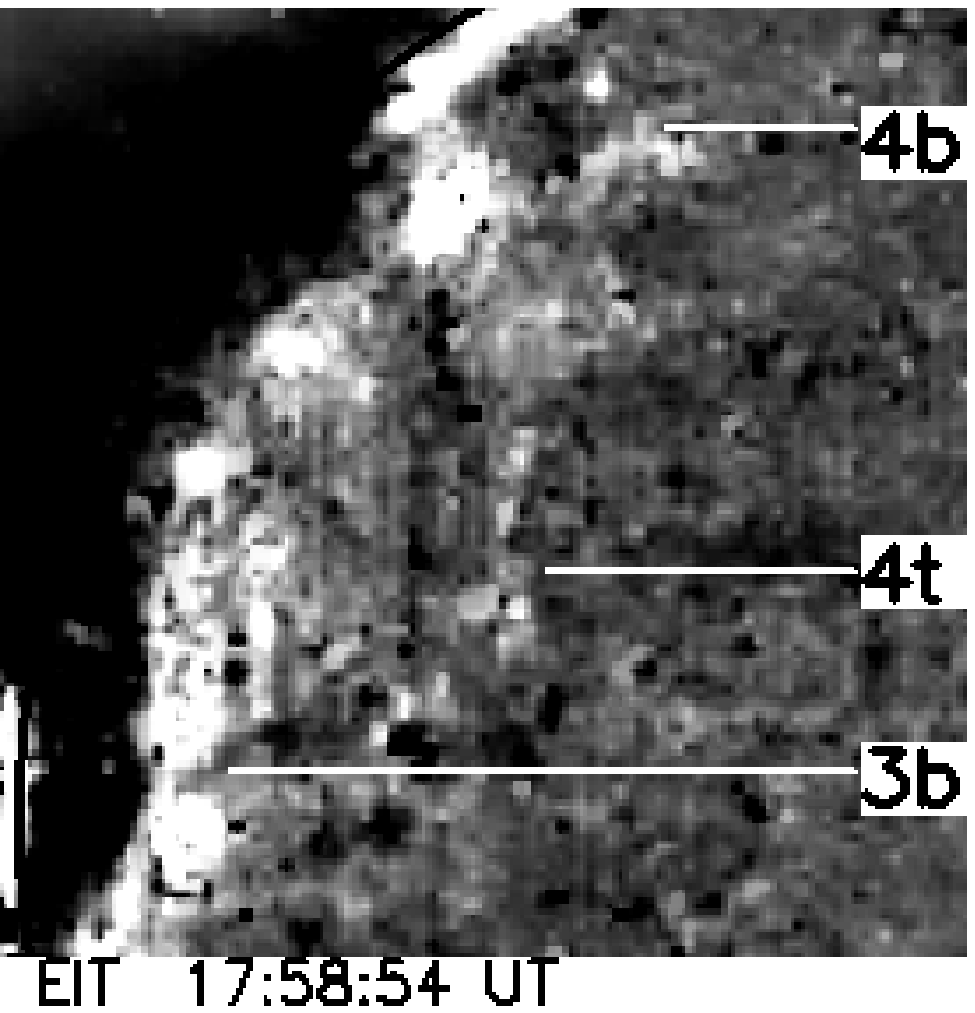}
\includegraphics[width=3cm]{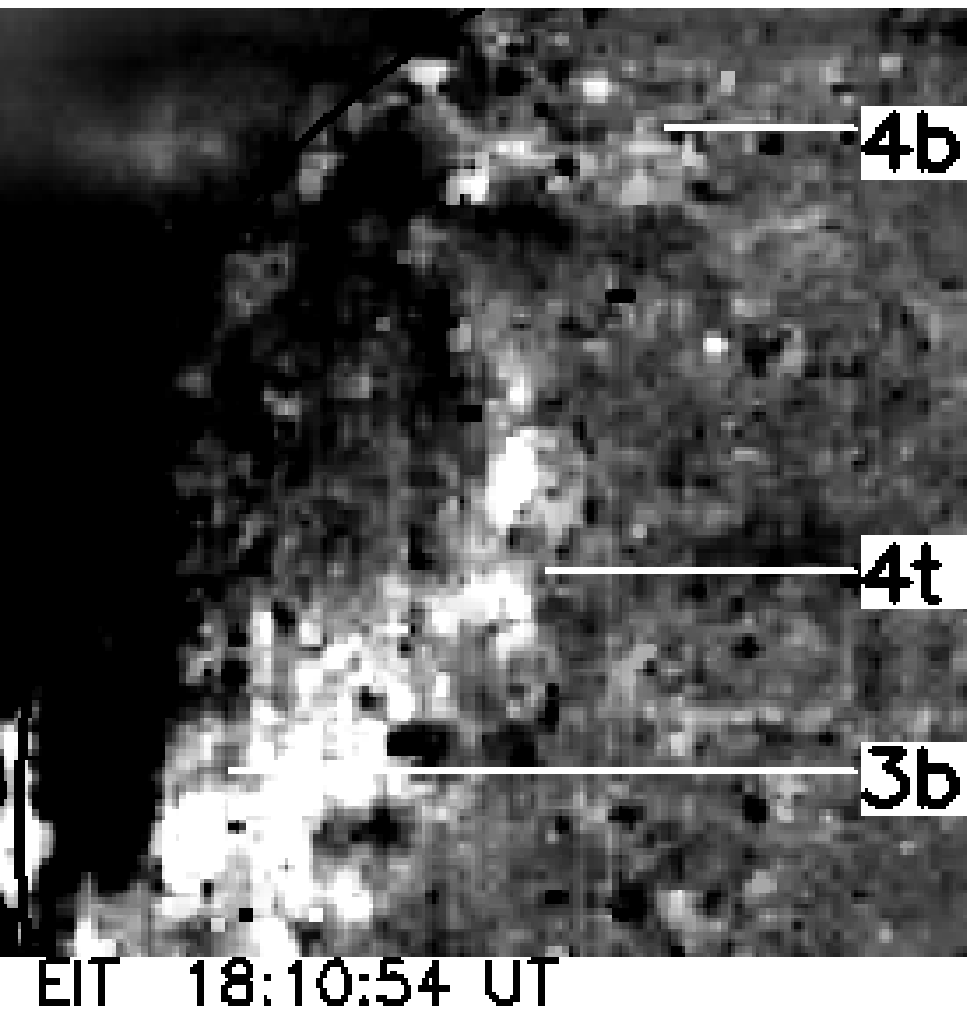}
\includegraphics[width=3cm]{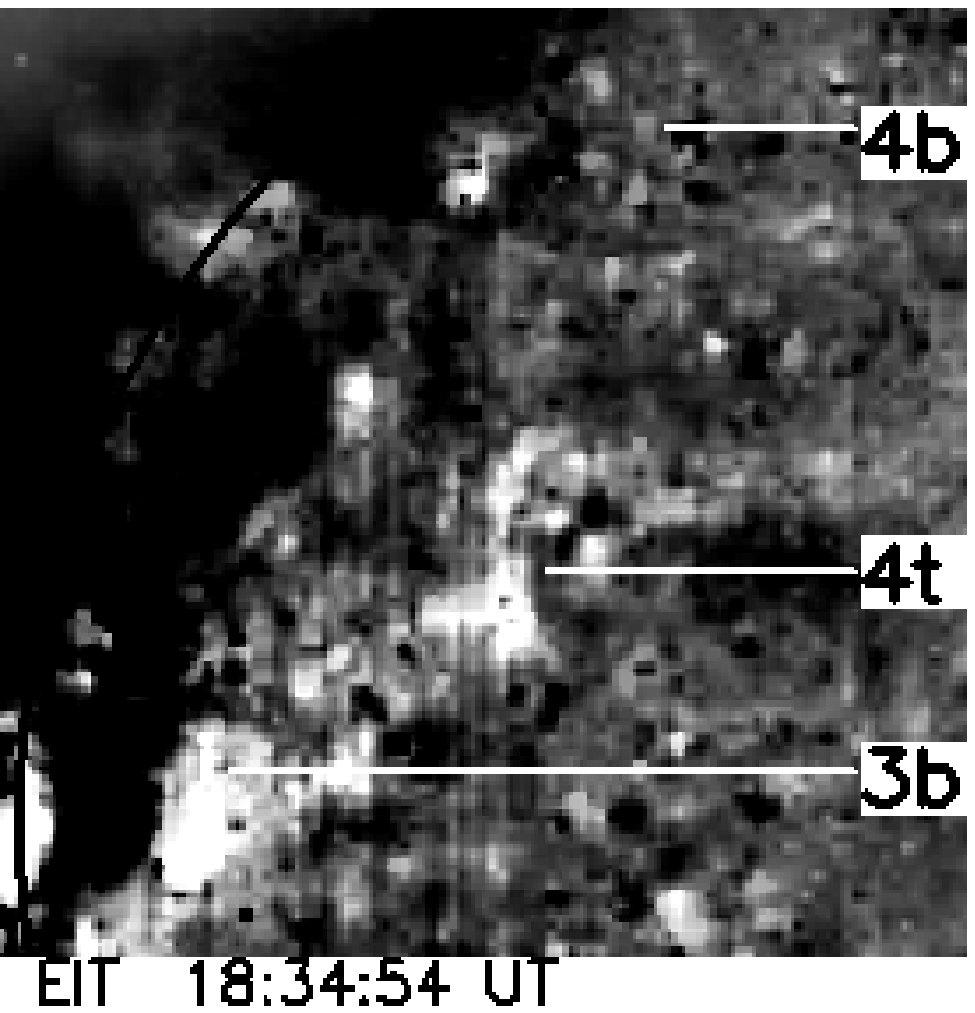}
\includegraphics[width=3cm]{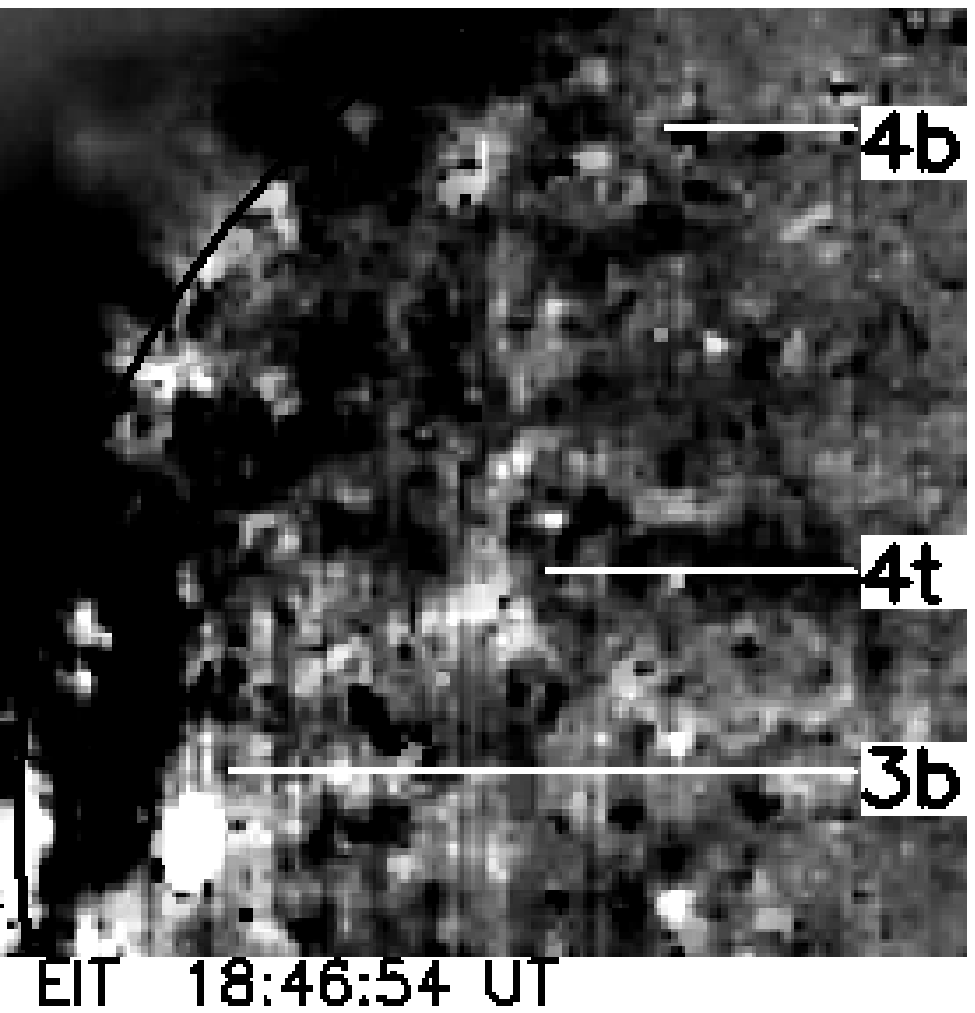}
\includegraphics[width=3cm]{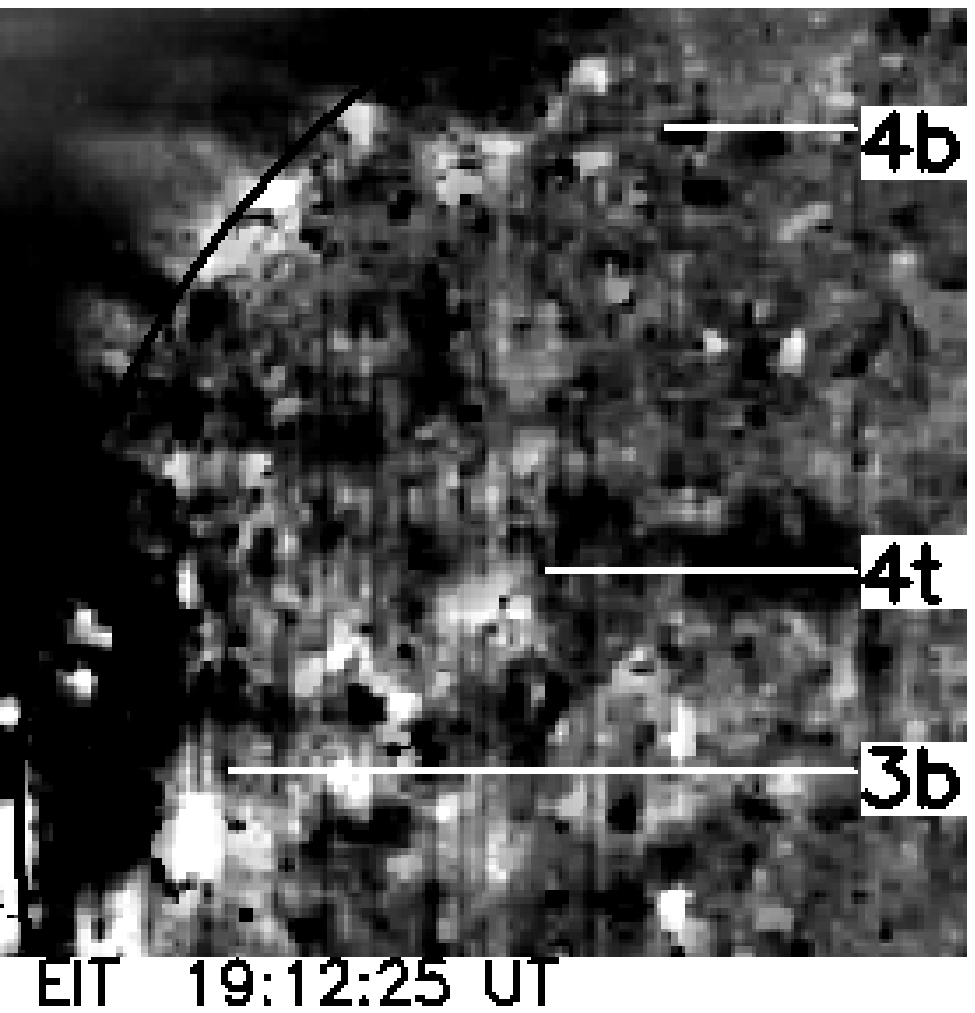}
\caption{Sub-field of view, shown in the black rectangle in the last panel of the Fig. 
\ref{figure2}, focusing on the on-disc structures appearing during the wave front passage observed in 195 \AA.}
\label{figure2quater}
\end{figure*}

\begin{figure*}
\includegraphics[width=2.98cm]{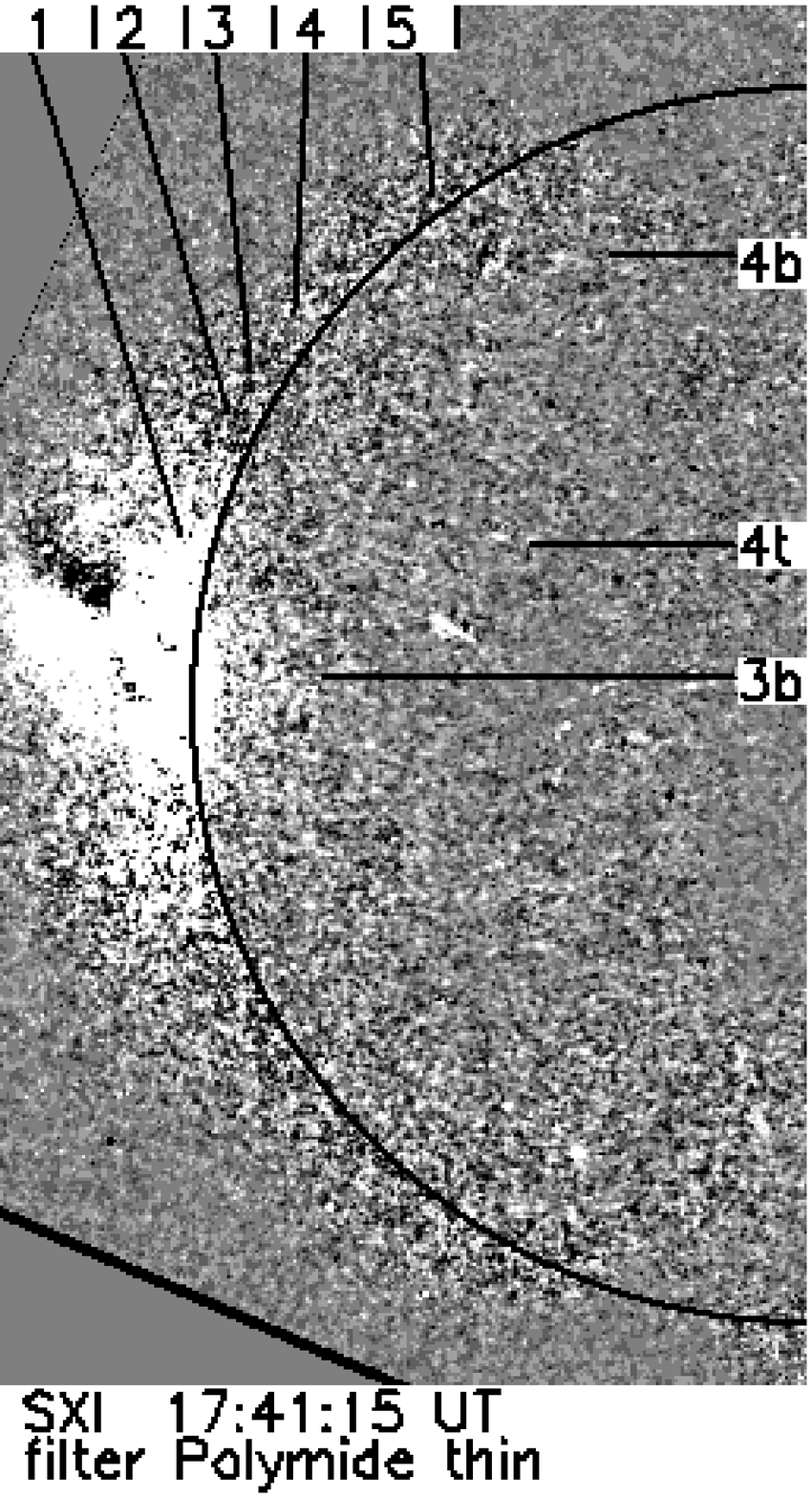}
\includegraphics[width=2.98cm]{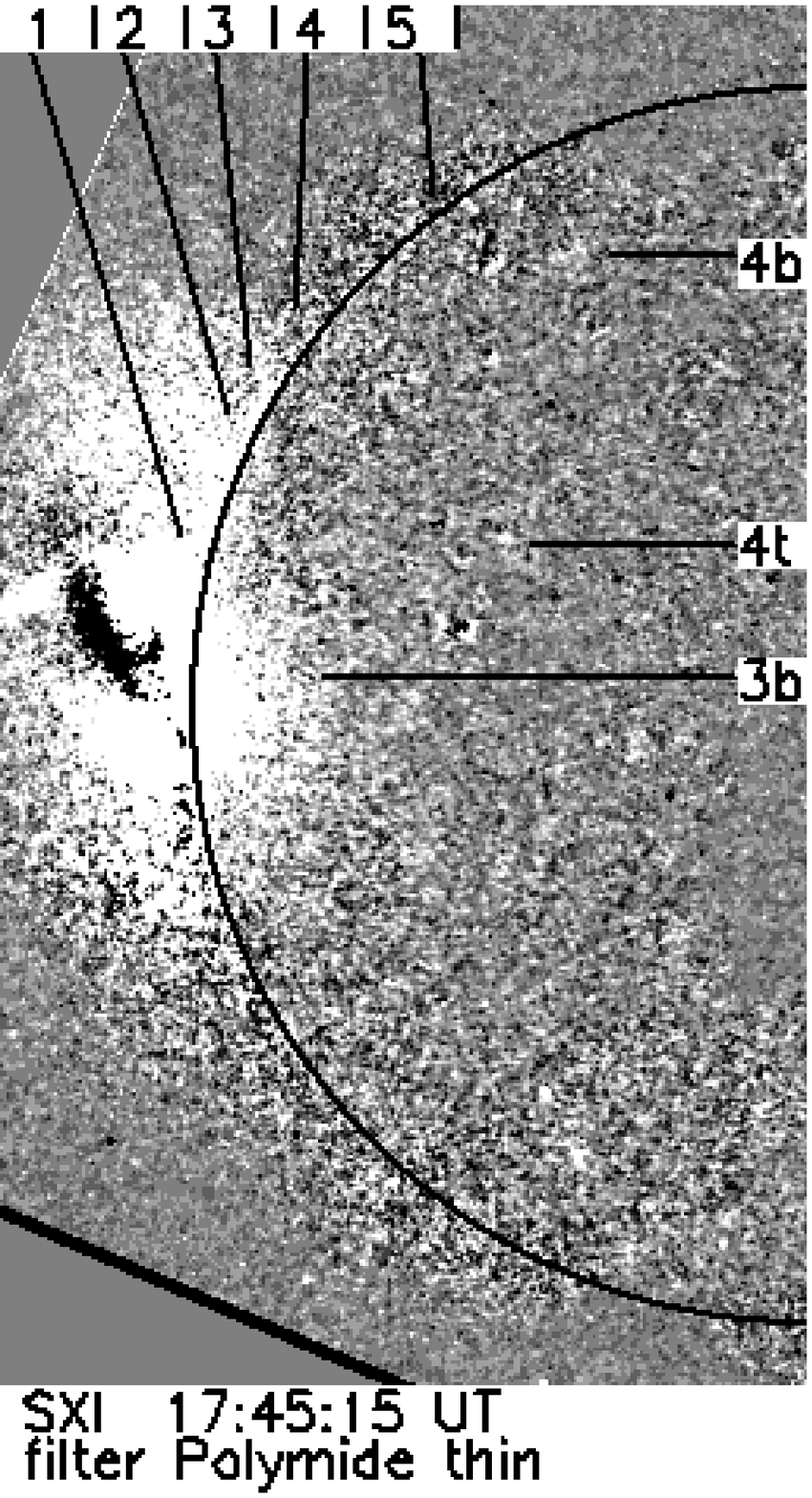}
\includegraphics[width=2.98cm]{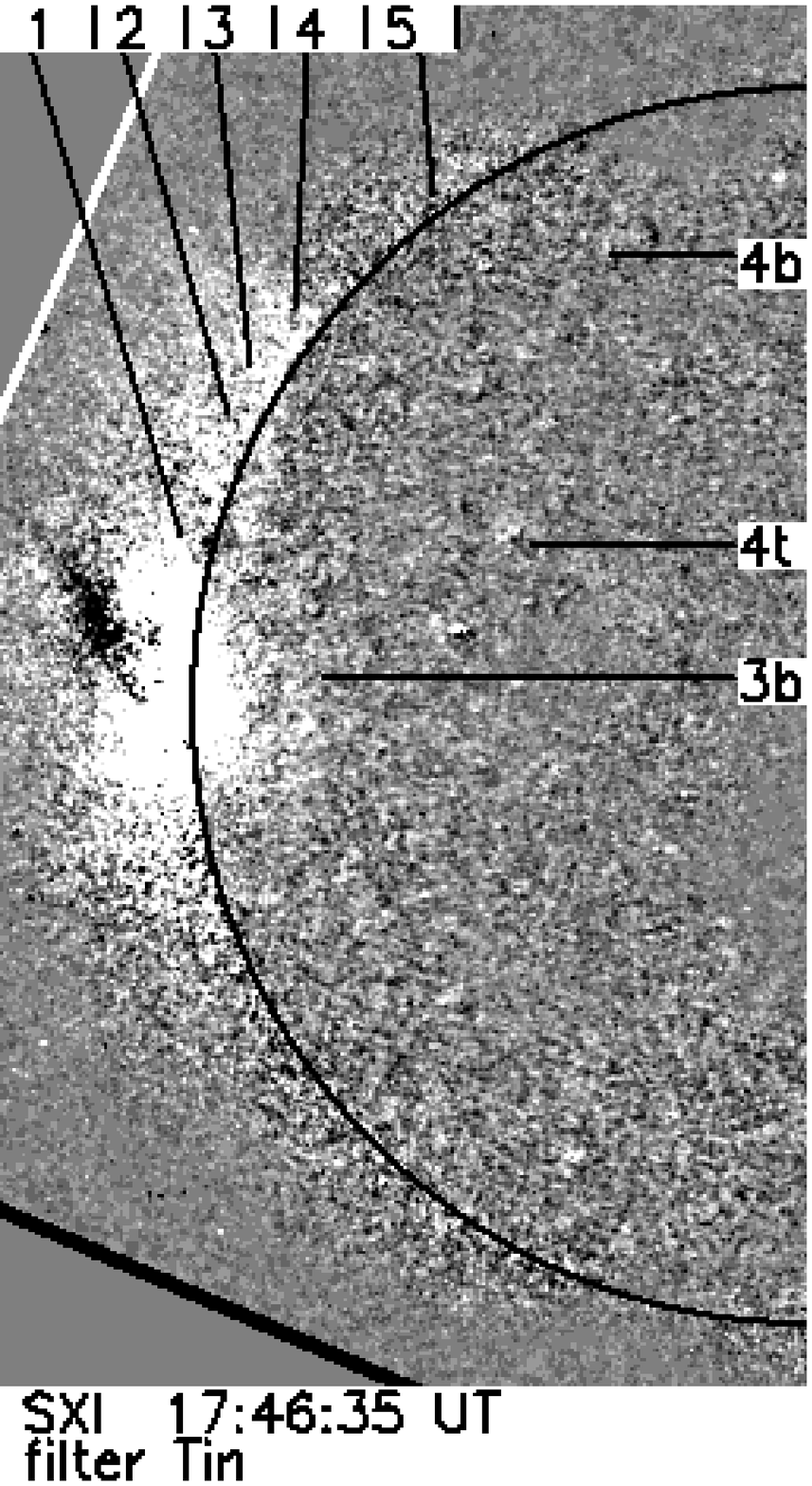}
\includegraphics[width=2.98cm]{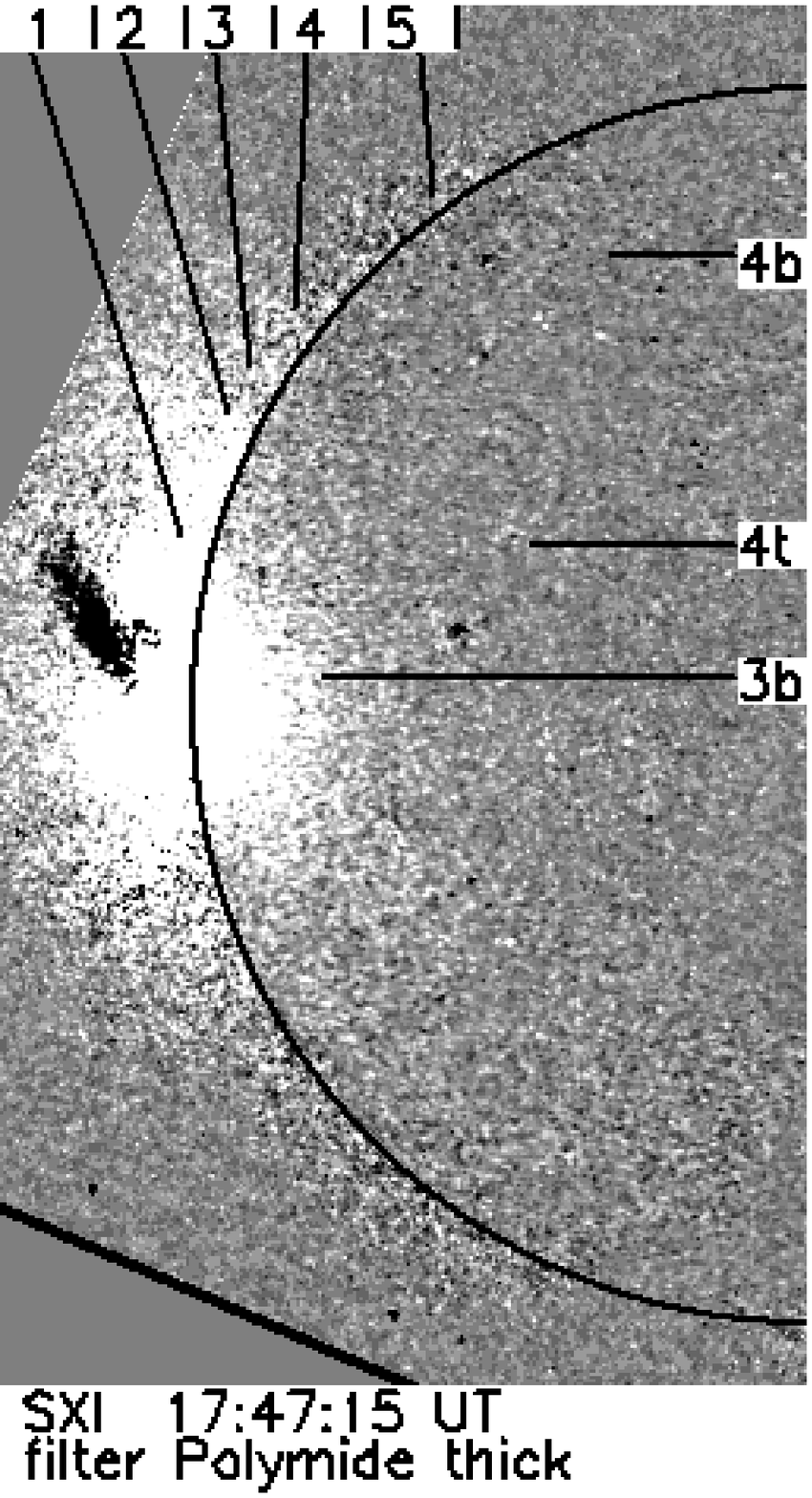}
\includegraphics[width=2.98cm]{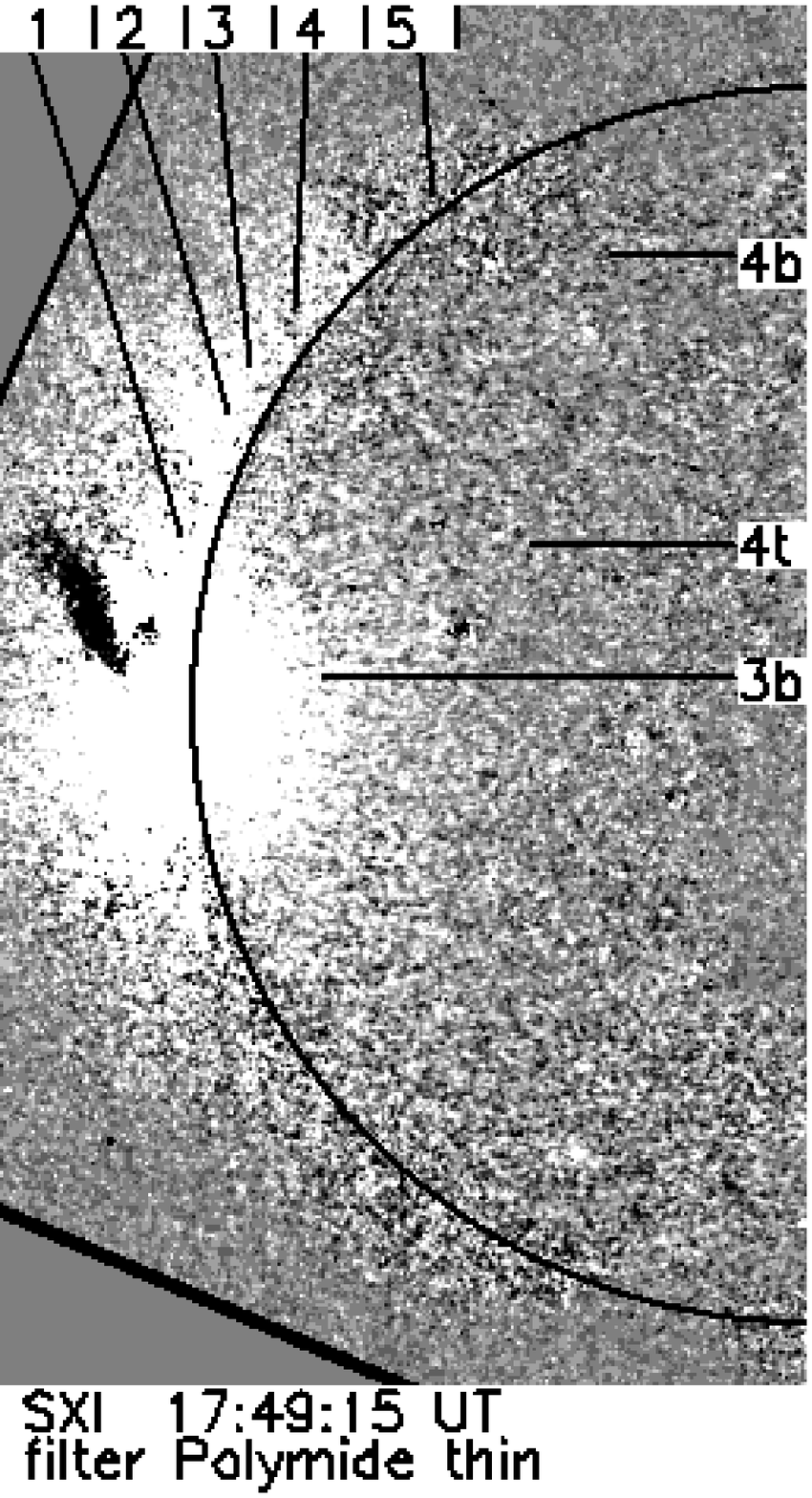}
\includegraphics[width=2.98cm]{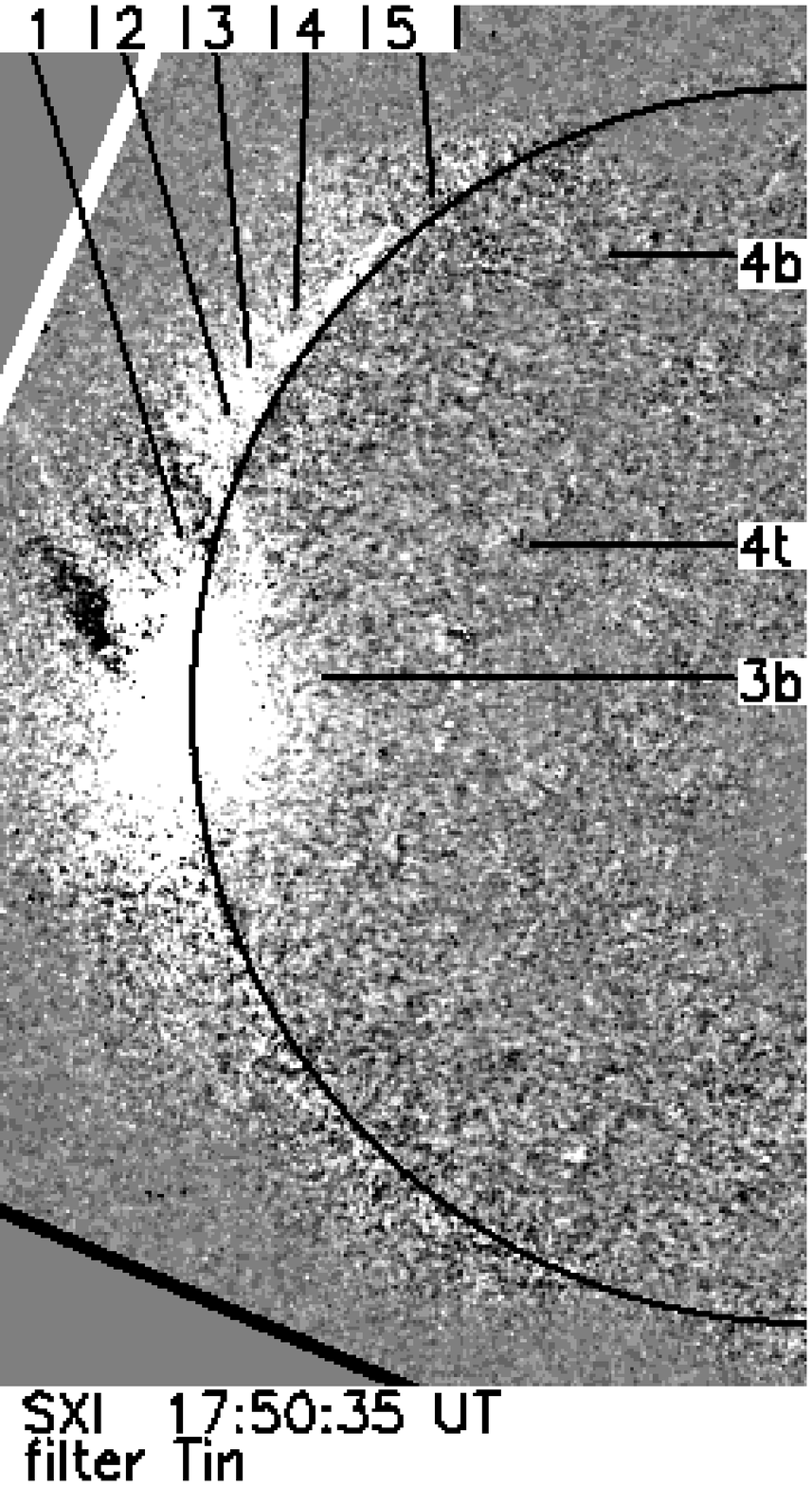}
\includegraphics[width=2.98cm]{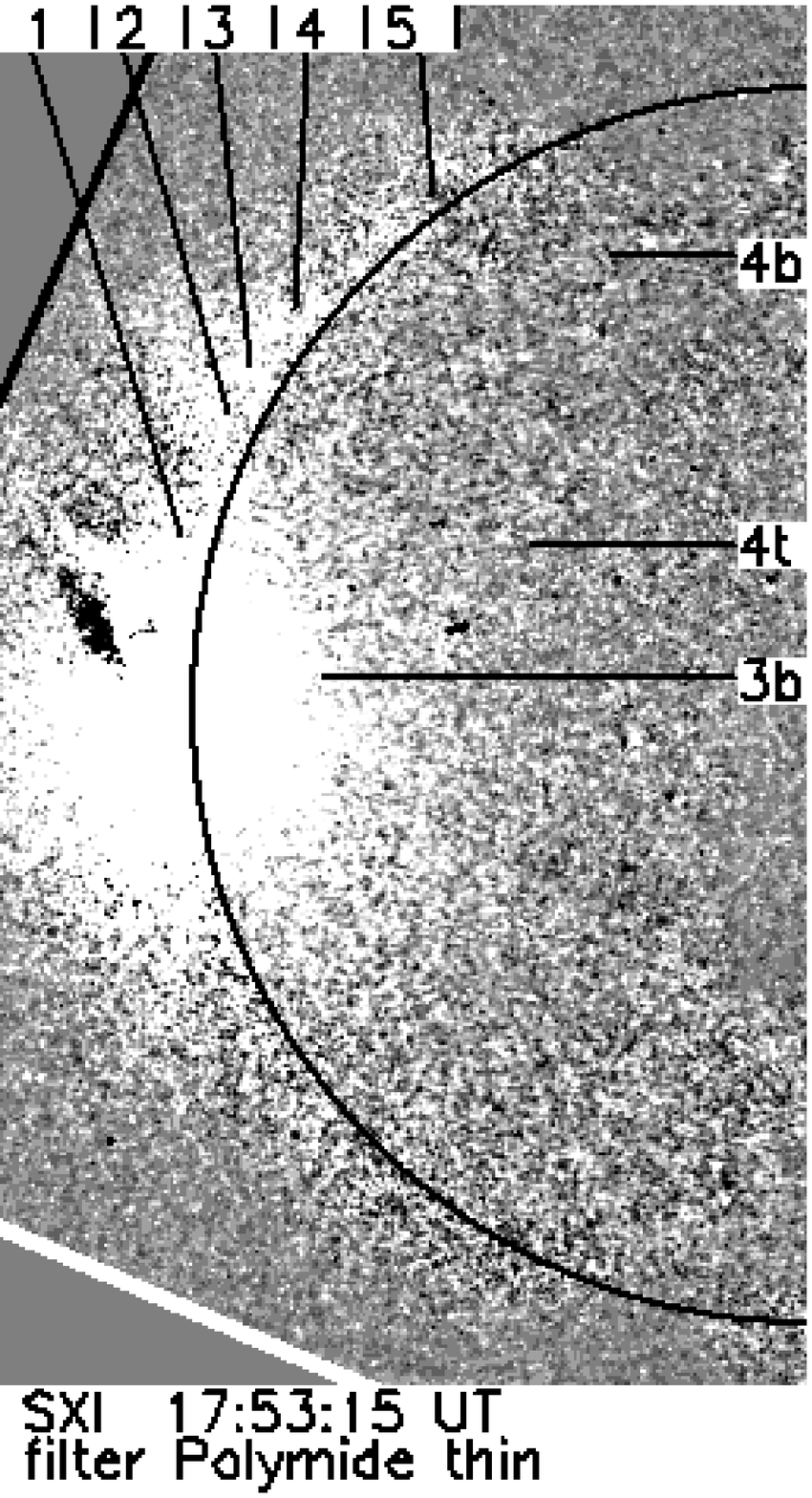}
\includegraphics[width=2.98cm]{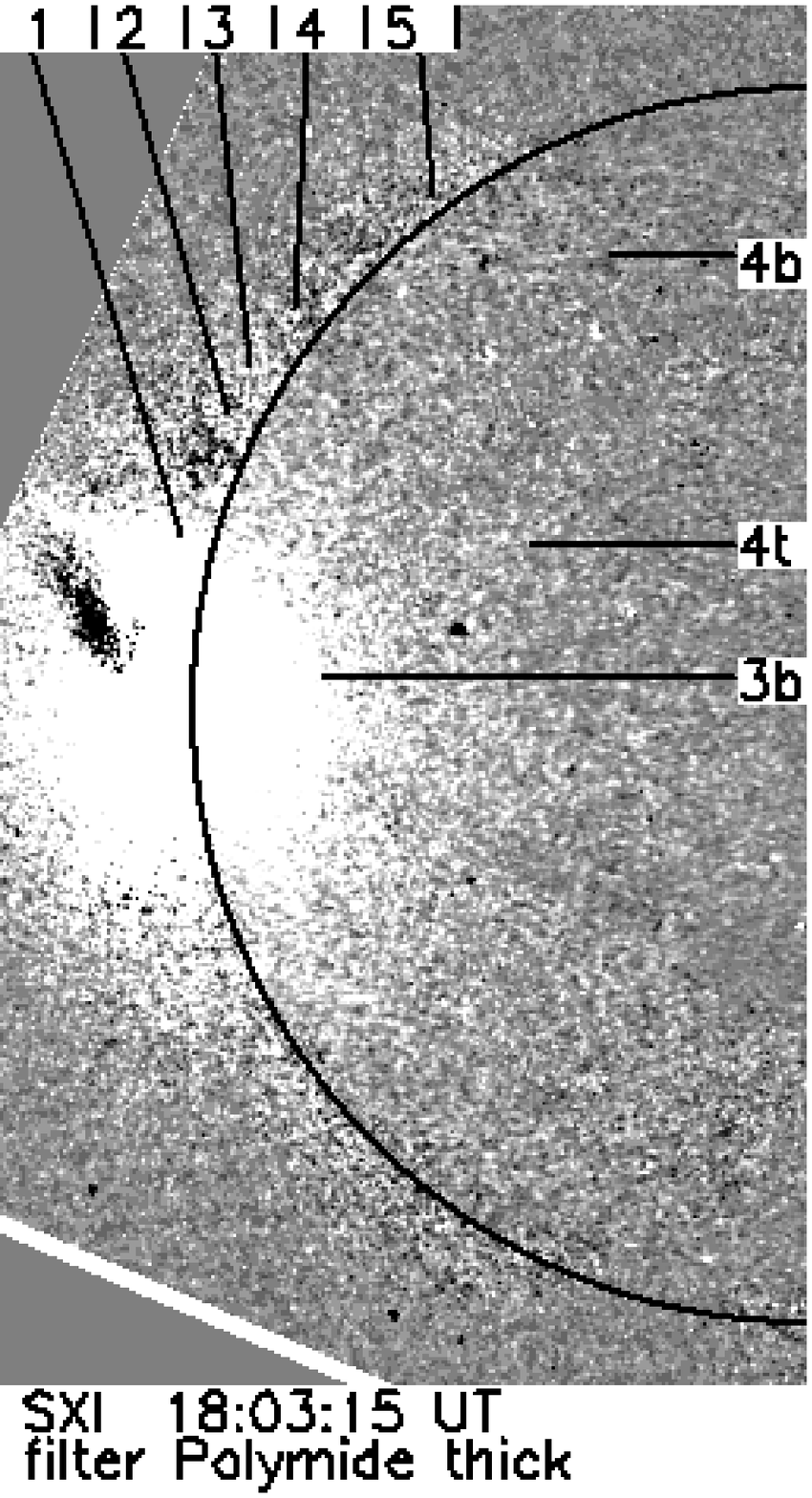}
\includegraphics[width=2.98cm]{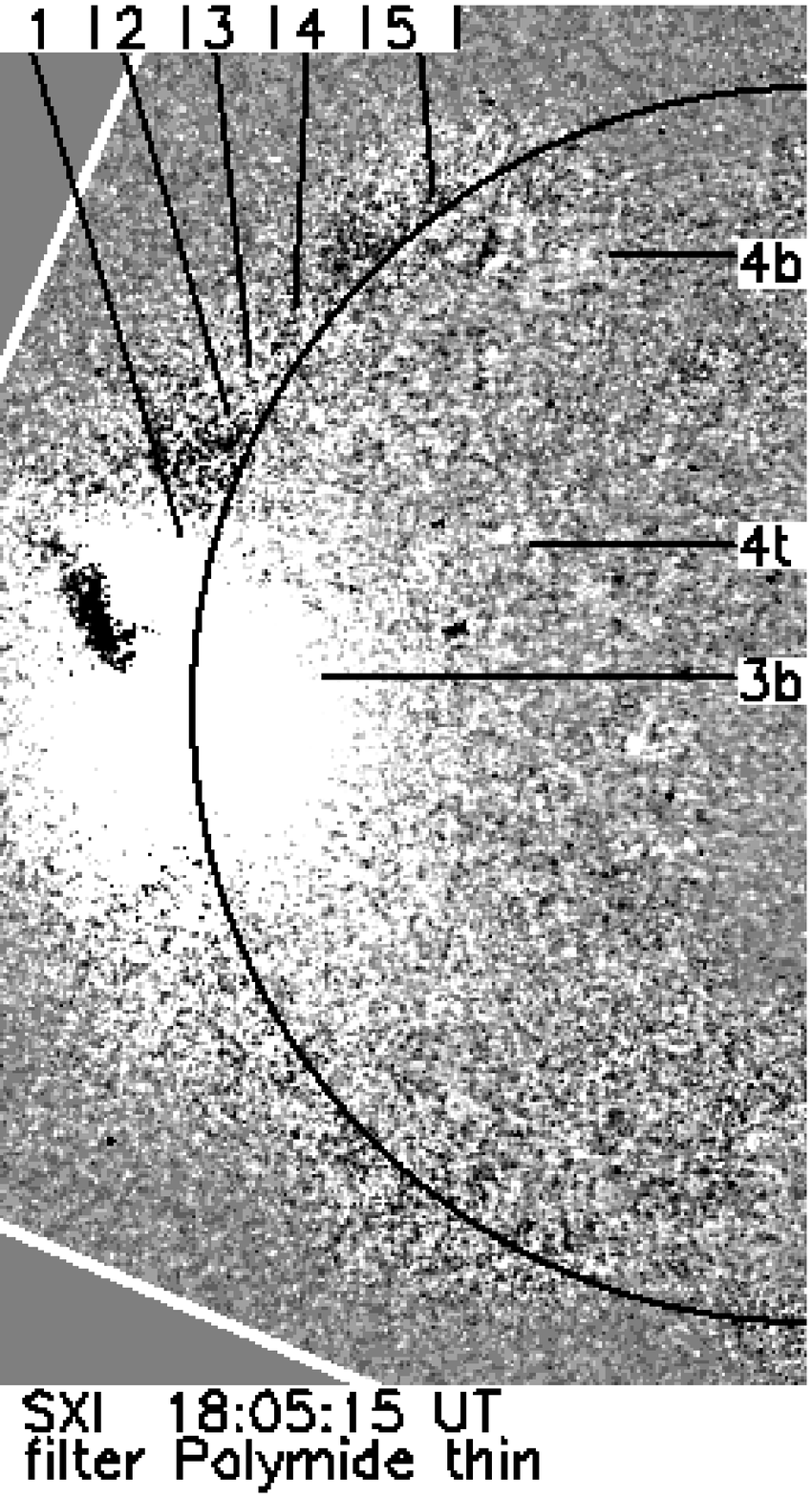}
\includegraphics[width=2.98cm]{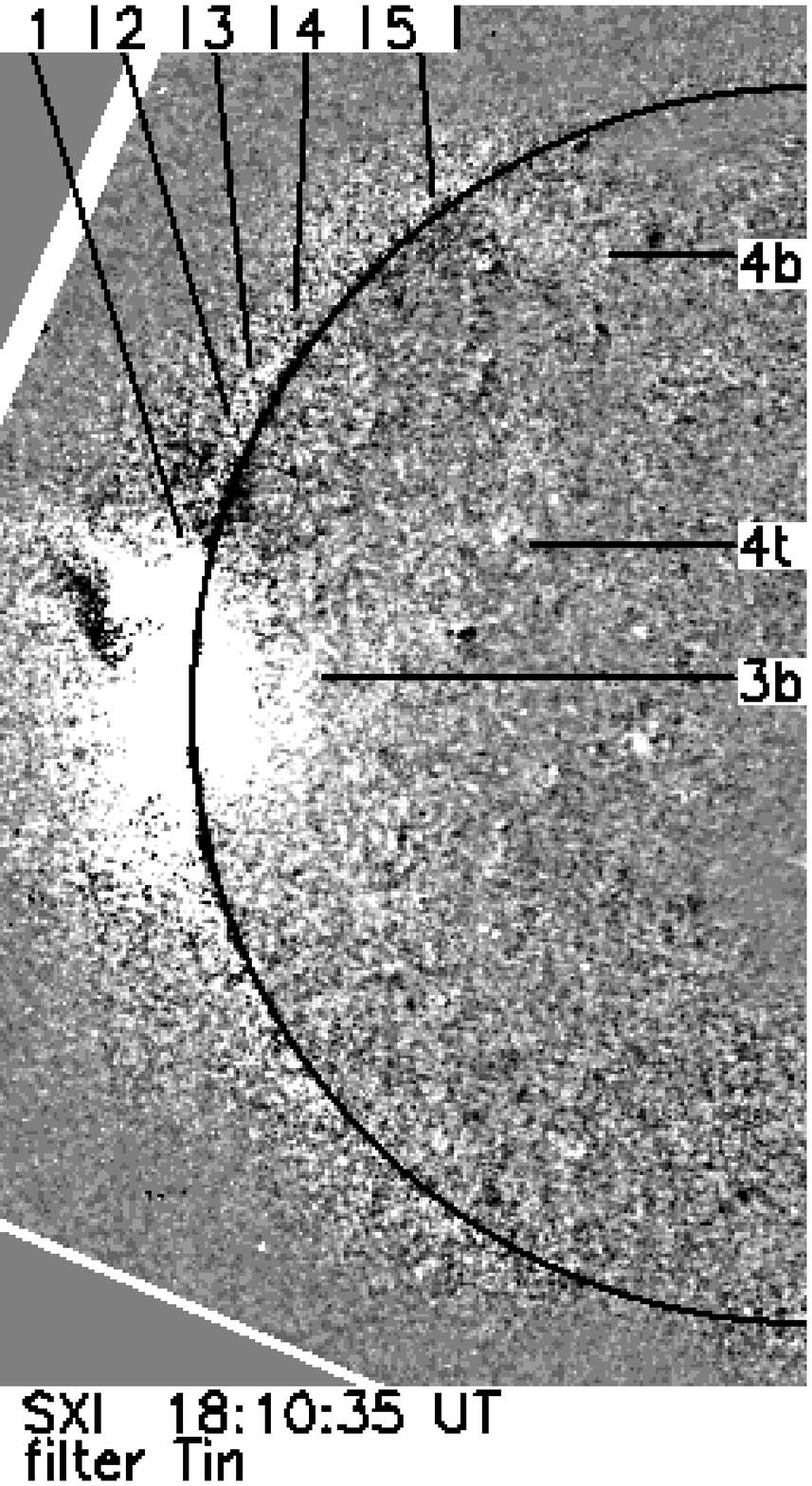}
\includegraphics[width=2.98cm]{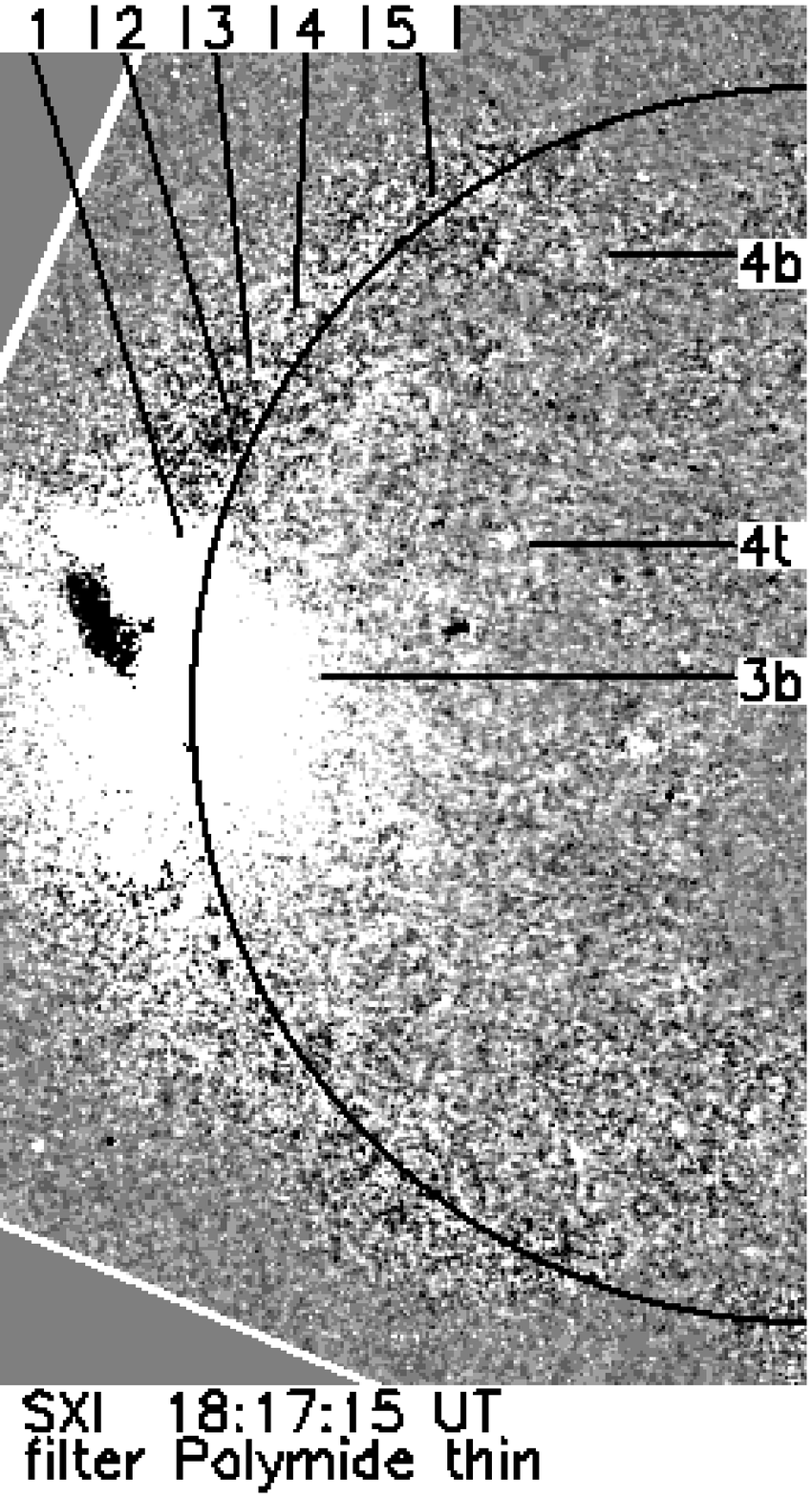}
\includegraphics[width=2.98cm]{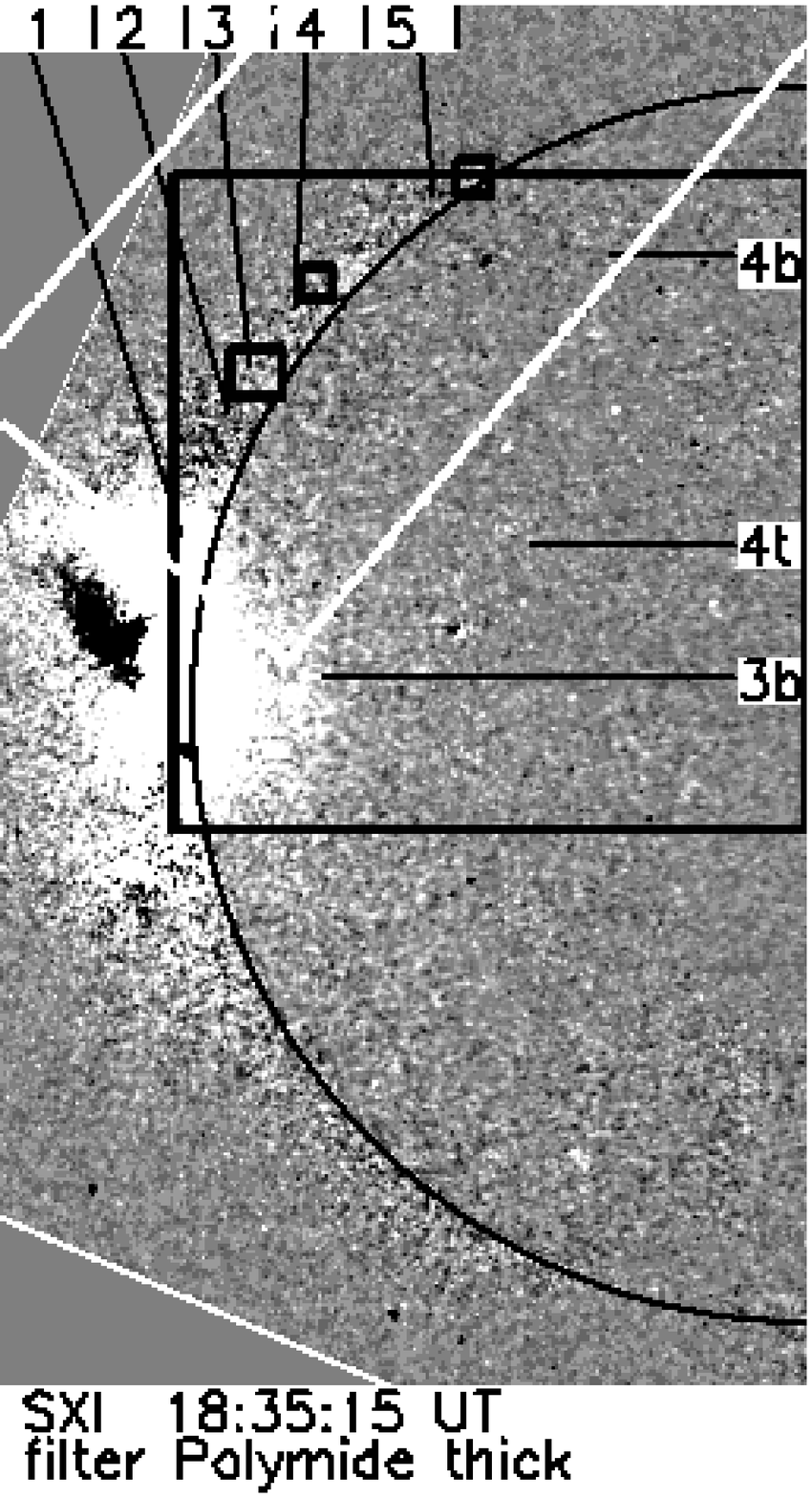}
\caption{DBDIs obtained using SXI. The names of the filters used are 
written on each image. Remark that the brightness of the structures 
is slightly different in the different images. The numbers indicate the successive 
appearance of the wave. The field of view contains half of the solar disc and is overlaid to the field of view of images in Figs. \ref{figure1} and \ref{figure2}. The small squares on the last panel show 
the area over which the mean brightness is measured and presented 
in Fig. \ref{figlightcurve}. The large white rectangle on the last panel
shows the sub-field of view of the images displayed in Fig. \ref{figure3c}. The large black rectangle on the last panel shows the sub-field of view of the images displayed in Figs. \ref{figure3g} and \ref{figure3h}. The SXI wave is observed progressing 
from an active region located slightly behind the limb to the northern 
coronal hole and to an eastern location where it stops. All the positions taken by the wave, 
comprising its latest position, stay very bright until 18:26 UT in 
most of the SXI filter.}
\label{figure3}
\end{figure*}

\begin{figure*}
\includegraphics[width=3cm]{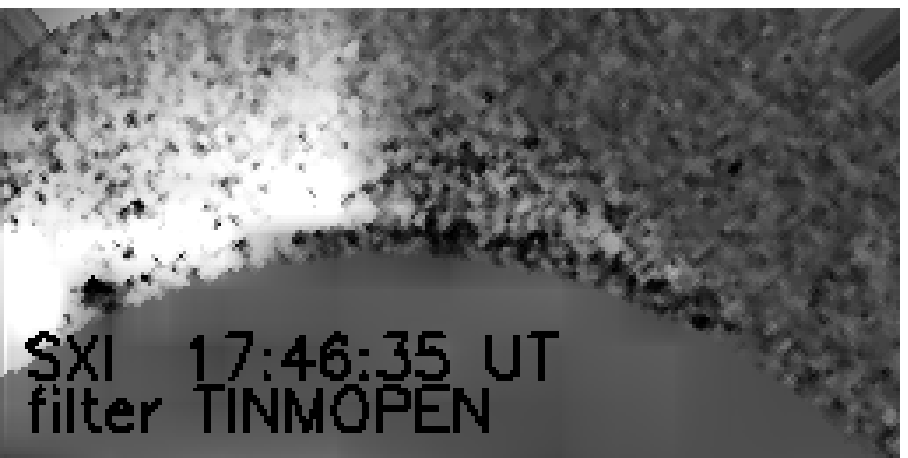}
\includegraphics[width=3cm]{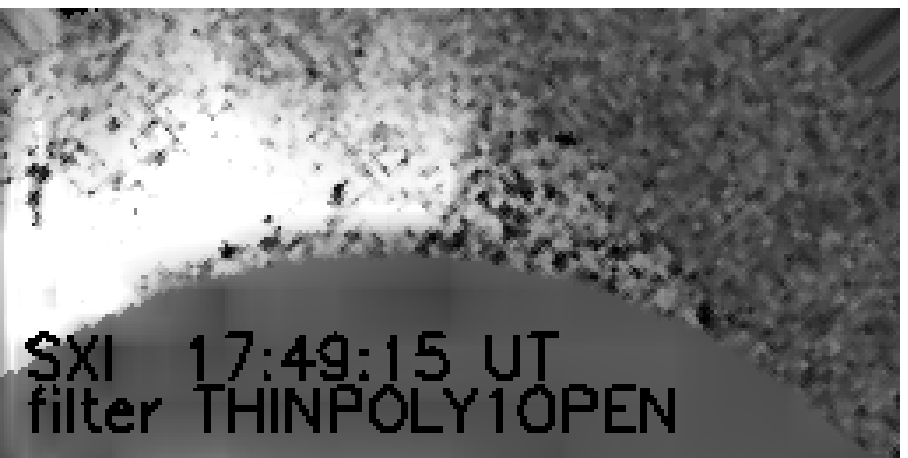}
\includegraphics[width=3cm]{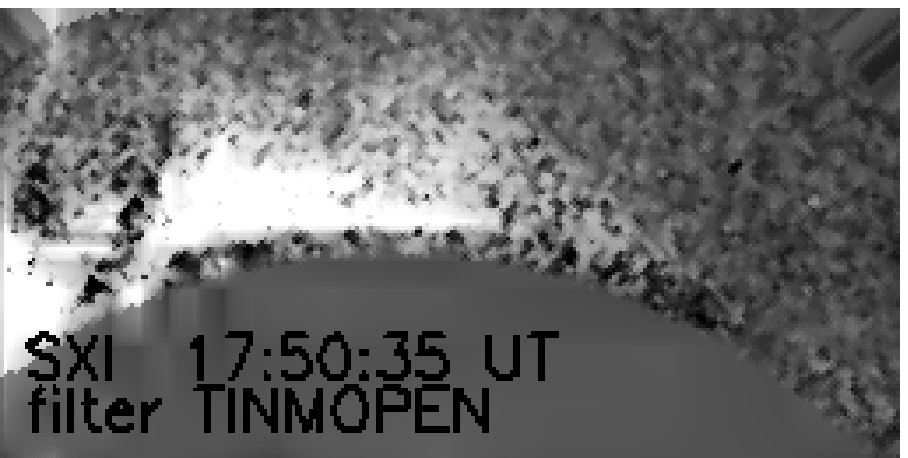}
\includegraphics[width=3cm]{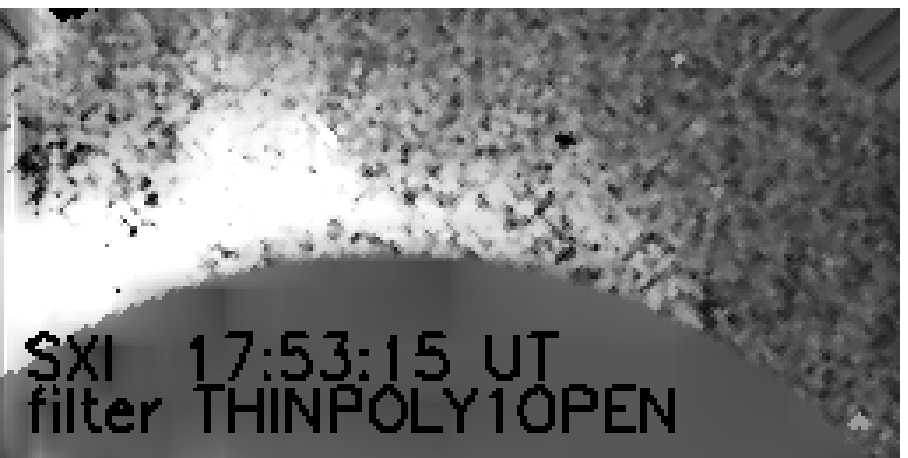}
\includegraphics[width=3cm]{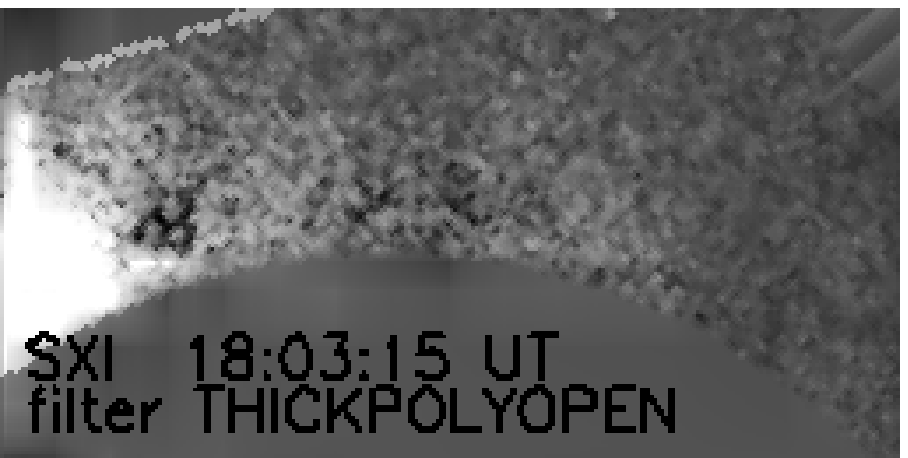}
\includegraphics[width=3cm]{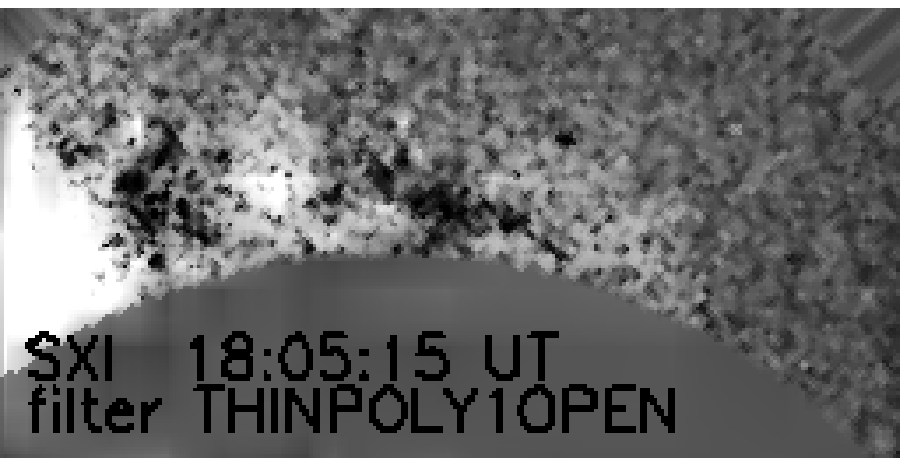}
\includegraphics[width=3cm]{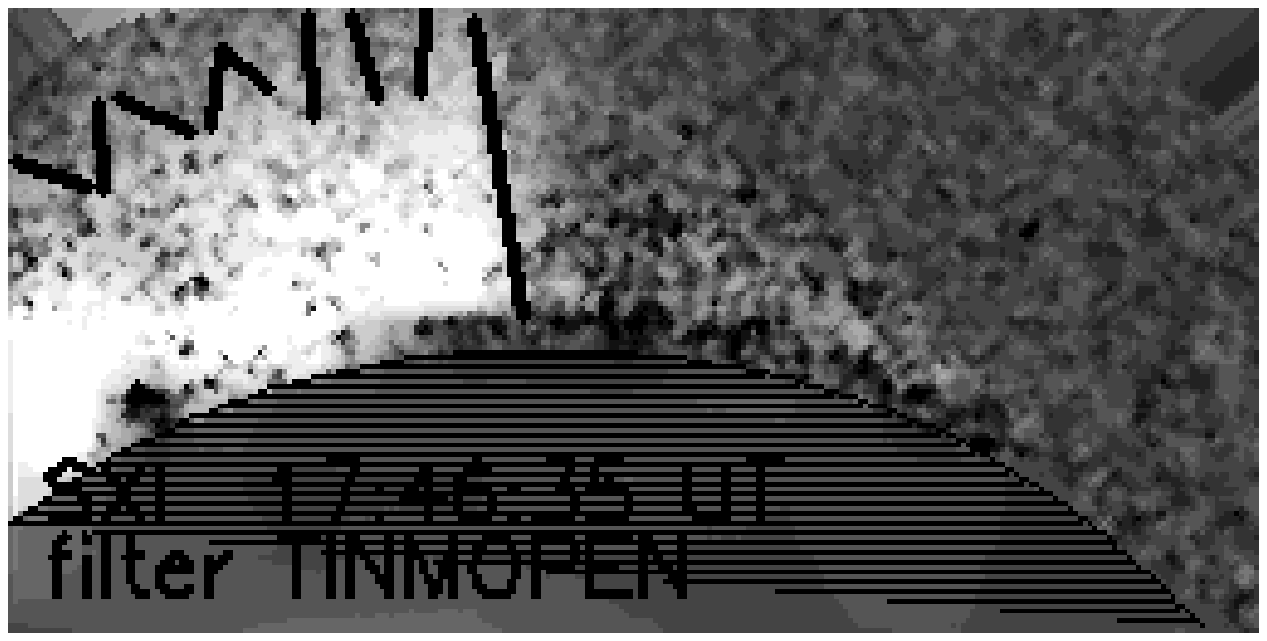}
\includegraphics[width=3cm]{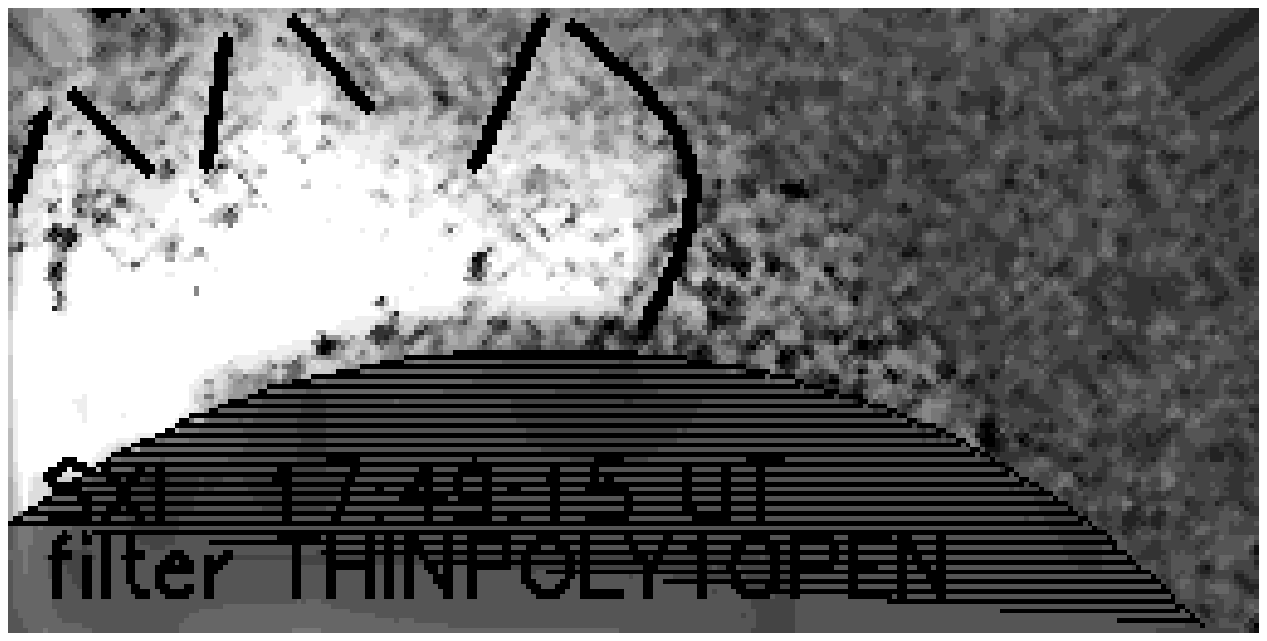}
\includegraphics[width=3cm]{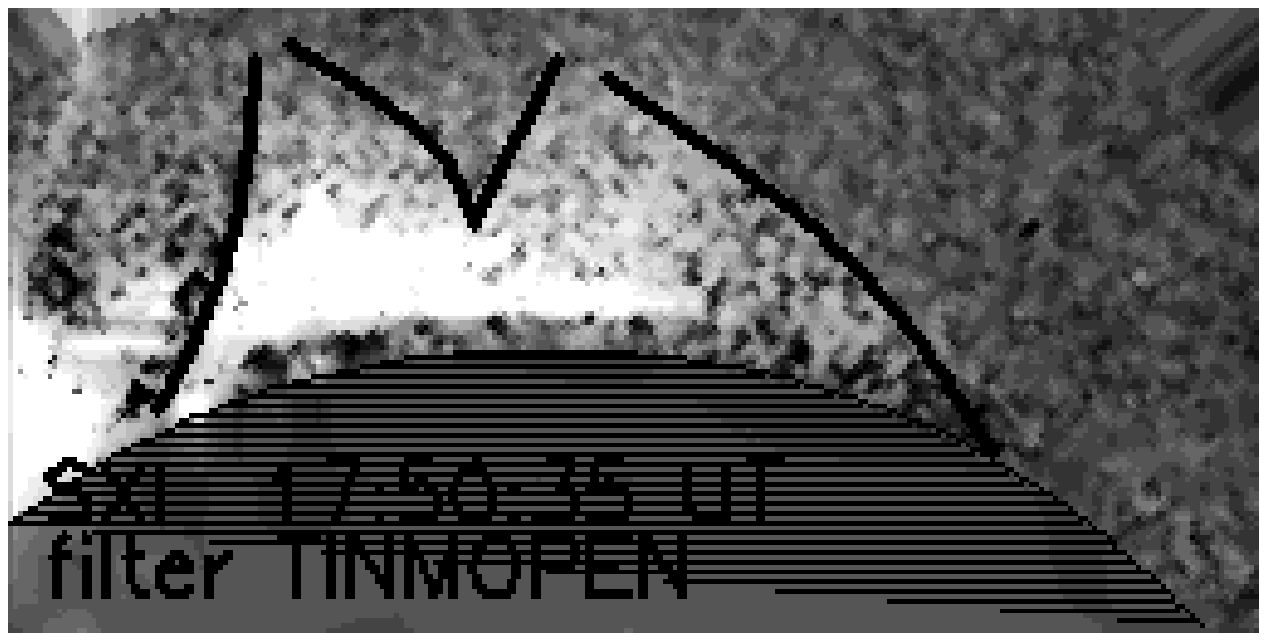}
\includegraphics[width=3cm]{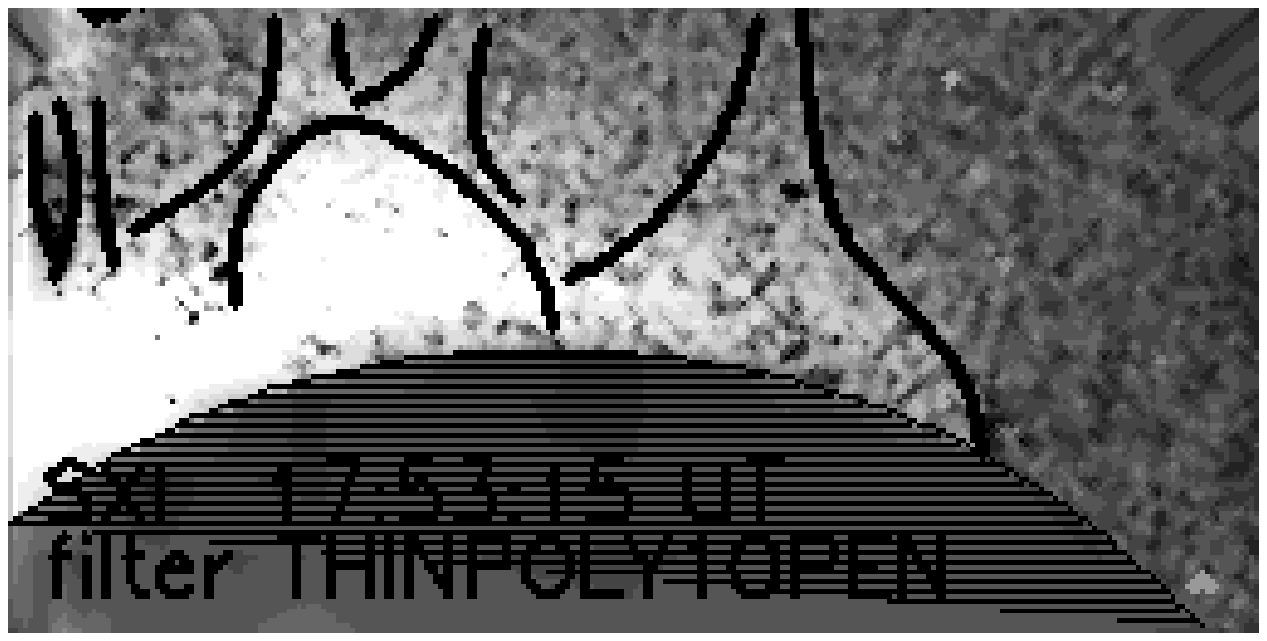}
\includegraphics[width=3cm]{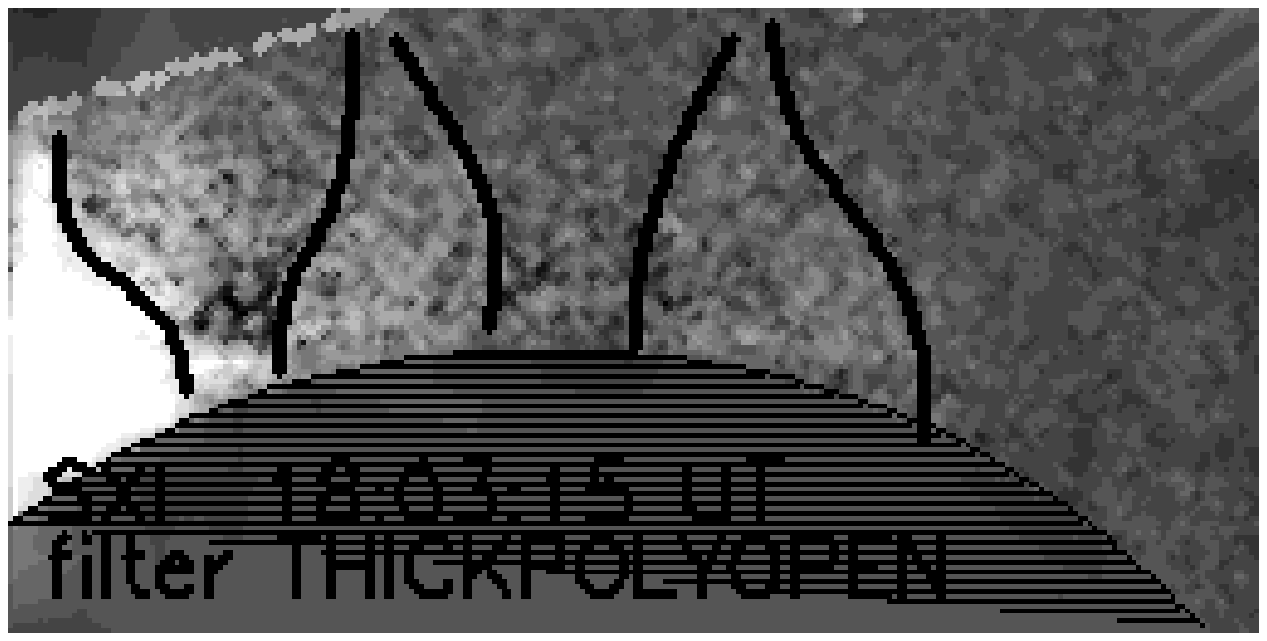}
\includegraphics[width=3cm]{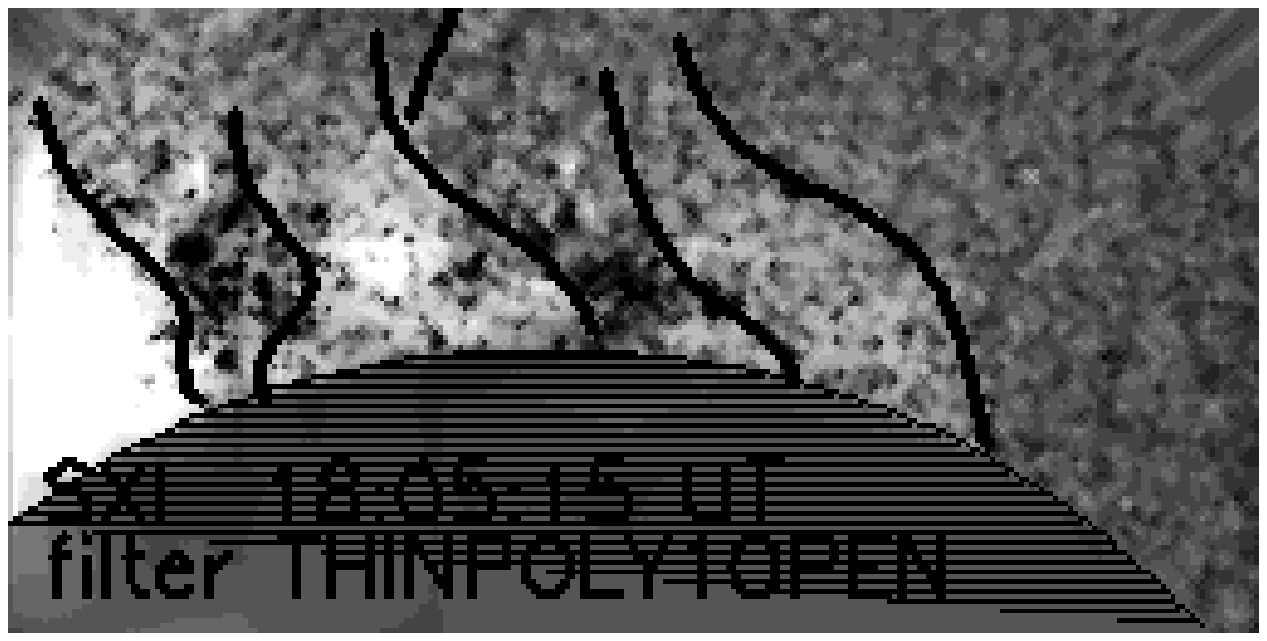}
\caption{Same sub-field of view as in Fig. \ref{figure2ter}, focusing on 
the over-limb structures observed in SXI. The images are highly processed (see the text). The second line show the same images as in the first line but with some over-limb structures highlighted with black drawing. The name of the filters are THINPOLYOPEN for Polyimide thin, THICKPOLYOPEN for Polyimide thick, TINMOPEN for Tin.}
\label{figure3c}
\end{figure*}

\begin{figure*}
\includegraphics[width=3cm]{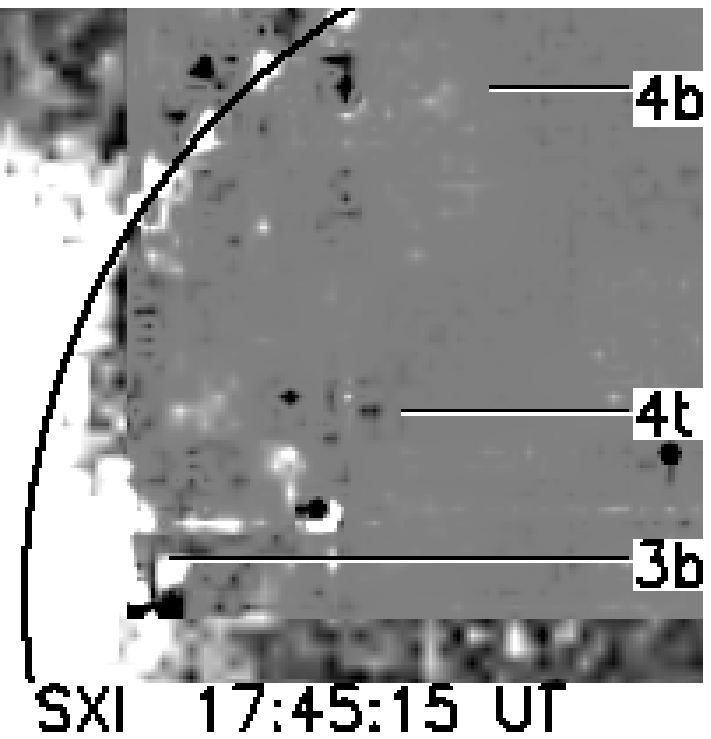}
\includegraphics[width=3cm]{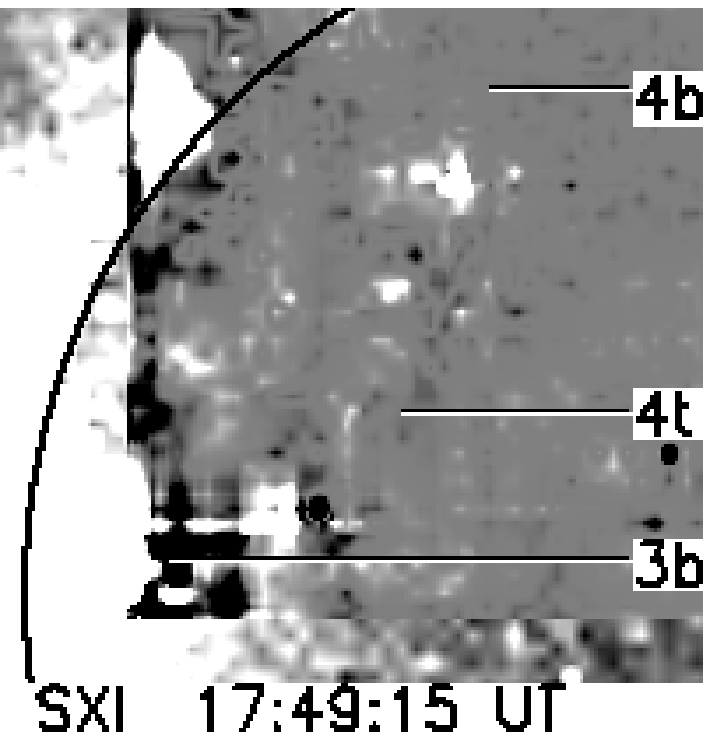}
\includegraphics[width=3cm]{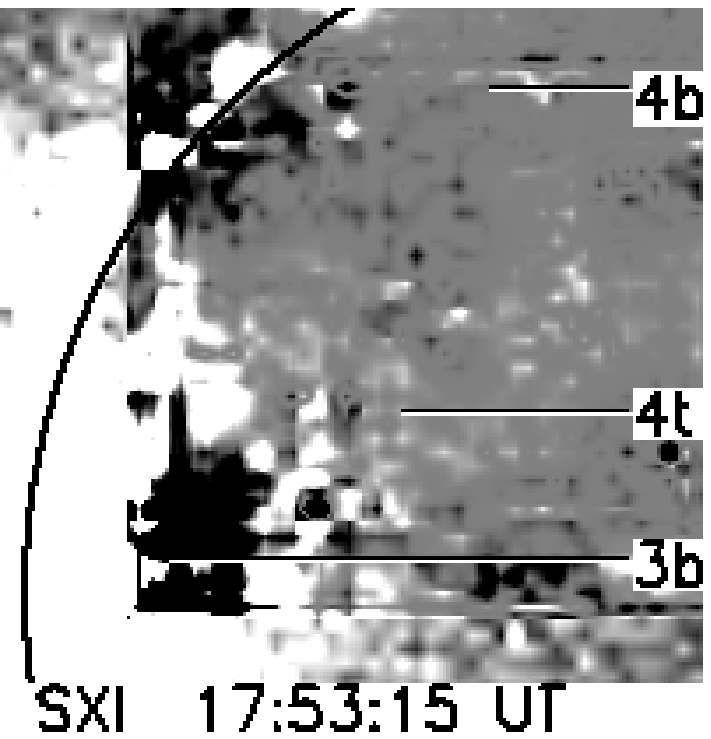}
\includegraphics[width=3cm]{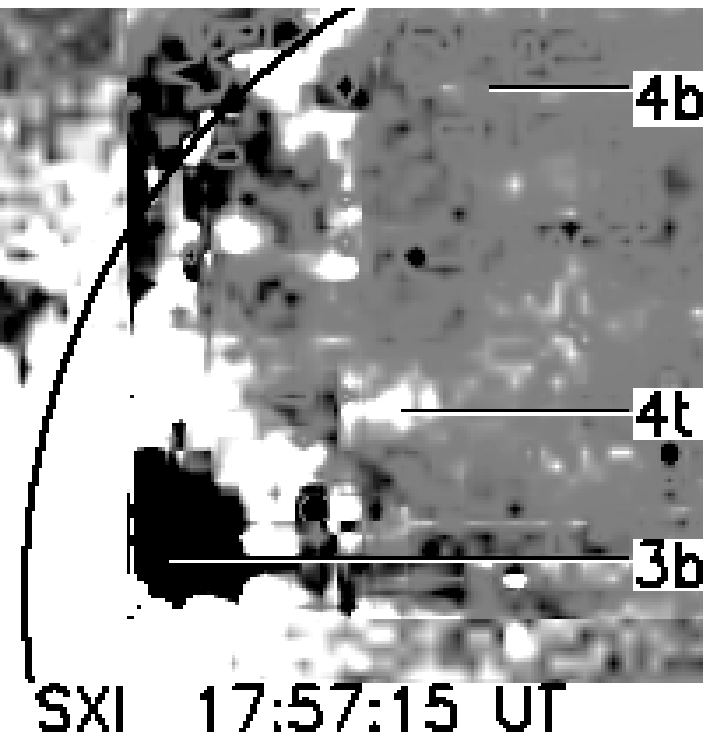}
\includegraphics[width=3cm]{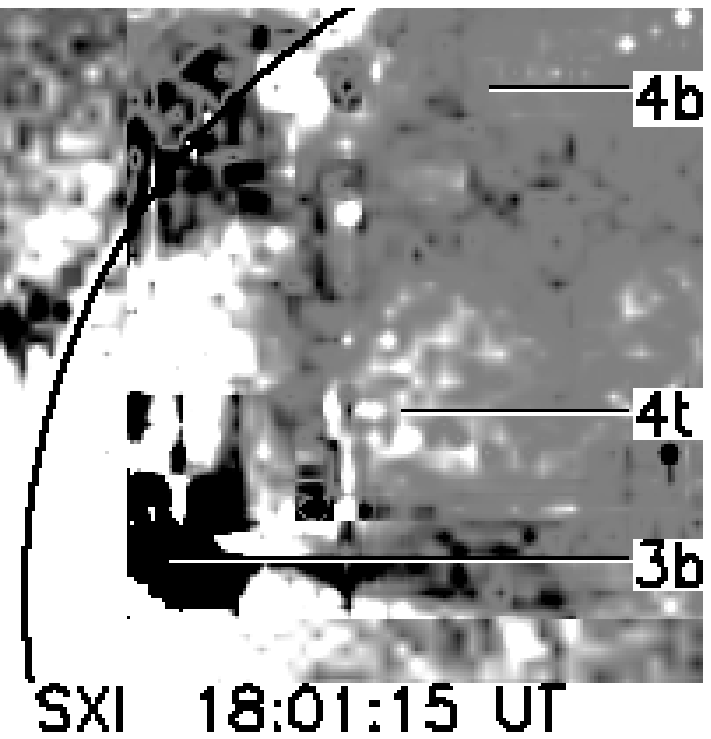}
\includegraphics[width=3cm]{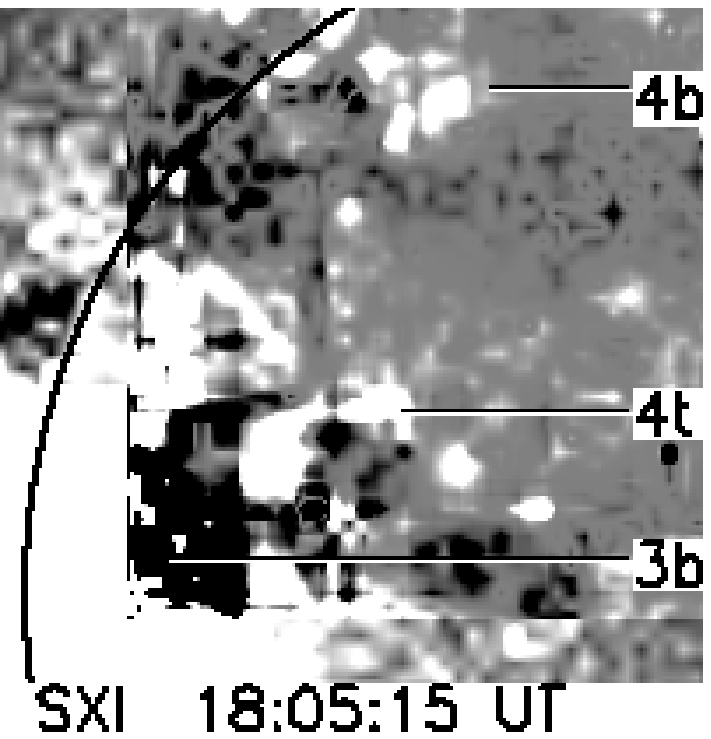}
\includegraphics[width=3cm]{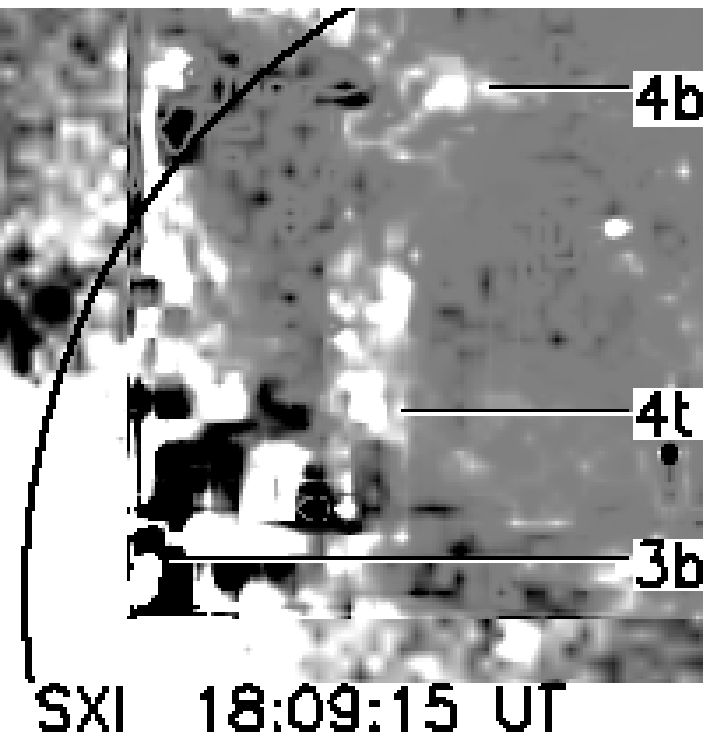}
\includegraphics[width=3cm]{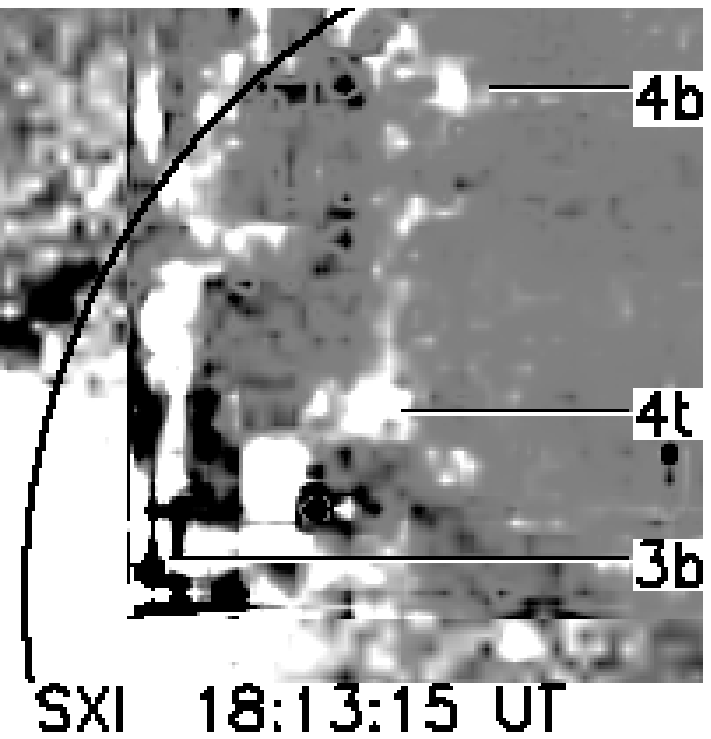}
\includegraphics[width=3cm]{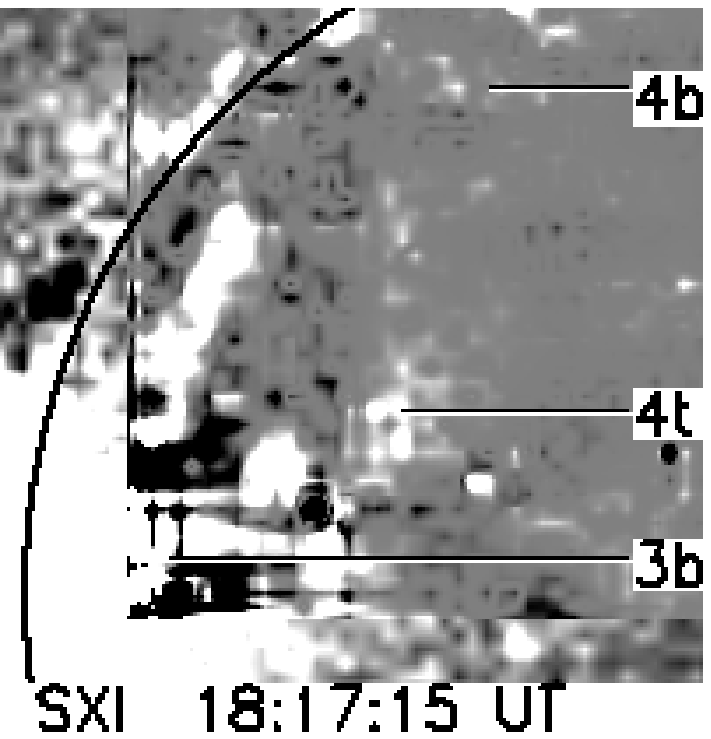}
\includegraphics[width=3cm]{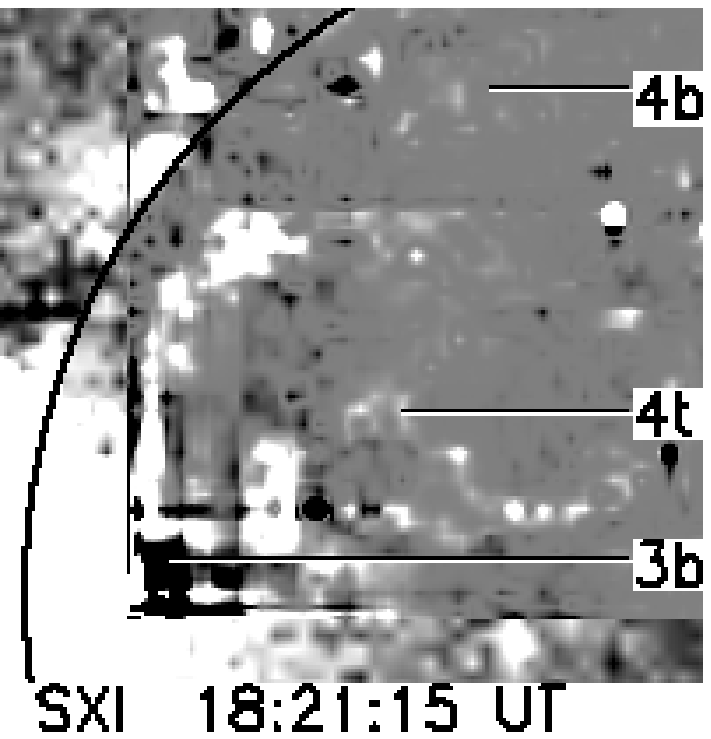}
\includegraphics[width=3cm]{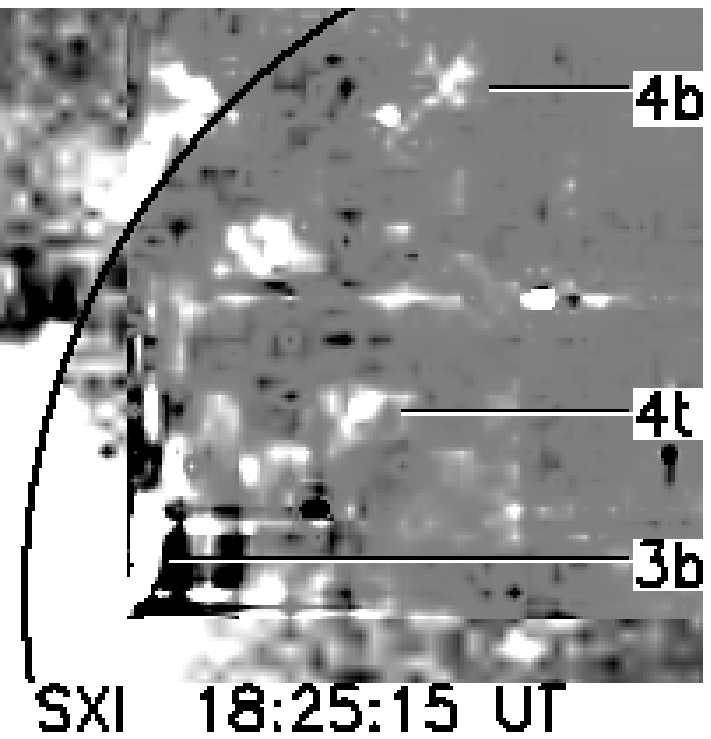}
\includegraphics[width=3cm]{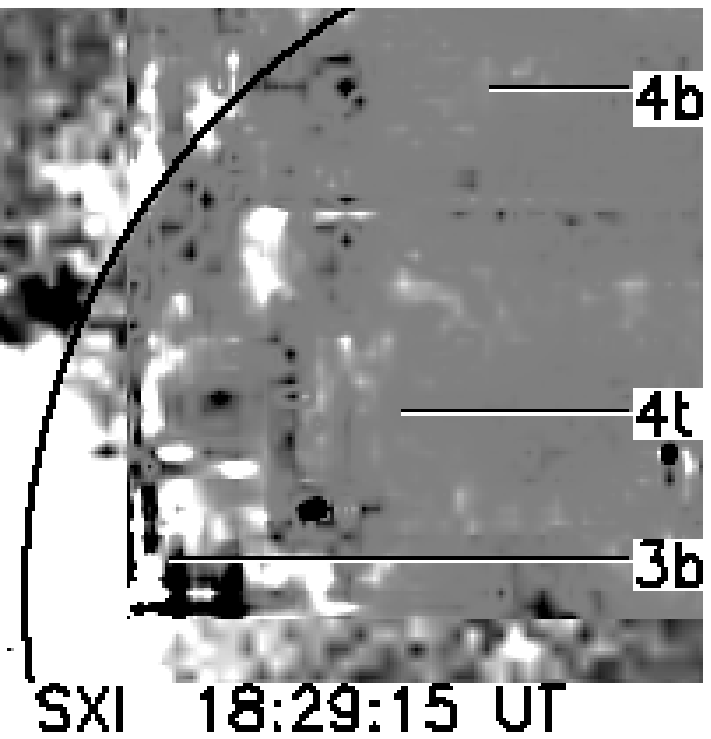}
\includegraphics[width=3cm]{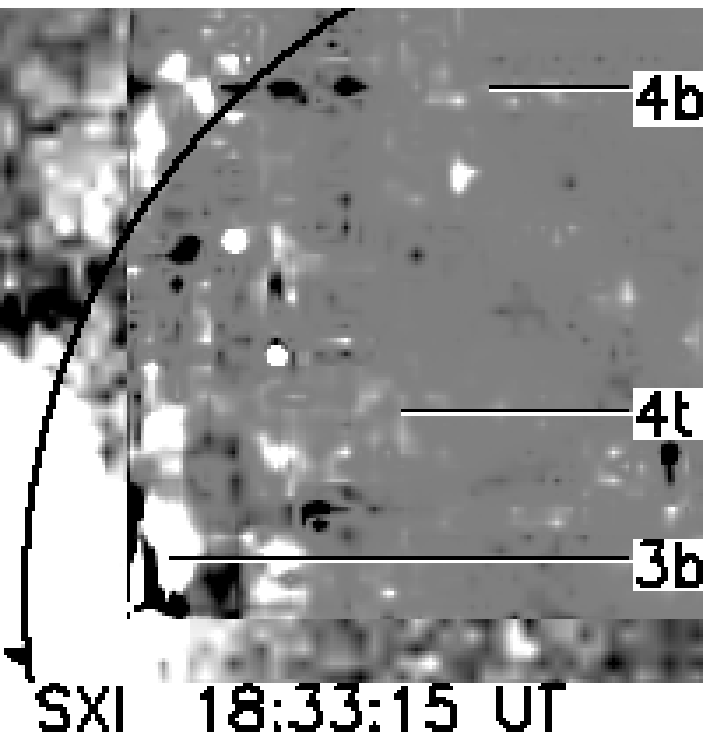}
\includegraphics[width=3cm]{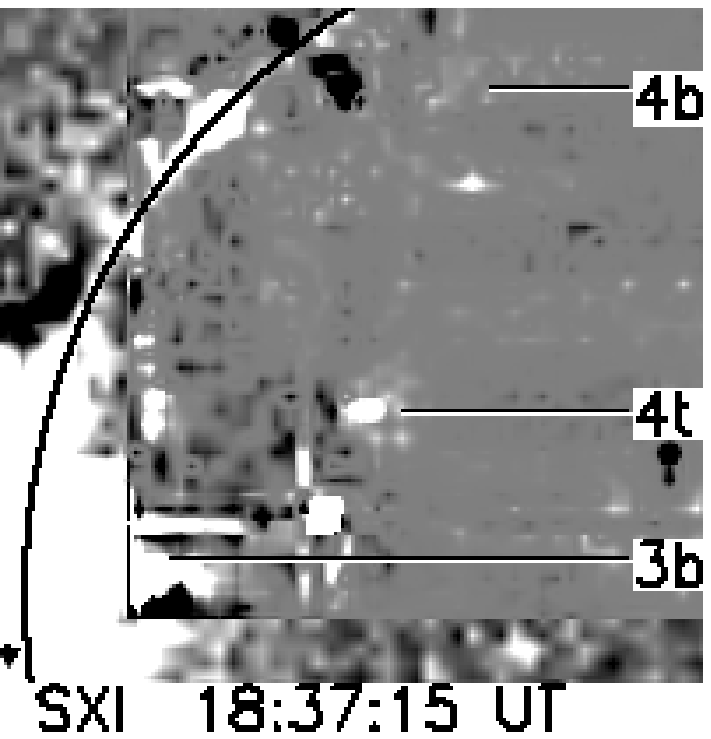}
\caption{Same on disc sub-field of view as in Fig. \ref{figure2quater}, 
focusing on the stationary brightenings labeled 3b, 4b and 4t, through the
"Polyimide Thin" filter. The images are highly processed to highlight the brightenings 
(see the text).}
\label{figure3g}
\end{figure*}

\begin{figure*}
\includegraphics[width=3cm]{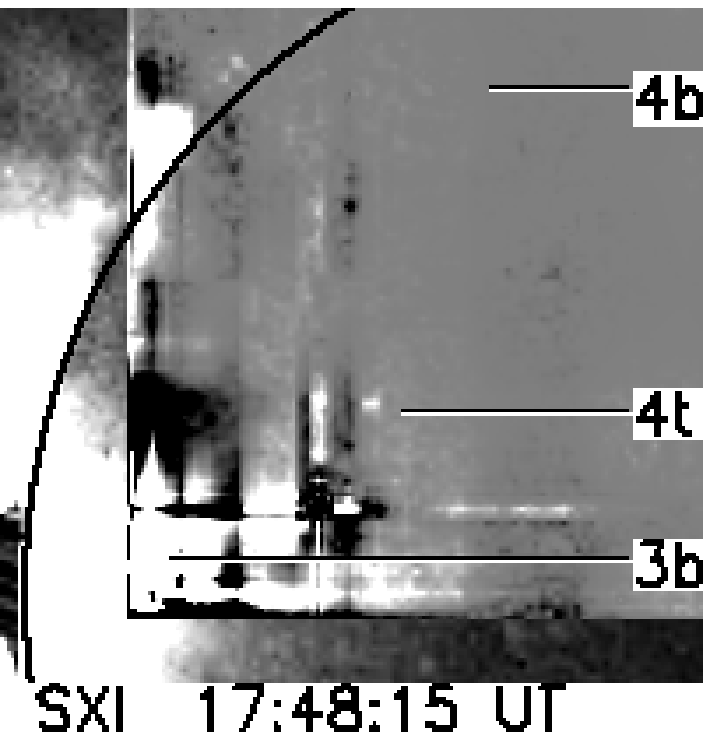}
\includegraphics[width=3cm]{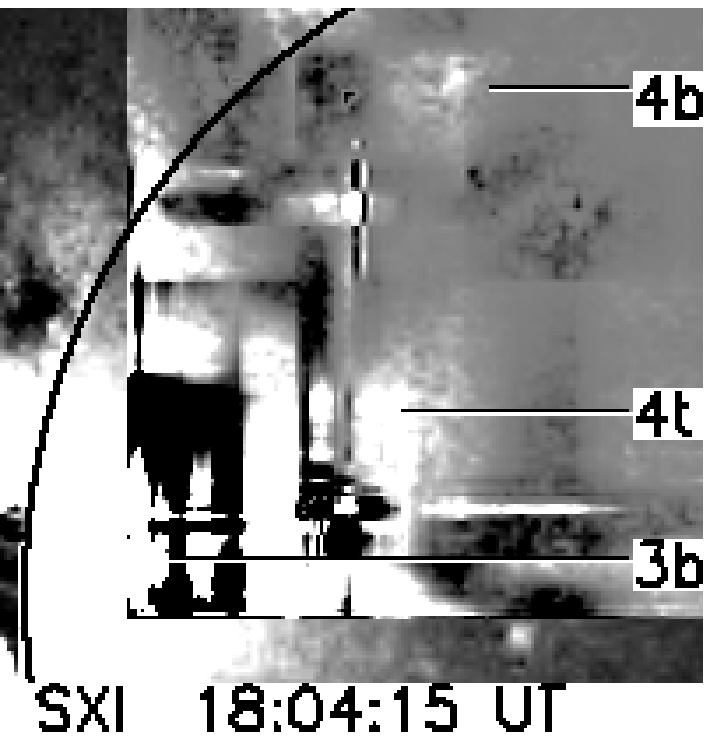}
\includegraphics[width=3cm]{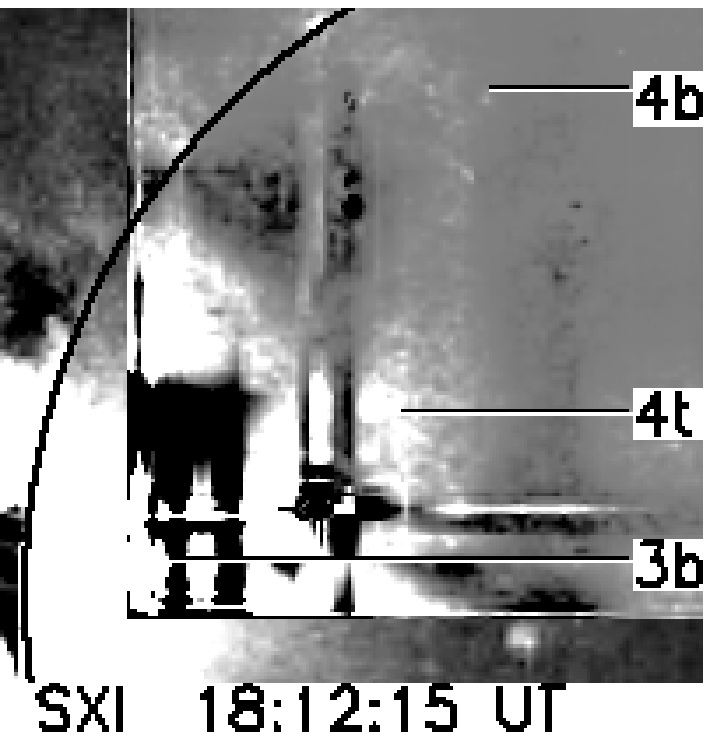}
\includegraphics[width=3cm]{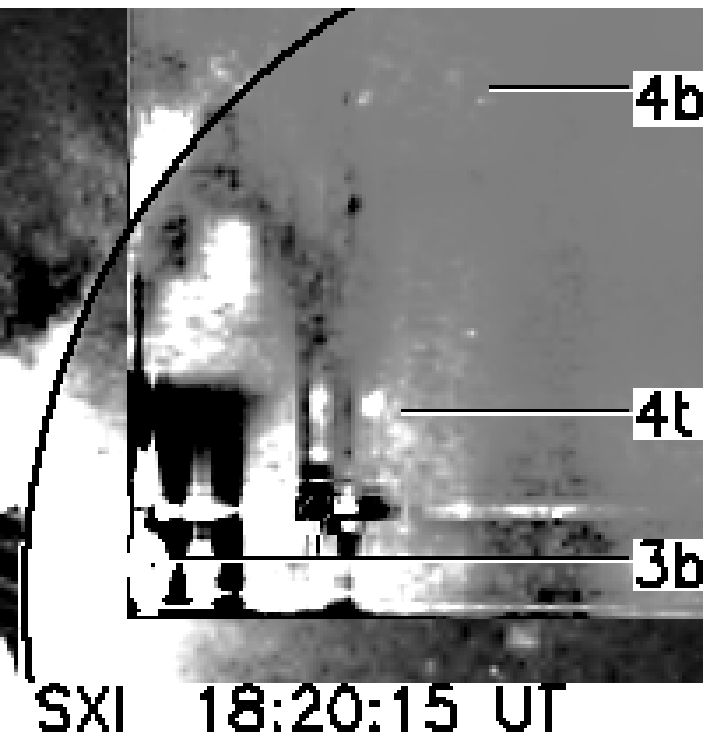}
\includegraphics[width=3cm]{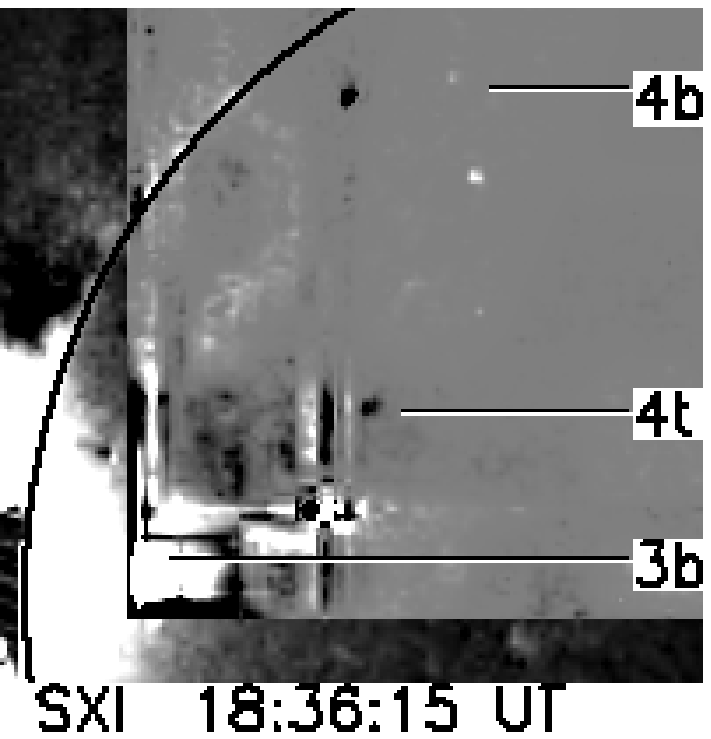}
\caption{Same on disc sub-field of view as in Fig. \ref{figure2quater}, 
focusing on the stationary brightenings labeled 3b, 4b and 4t, through the "Polyimide Thick" filter. The images are highly processed to highlight the brightenings (see the text).}
\label{figure3h}
\end{figure*}

In EUV, the EIT wave front defines the border of a dimming 
(Fig. \ref{figure2} and the EIT movie). This front is large, wide and diffuse. It propagates from the flare site to the northern pole and on the disc from the eastern limb to the central meridian (see the bright arch in panels of Fig. \ref{figure2quater}). On the disc, it stops at locations 
reached at 18:10 UT staying brighter than before the eruption 
for about one hour (see especially the bright diffuse arch linking the brigth points labeled 3b, 4b and 4t on the disc in Figs. \ref{figure2} and \ref{figure2quater}).
Above the limb, the wave front is formed of elongated 
in altitude structures (Fig. \ref{figure2ter}, image at 17:46 UT). The last observed position of the front is obtained at 17:58 UT at label 5 (Figs. \ref{figure2} and \ref{figure2ter}). From 18:10 UT to at 
least 19:34 UT, the location 5 is dimmer than prior to the 
flare. Between the locations 2 and 5 (Figs. \ref{figure2} and \ref{figure2ter}) and after 18:10 UT, a very faint brightness illuminates coronal loops above the limb, that are dimmed in the image at 17:58 UT. This brightness morphology does not resemble to the wave front that passed on its way from the south to 
the north: its wideness (in the range of 250-300 Mm at the limb at 18:34 UT 
and 18:46 UT) is too large (the wave front wideness is 87 Mm and 131 Mm at 17:46 UT and 17:58 UT resp.) and it does not move between 18:34 UT and 19:12 UT. It could be very hardly the reflection of the wave front on the northern coronal hole. Between the wave and the flare sites a large dimmed area develops.

The bright front observed in SXI is also made of structures elongated in altitude (see the almost radial structures above the limb in Figs. \ref{figure3} and \ref{figure3c}). Every location reached by the SXI wave on the limb stays very bright until 18:26 UT. Therefore, the wave front is wider in SXI ($\approx 246$ Mm) than in EIT ($\approx 87$ Mm) at early time (see the elongated structures above the limb at 17:46 UT in 195 \AA~ in Fig. \ref{figure2ter} and in soft X-ray in Fig. \ref{figure3c}). Later, 
the wideness of the wave front becomes of the same order in the two band 
passes ($\approx 158$ Mm in SXI and $\approx 131$ Mm in EIT, at 
17:58 UT at the location 5). As every locations reached by the wave front remain bright for few minutes at the early stage, a dimmed region between the flare and the wave front, takes few minutes to develop: it appears after 17:57 UT between locations 5 and 4 and extends to reach location 2 at 18:05 UT. This dimming is overlaid to the one observed in EIT, but it is less deep and uniform than in EIT.

\begin{figure*}

\includegraphics[width=3cm]{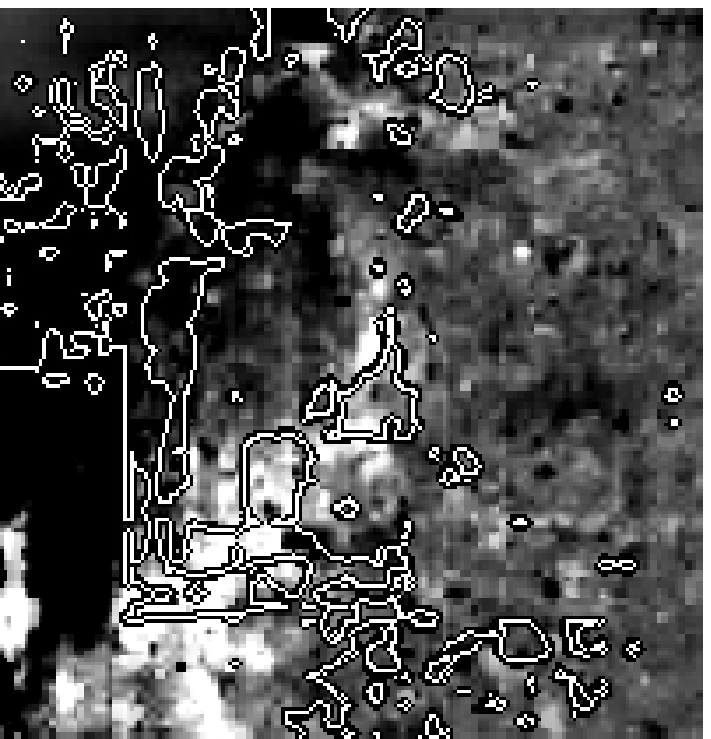}
\includegraphics[width=3cm]{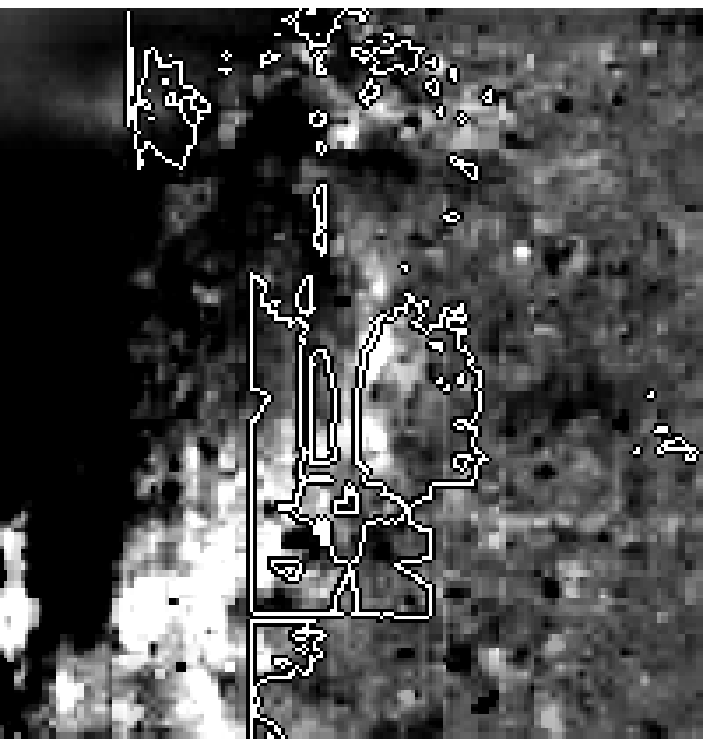}
\caption{Right (left) image is an overlay of the Fe{\sc xii} at 18:10 UT and contours of the emission through the "Polyimide Thin" at 18:13 UT ("Polyimide Thick" at 18:12 UT, resp.).}
\label{figure3i}
\end{figure*}

The SXI wave front on the disc is not clearly visible as it is more diffuse and appears in a more noisy environment due to the presence of many bright points. At 17:45 UT the wave front passes on the disc. The bright point at location 3b is illuminated since 17:45 UT (see Fig. \ref{figure3}) and is fully co-spatial to the one observed in Fe{\sc xii} at 17:46 UT. The same location appears dark in Figs. \ref{figure3g} and \ref{figure3h}, because the polynomial fitted surface of the scattered light that is removed as a background from the images, oscillates close to the border of the selected sub-field of view. This oscillation is not real and the dark appearance of the location 3b is an artifact of the processing of image. The wave front appears as a faint arch that can be overlaid to the Fe{\sc xii} wave front, having the same shape and behavior (Fig. \ref{figure3i}). It illuminates some bright points on it passage, for example 4t appears at 17:49 UT and 4b at 18:05 UT, both last until 18:37 UT in the «Polyimide Thin» data. 4b is almost disappeared at 18:20 UT in the «Polyimide Thick» data. 

\begin{figure*}
\includegraphics[width=6.2cm]{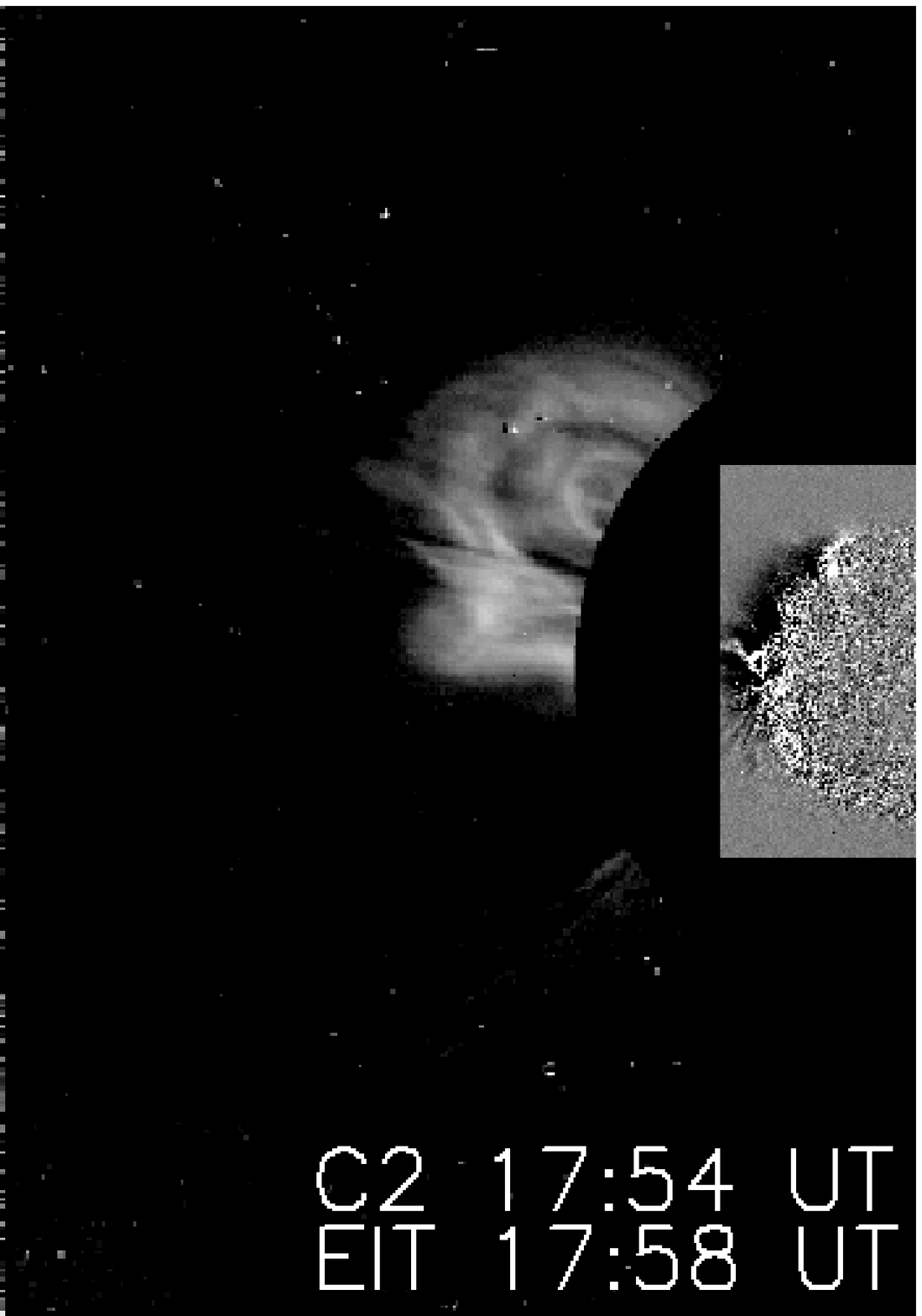}
\includegraphics[width=6.2cm]{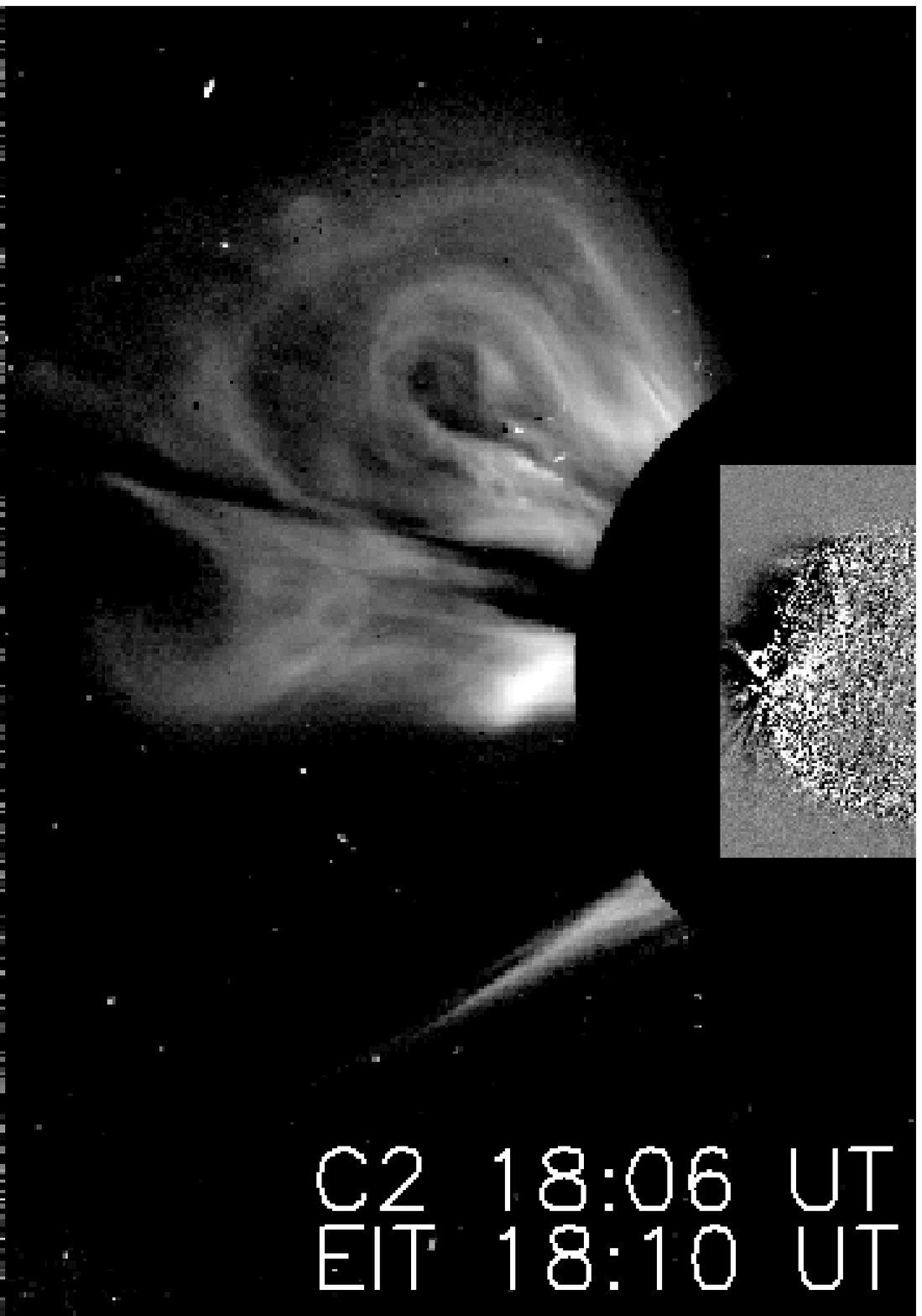}
\includegraphics[width=6.2cm]{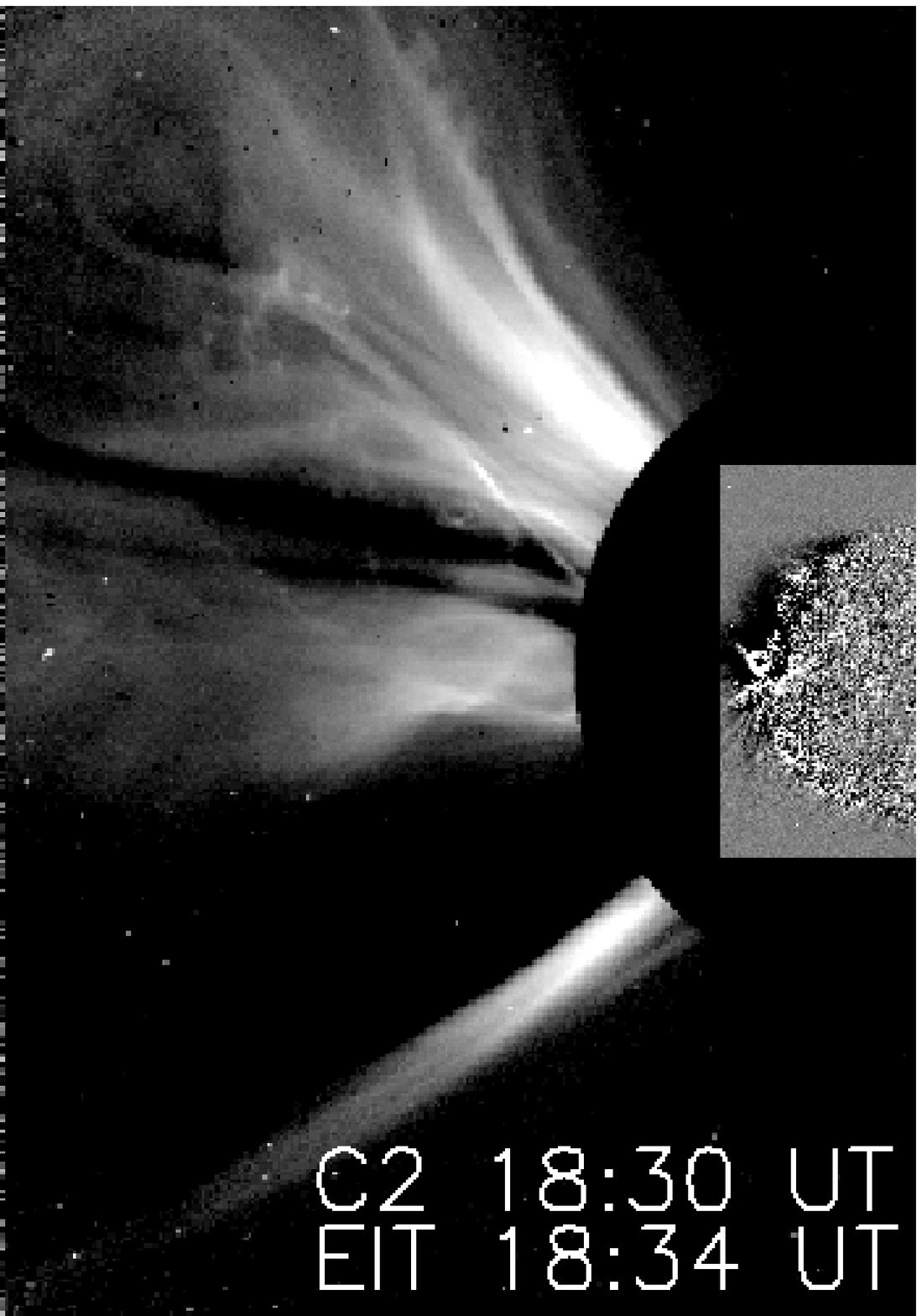}
\includegraphics[width=6.2cm]{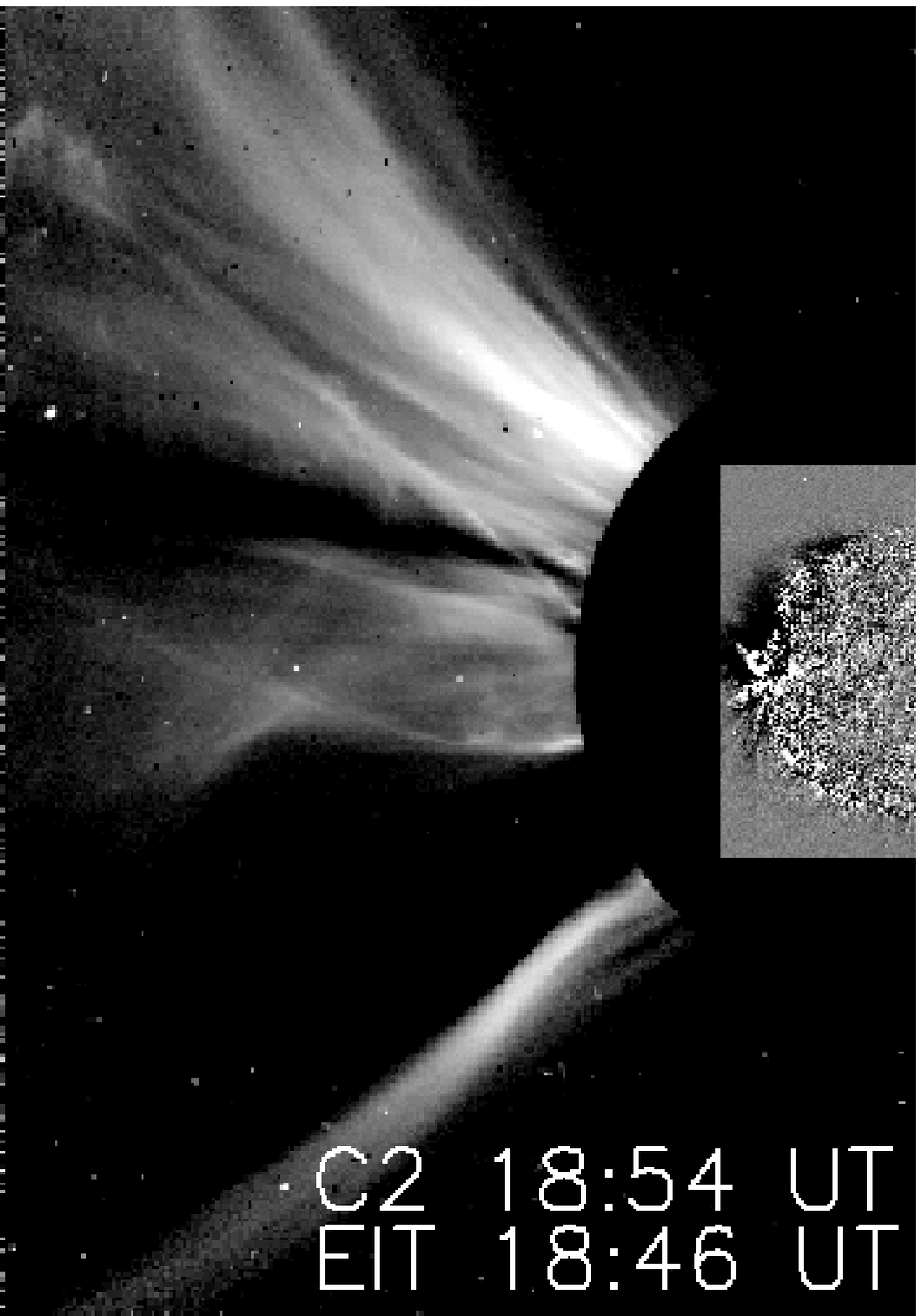}
\includegraphics[width=6.2cm]{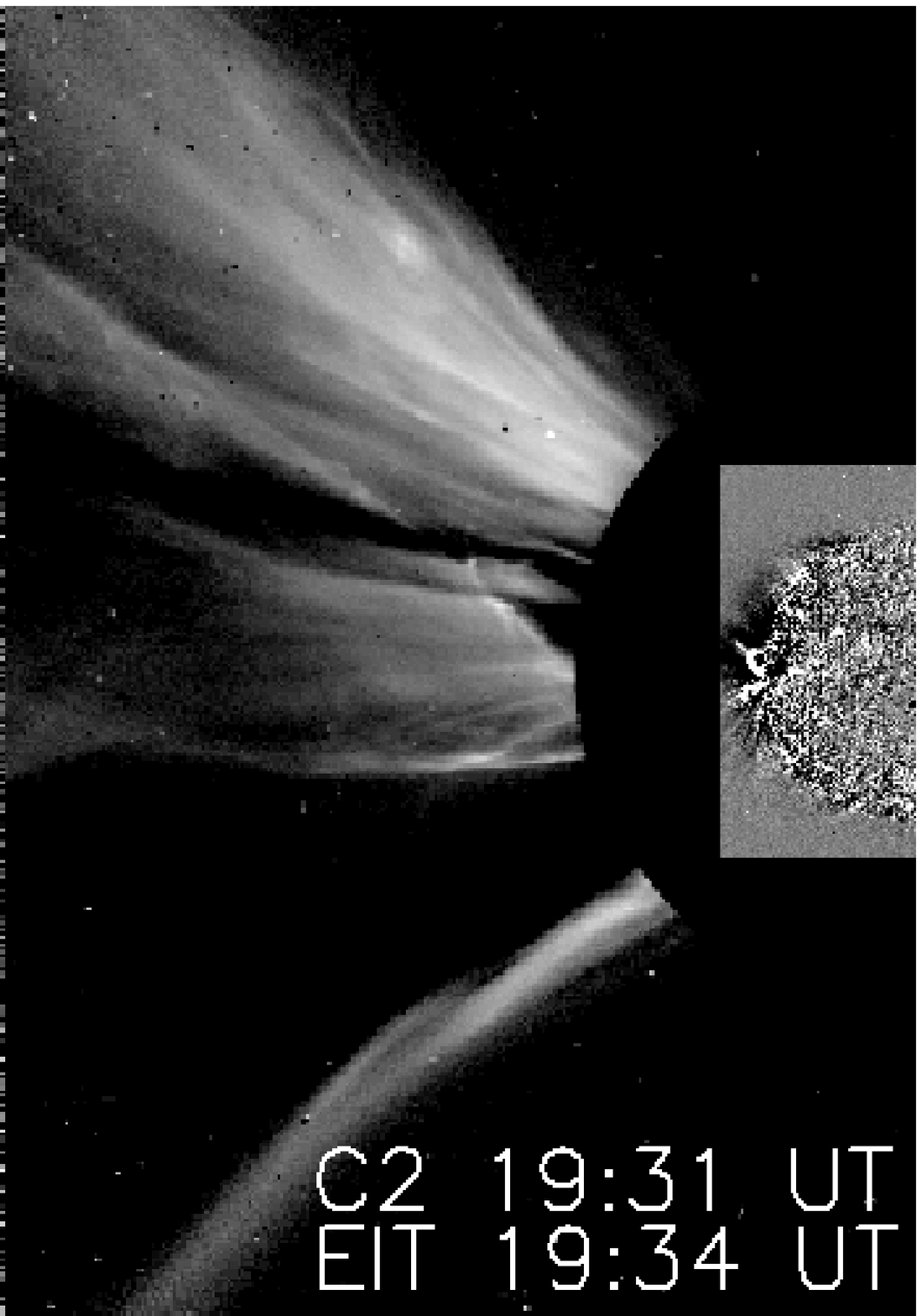}
\caption{Overlay of the same DBDIs obtained with the EIT data and 
BDIs obtained with LASCO C2. The dimming sector close to the solar 
disc can be extended to the sector covered by the brightness in LASCO C2. 
The northern leg of the CME in the first image at 17:58 UT corresponds to the 
brightness appearing where the EIT and SXI wave stops in its farther 
north and last location.}
\label{figure4}
\end{figure*}

The LASCO C2 images show the CME related to this event 
(Fig. \ref{figure4}). The CME appears in the northern east 
quadrant of the corona at 17:54 UT. On the combined 
LASCO/EIT images, the sides of the CME can be drawn back to the 
limb and intersects the most in the north location of the EIT wave 
front at 17:58 UT (location 5 in Fig. \ref{figure3}). In EUV, this 
location then dims. After 18:10 UT some loops are brightened 
(see label 4 in Fig. \ref{figure3}). Again the CME leg can be 
drawn back to the solar limb and intersects those brightened 
loops. The southern CME leg is related to the southern edge of 
the flare. I note that a jet 
located in the southern hemisphere appears during the CME. This jet 
is progressing at the same speed than the CME. Its shape seems 
deformed and enlightened by the progression of the CME.

\subsection{Some understanding in light of magnetic field topology}
\begin{figure*}
\includegraphics[width=9.cm]{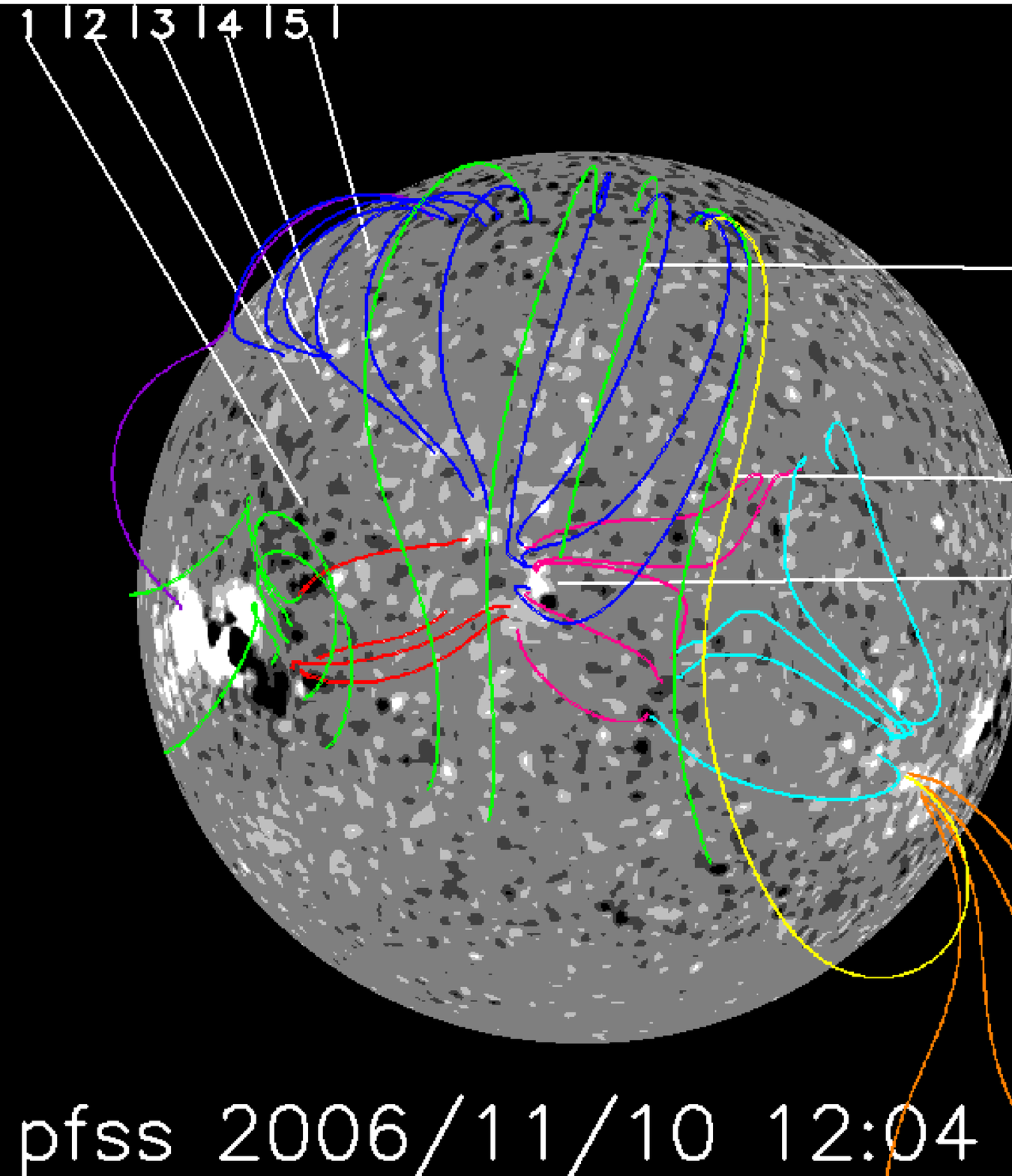}
\includegraphics[width=9.cm]{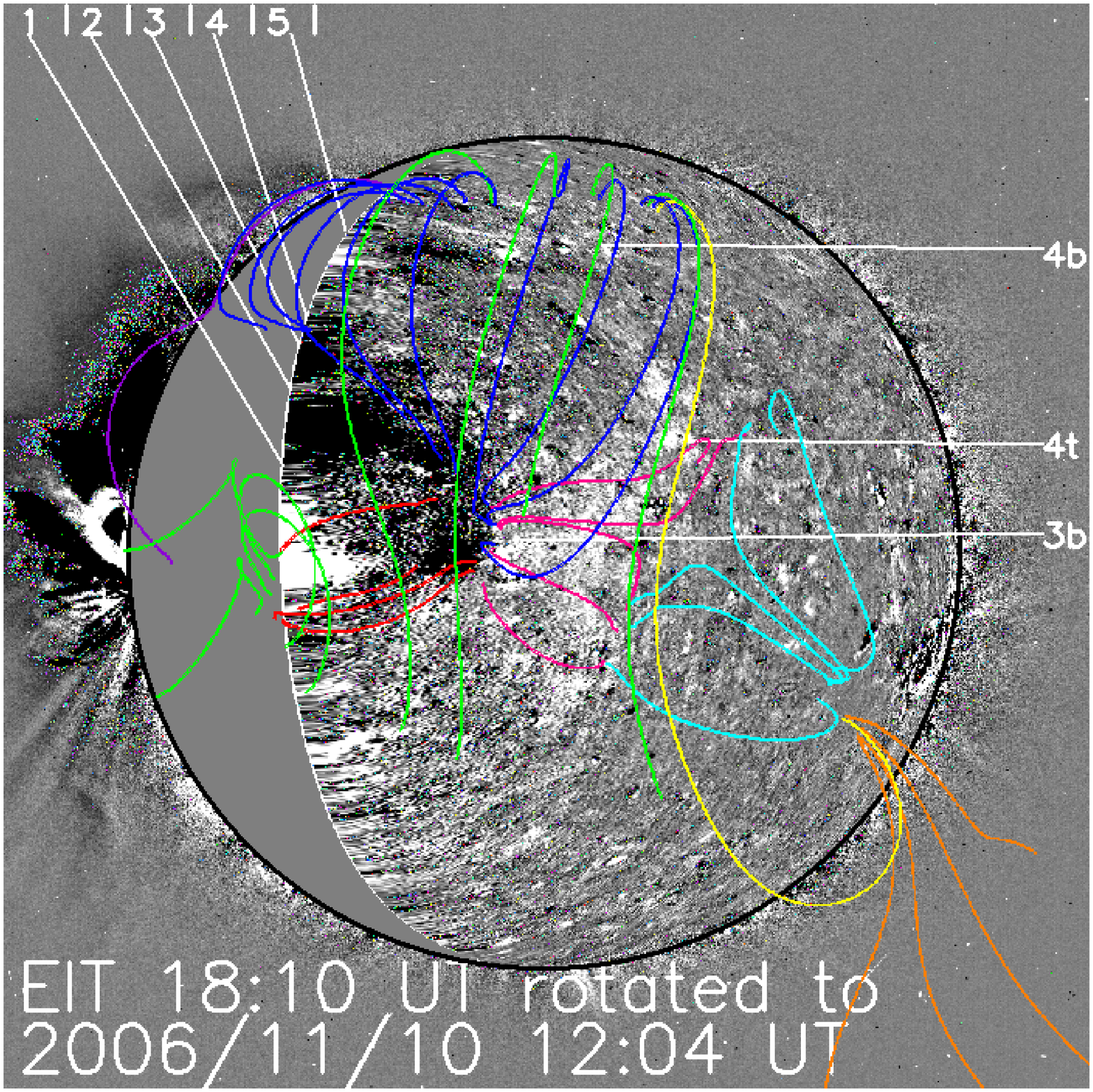}\\
\includegraphics[width=9.cm]{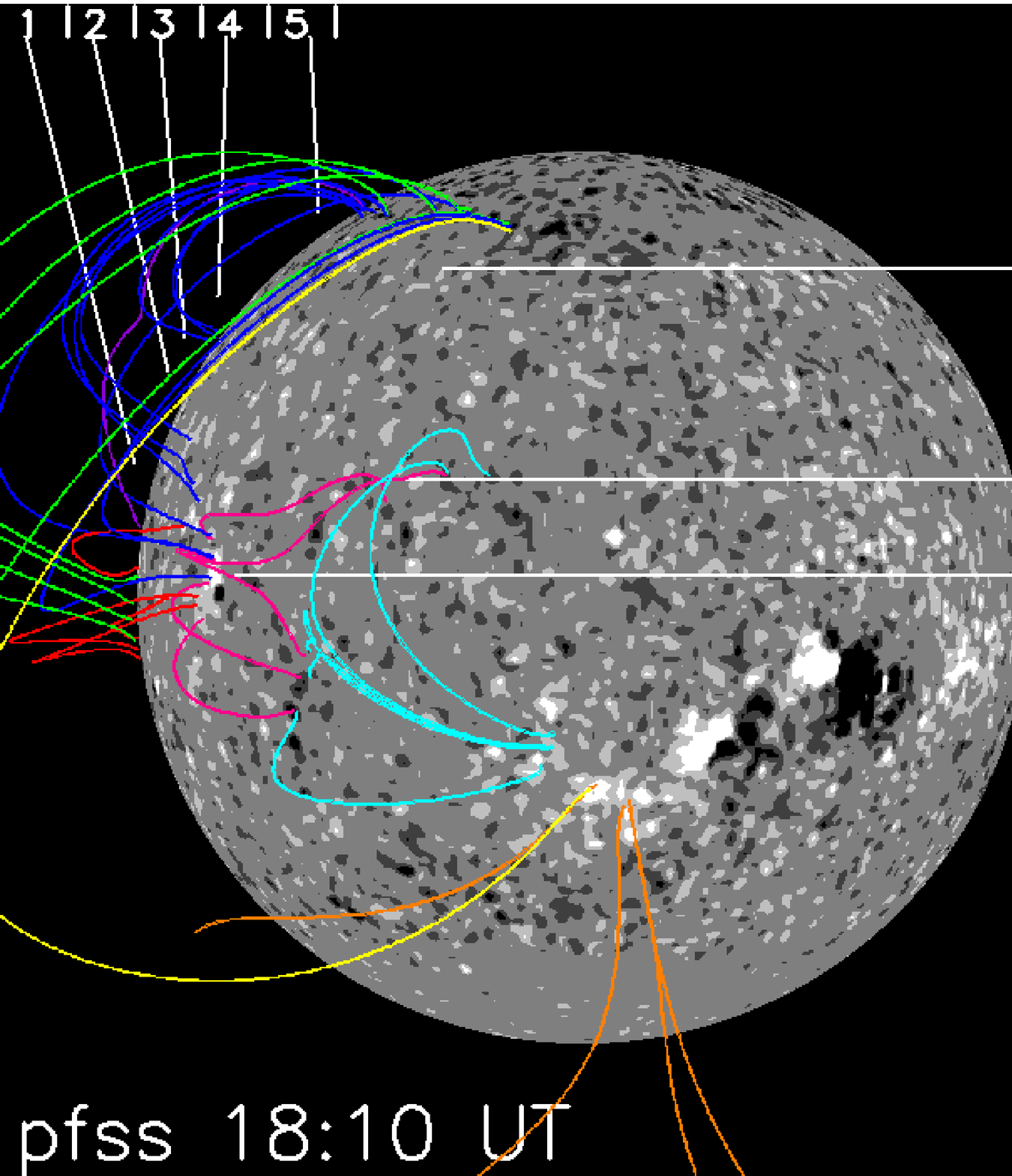}
\includegraphics[width=9.cm]{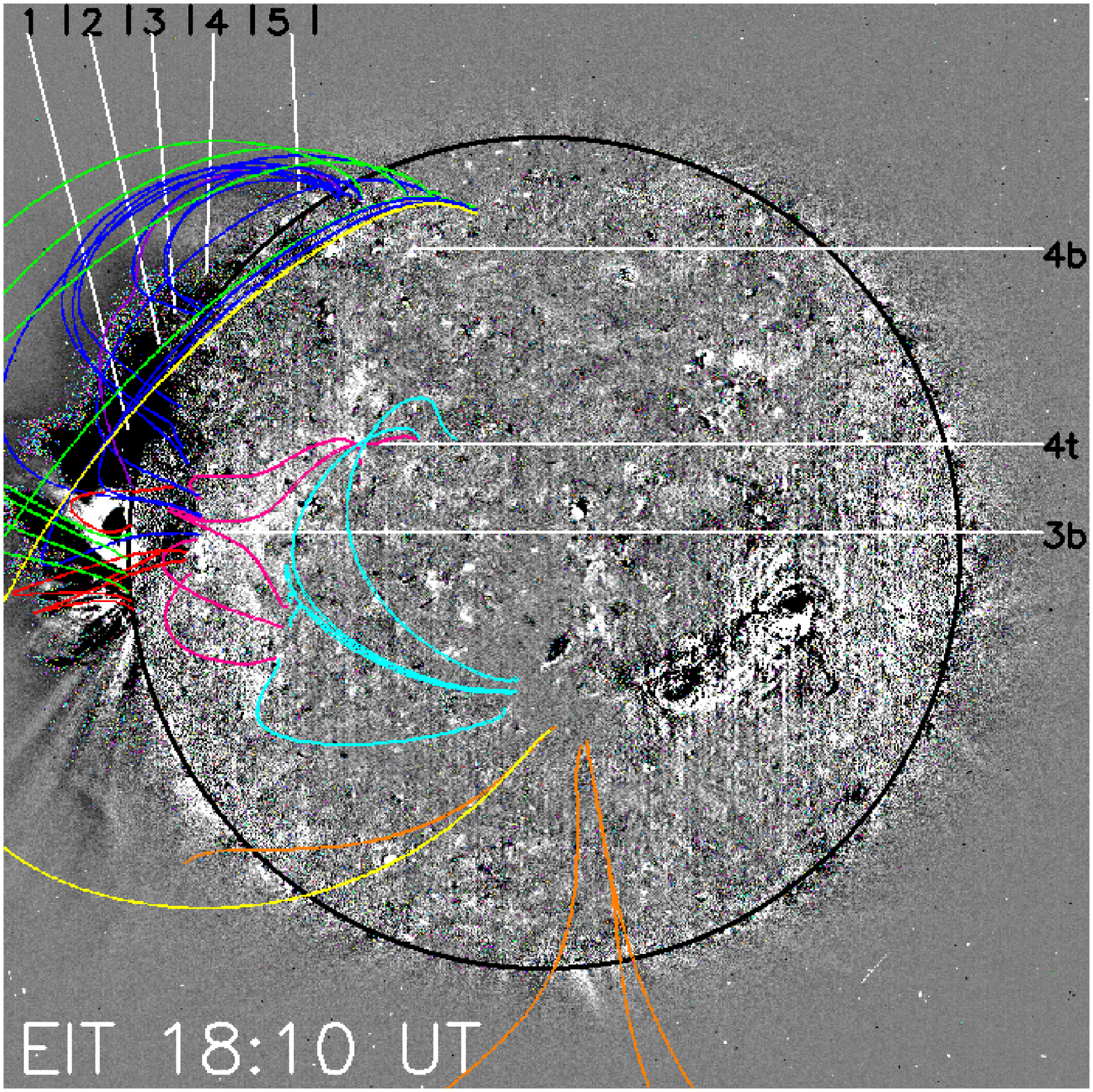}\\
\caption{Magnetic map (left) and BDBI at 195 \AA~ (right) overlaid with
some magnetic field lines that define some magnetic jumps of connectivity
rotated to November, 10 2006 at 12:04 UT (upper) and to November, 6 2006
at 17:25 UT (lower). The stationary brightenings are lying in 
magnetic field lines jumps of connectivity.}
\label{figure5}

\end{figure*}

The pfss extrapolation of the region involved in the event 
gives some understanding of those structures (Fig. \ref{figure5}). The magnetic field lines are drawn in different colors to highlight the magnetic field topology. Despite the fact that the projection on November 6 (lower left panel of Fig. \ref{figure5}), makes the green magnetic field lines appearing closed, connecting the northern pole to the active region, the projection on November 10 (see upper panel of Fig. \ref{figure5}) show that they are not connected at all. Those green lines are all opened (i.e. they connect the solar surface to the source surface of the high corona), they have the same orientation, entering inside the sun. The orange ones are also opened but in the inverse orientation, connecting the source surface to the positive magnetic polarity of AR 0922 located S09 W13. The red ones connect the location 3b to the flaring region. The blue ones connect the location 3b to the northern coronal hole. The pink ones connect the location 3b to the location 4t. The cyan ones connect the southern coronal hole in AR 0922 to the location 4t.

A coronal hole lies between the eastern limb and AR 0923 (see the open green 
magnetic field lines on the left upper panel of 
Fig. \ref{figure5}). I checked the stability of this 
coronal hole. The extrapolation of the magnetic field shows this 
coronal hole from November 5 to November 13 2006, even if the 
magnetic polarity in AR 0923 changes during those 8 days. As 
frequently observed the waves are not crossing coronal holes 
(Thompson et al. 1999). So, I believe that the eruption takes 
place close to red magnetic field lines drawn in the southern hemisphere on the magnetic map of November, 10 2006. The EUV dimming located between the flare site and the locations 1 and 3b is under these red lines indicating that some of those magnetic field lines are expanding during the eruption.

The stationary brightenings 3b, 4b and 4t are lying in drastic jumps of magnetic field lines connectivity: 3b is at the footpoints of the red and pink magnetic field lines, 4t at the footpoints of the pink and cyan ones, 4b at the footpoints of the blue and green ones. A fainter brightness, visible in EIT on the disc in the southern east quadrant, lies in the pink magnetic field lines (see the lower right panel of Fig. \ref{figure5}). This brightness does not cross the pink magnetic field lines footpoints which are close to the cyan ones. The pink magnetic field lines define the limit of a topological domain. The blue magnetic field lines also show the approximate location of the limit of a magnetic topological domain. If the red magnetic field lines are opening, they produce the perturbation of these limits (blue, pink magnetic field lines and at location 3b, 4b and 4t) in which the plasma can be heated, and brightened in soft X-rays. The blue and pink magnetic field lines are possibly not opened during the eruption. When 
the blue magnetic field lines are viewed as on November 6, 2006, 
they are all superposed onto the plane of the sky, therefore, during the eruption they will be observed as bright elongated structures or loops that are very close to each other, as it is in fact observed in SXI. 
The stationary brightenings visible on the limb in SXI at label 5, and on the disc between the label 5 and 4b are lying at the edge of the northern coronal hole, at the footpoints of the blue and green magnetic field lines. Again, during the eruption, this limit of magnetic field domain is perturbed producing the heating of the plasma that is observed as bright points.

The orange magnetic field lines projected onto the plane of the sky on November, 6 2006 has the shape of the perturbed streamer observed on the 
LASCO C2 images (lower left panel of Fig. \ref{figure5}). Therefore, the distortion of those orange magnetic field lines is possibly due to the expansion of the red ones that push them. As those magnetic field lines also define the limit of a magnetic field line connectivity, the plasma is heated, producing its brightening.

I here note that the described phenomenon resembles to the one in Terradas and Ofman (2004). In this article, the top of the loops present over density and are compared to the brightened loops that enter in oscillation while an observed wave passes through them. This phenomenon cannot be applied to explain the observed structures in the present study as none loops present a brightened apex, on the contrary, or they are brightened all along their length or only their footpoints are brightened. The reason is that to produce the kind of oscillation presented in Terradas and Ofman (2004), the motion of the perturbation has to be transverse to the loop direction, and in our observations, the pfss extrapolation reveals that the motion of the perturbation is along the magnetic field lines.

\subsection{Temperature}

\begin{figure*}
\includegraphics[width=18.0cm]{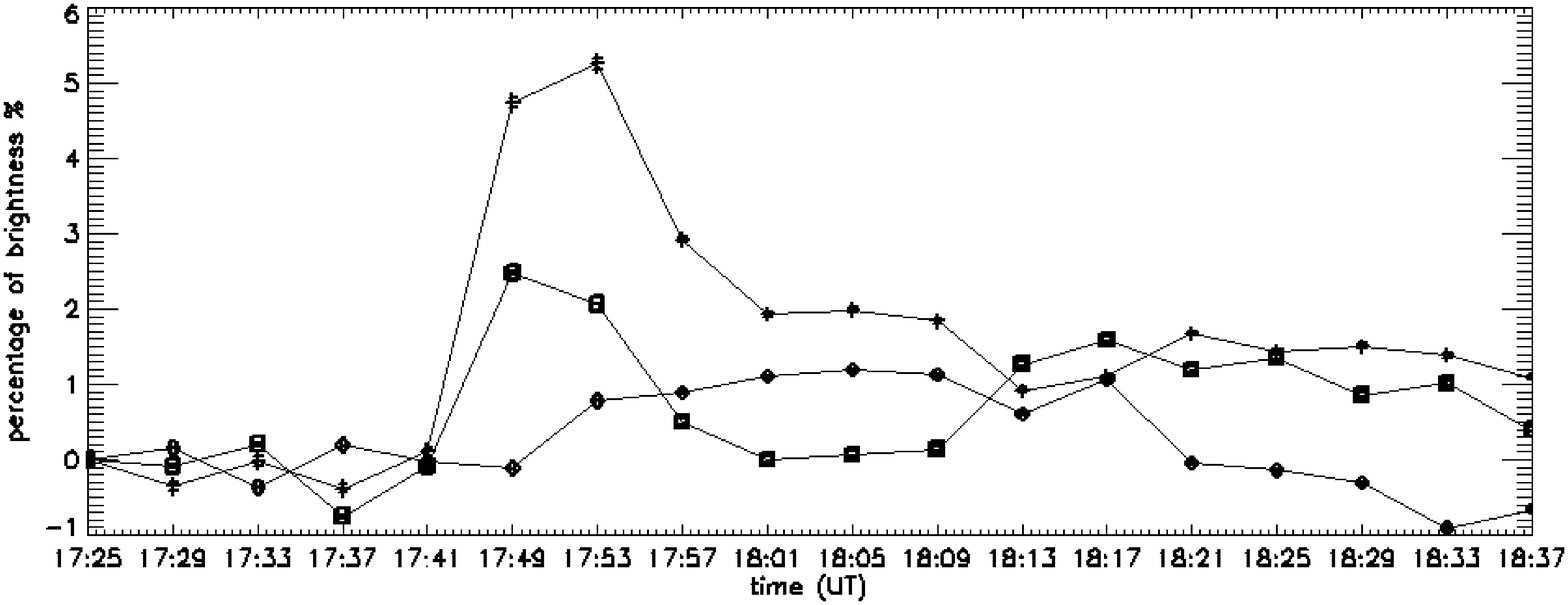}
\includegraphics[width=9.0cm]{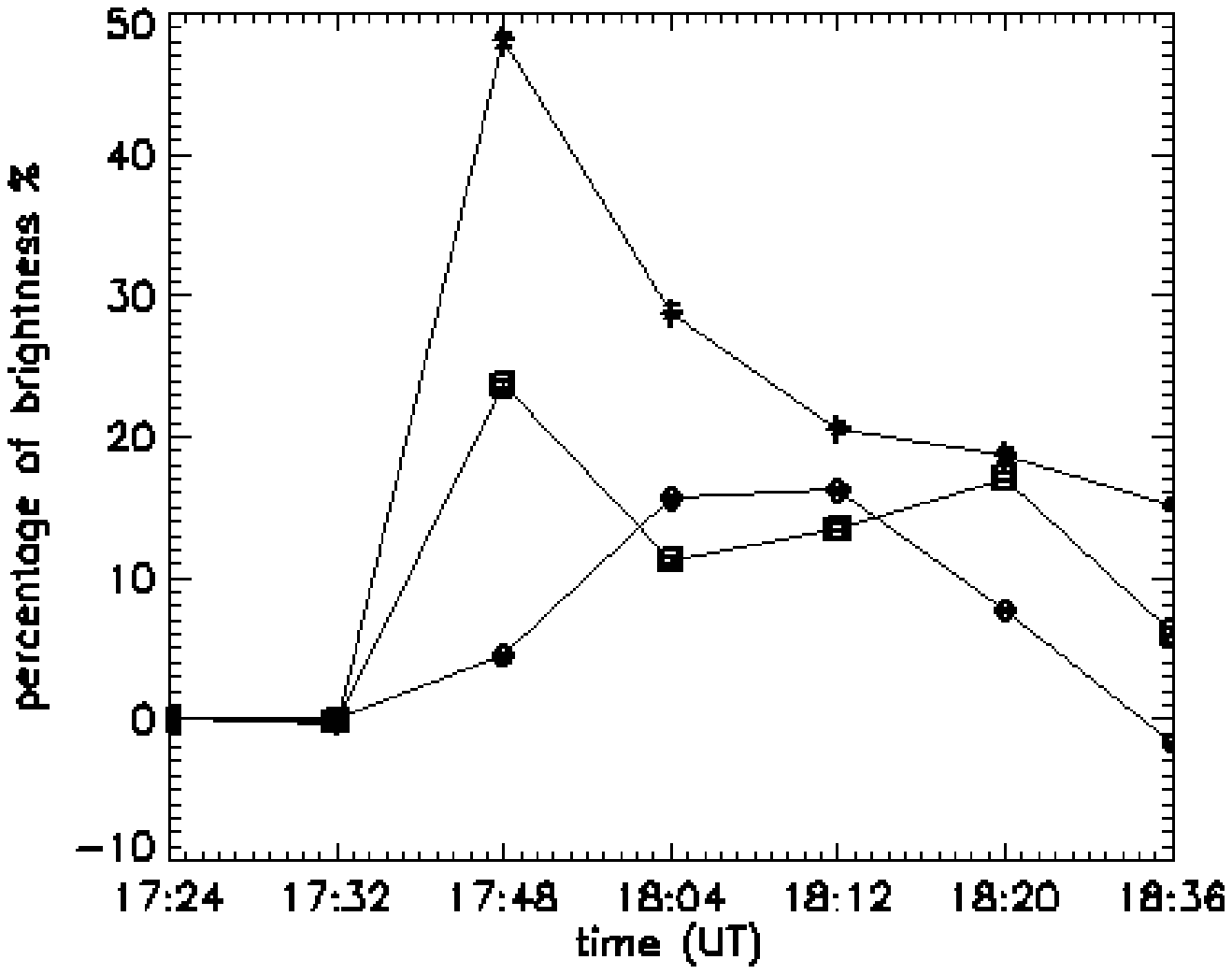}
\includegraphics[width=9.0cm]{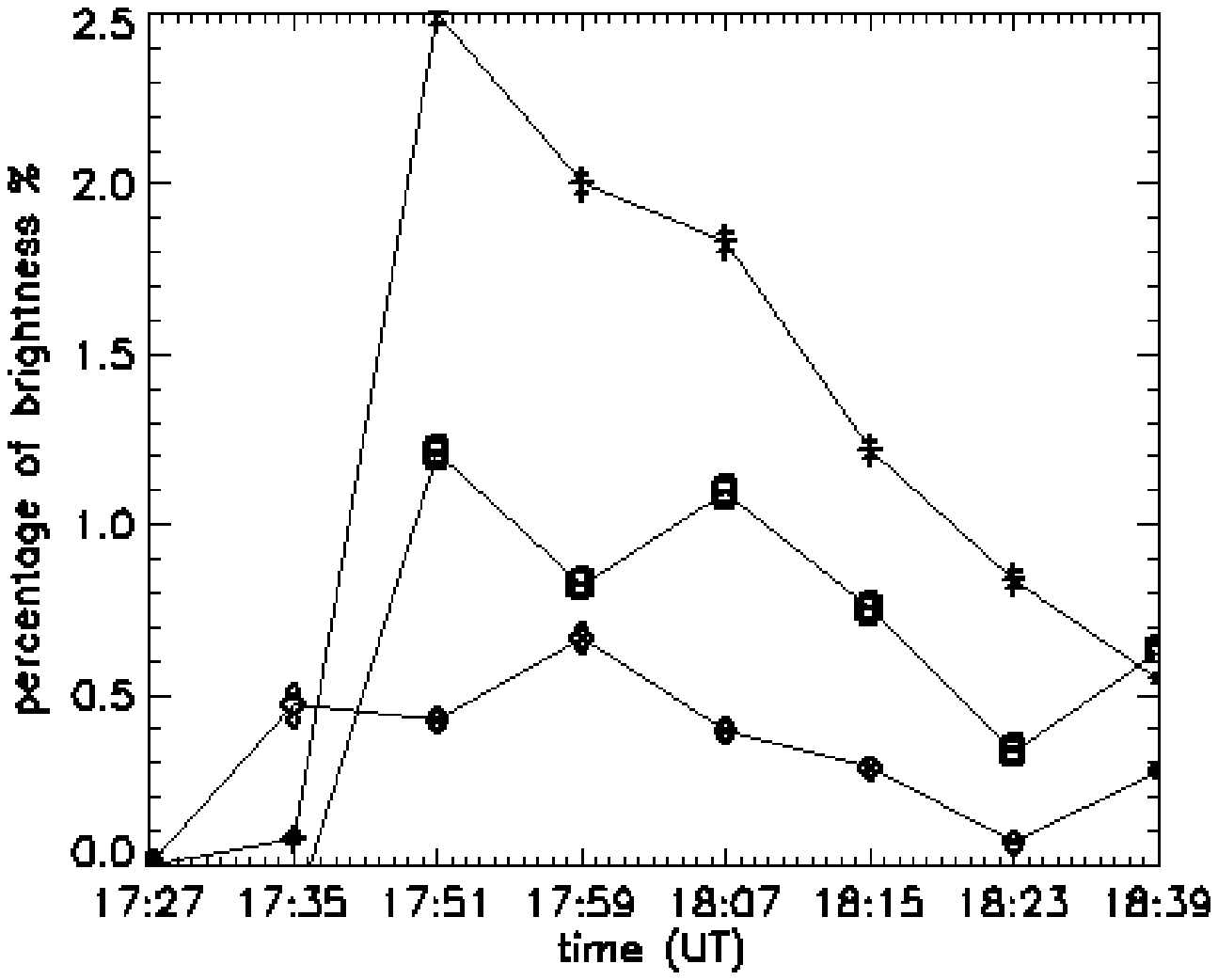}
\caption{Curves of the percentage increase of the mean
intensity in the squares drawn in the image at 18:36 UT 
in Fig. \ref{figure3} with the initial mean intensity 
in the same areas plot against the time, using the "Polyimide Thin" 
filter (top panel), the "Polyimide Thick" filter (low left), and the 
"Beryllium ThinA" filter (low right). Crosses are for the area 2, squares for 4 
and diamonds for 5. Superimposed points
give the error made on their values.
The light curves reveal the propagation 
of the wave front above the limb, each reached location 
remaining bright for a while.}
\label{figlightcurve}
\end{figure*}

I now estimate the temperature of the wave front, focusing on the elongated structures that have their footpoint at locations 2, 4 and 5. As the coronal wave is almost not visible in H$\alpha$, it is made of hot plasma. 
The most in the north location of the wave front, labeled 5, and the location  4 (Fig. \ref{figure3}), emits light that passes through the "Polyimide Thick" filter from 18:04 UT 
to 18:12 UT. In the same interval of time, the same locations are very 
dimmed in EUV (see image at 18:10 UT in Fig. \ref{figure2}). The temperature range of the emission in SXR is from 1 to 10 MK and in EUV the temperature 
of the principal spectral emission line Fe{\sc xii} is 
about 1.5 MK. Therefore, the SXR brightness located at the 
labels 5 and 4, is due to plasma with 
temperature higher than $1.5\ 10^6$ K. I estimate the mean 
emission over areas located above the limb near the labels 2, 4 and 5 
from 17:24 UT to 18:36 UT (see the squares 
on the image at 18:36 UT in Fig. \ref{figure3}). The 
percentage increase of the mean light intensity over the 
initial mean light intensity in each filter and in each area 
is plotted against the time in 
Fig. \ref{figlightcurve}. The signal of the wave in the "Beryllium Thin A" 
filter is not detectable which is confirmed by the DBDI images (see the movie) - the increase of emission through this filter is due to the scattered light of the flare, proved by the fact that the emission decreases from position 2 to position 5 at the same time and also monotonic decreases with the time at each location - therefore the coronal wave temperature is less than 10 MK. 

Through the "Polyimide Thin" and "Polyimide Thick" filters
the temporal intensity variations show an abrupt increase at 17:49 UT (17:48 UT) at location 4 and at 17:53 UT (18:04 UT, resp.) at location 5, revealing the propagation of the wave front and also that the position 5 is weakly affected by the scattered light of the flare. All light curves verify that each location remains bright for 
several minutes. At the location 5, the brightness continues to increase for a 
while, then decreases slowly. At location 4, the intensity through the "Polyimide Thin"' filter is less than prior to the eruption between 17:57 UT and 18:09 UT, corresponding to the formation of the dimmed region, but then increases and stays quite high until the end of the sequence of observations. 

The intensity variations are very small, of order of $0.5-2.5 \%$ more than prior to the eruption through the "Beryllium Thin A" filter, of $1-5.5 \%$ through the "Polyimide Thin" filter and of $10-30 \%$ through the "Polyimide Thick" filter. 
The errors made on these estimates are due to a statistical enlightening of
a pixel hit by a photon that does not come from the pointed 
structures. The standard deviation $\sigma$ of a sample of N 
pixels with $x_i$ brightness is:
\begin{equation}
\sigma=\sqrt{\frac{1}{N}\sum^N_{i=1}{(x_i - \overline{x})}}
\end{equation}
The statistical error is at most 2 \% 
through the "Beryllium Thin A" and "Polyimide Thin" filter, and 20 \% through the "Polyimide Thick" filter. These errors are quite
small due to rather large surface of integration of the brightness, which
enlarge the sample of pixels and, therefore, reduces the statistical error.
Another source of error is the difficulty to locate
the border of the illuminating structure. Therefore the same quantities
are computed over the same area enlarged at border by 3 pixels along the 
equator direction, then reduced by 3 pixels along the equator direction.
The quantities are shown to fluctuate by at most 1 \% through 
the "Beryllium ThinA" and "Polyimide Thin" filters, and 16 \% through the "Polyimide Thick" filter. 
This fluctuation is reduced if the area of integration is enlarged, and
increased if the area of integration is reduced. This confirms that
the border of the chosen  area to integrate the brightness is very close
to the border of the studied structures and explains the small
error made. The propagated error of the variation of
average brightness, $\overline{x}(t)$, against the time, t, is given by:
\begin{equation}
	\Delta \left(\frac{\overline{x}(t)-\overline{x}(0)}{\overline{x}(0)}\right) = \frac{\Delta \overline{x}(t)}{\overline{x}(0)}+\frac{\overline{x}(t)^2}{\overline{x}(0)^4} \Delta \overline{x}(0)
\end{equation}
where
\begin{equation}
	\Delta \overline{x}(t)= \sigma(t) + \left|\overline{x}(t)-\max(\overline{x}_{-3}(t),\overline{x}_{+3}(t))\right|
\end{equation}
$\sigma(t)$ is given by equation 1 at each time of observation, 
and $\overline{x}_{-3}(t)\ (\overline{x}_{+3}(t))$ is the average brightness over the reduced 
(enlarge, respectively) area at each time of observation. The propagated error is also small 
(at most 2 \% through the 3 filters) due to the presence of the 
ratio between 2 quantities presenting similar errors. Therefore the
temporal variations of the increase of the average brightness presented
is Figure \ref{figlightcurve} can be trusted. However, these variations
at location 2 are affected by the light emitted by the flare and 
scattered in the instrument. This error is very difficult to estimate
as the pixels at the flare site are saturated, leading to the 
impossibility (up to now) to find a good fit of this scattered light.

Then I divide the mean emissions passing through the filters "Polyimide Thick" over the mean emission passing through the filter "Polyimide Thin". The ratios appear to be very constant over the time, except at 18:37 UT, and over the locations: $\approx 0.477$. Using the plots of the ratio of responses from various filters given in Lemen et al. (2004) as a
function of temperature, I find that the observed ratio corresponds to a temperature $\approx 7\ 10^6$ K for every brightenings. I note that the 
brightness of the location 2 is quite affected by scattered light, 
leading to a very poor confidence in the results at this location 
but this scattered light is very low at the location 5 leading 
to a more accurate estimate of the temperature.

The brightenings 
in SXI decrease after 18:26 UT to become even darker than prior 
to the eruption. As these new appearing dimmings are also visible 
at the same time in EUV, I suggest that the plasma is 
very depleted in this region, and that, as the plasma is depleted 
deeper with the time, the brightness due to the increase of 
temperature created on the passage of the wave can no longer 
persist. The depletion may have balanced the temperature increase 
producing the dimming in a wide range of temperature in the same region.
 
\section{Discussion}

The wave is barely visible or even invisible in H$\alpha$ which can be due to the fact
that the temperature of the wave is higher than the chromospheric
temperature. As it has been shown in other studies, the H$\alpha$, Fe{\sc xii}
and SXR wave fronts are co-spatial when observe during a same event.
I fully believe that the lack of emission of the studied wave front
is due to a particular physical parameter of this wave.
The plasma could be too hot but also it is possible that the morphology
of the wave front influences its observability. The Fe{\sc xii}
and the SXI emissions are very faint above the limb
because the structures are seen from their sides leading to a low accumulation
of their emission in their optically thin plasma. For the same reason of geometry, 
above the limb, the H$\alpha$ emission could be too faint to be detected. 
However, the part of the wave front propagating on the disc is also not 
visible, which could be due to the total
absence of the H$\alpha$ emission in this studied wave.

As many other cases of coronal wave observed conjointly in H$\alpha$
and in coronal emissions are presented in the literature, I ask the question:
what additional process may lead to the observability of these waves in H$\alpha$
that is missing in the present case. Several authors believing that the observed coronal waves
are magnetosonic waves suggest that the H$\alpha$ wave are different magnetosonic
wave mode than the coronal wave mode, therefore, in the present case, this magnetosonic wave 
mode should be not generated. Others suggest that the
coronal wave can sometime dig into the corona to produce the H$\alpha$
emission, therefore, in the present case, the wave should be too weak to 
dig in the chromosphere. However, I remind to the reader that several arguments
discredit the magnetosonic wave model for the observed wave, leading me to
disregard this model and the interpretation that the H$\alpha$ emission is not a different wave than the coronal one. I know other coronal structures that can emit in H$\alpha$ (prominences, post-flare loops), I here suggest that the additional process could be a thermal cooling instability that appears only in some particular plasma conditions of temperature and density. This suggestion is already proposed by Delann\'ee et al. (2008) and Balasubramaniam et al. (2007). It supports also the observations obtained by Delann\'ee et al. (2007). In this latter study,
the H$\alpha$ wave front is slightly shifted inside the Fe{\sc xii} wave front, which
could indicate that the wave front emits at coronal wavelength and takes 
some few time to emit in H$\alpha$, as it would be in a case of thermal cooling 
instability.

The coronal wave front is found to be footprint of the CME leg.
This is in contradiction with Vr{\v s}nak et al. (2006) and Patsourakos 
and Vourlidas (2009), who both conclude that 
the coronal wave front is ahead of the CME leg. However, the 
DBDI method is not used in the latter analysis which instead use running 
difference images. This method is not suited to detect slowly inflating 
brightenings like the ones found in our set of observations. Doing running 
difference images of a fading bright slowly inflating structure
produces a dark feature enclosed by a bright and sharp border.
Therefore, the brightenings produced by a coronal wave cannot be 
well represented using the running difference image process and 
can produce the discrepancy between the wave front location 
and the CME leg as the one shown in the Fig. 4c in Vr{\v s}nak et al. (2006) 
and in the Figure 4d in Patsourakos and Vourlidas (2009).

\section{Conclusion}

I present conjoint observations in H$\alpha$, in soft X-ray and in 
195 \AA~of a coronal wave. The bright front
observed in the two last wavelengths are co-spatial when observed 
together, so it seems very unlikely that they are different 
structures. The wave does not emit in H$\alpha$.

The observability of the coronal wave in soft X-ray indicates that 
it is a hot coronal structure with temperature $\approx 7$ MK. 
This means that heating processes are at work to form the coronal 
wave. This conclusion is coherent with the one given by Wills-Davey 
et al. (2000) from the observation of a coronal wave in Fe{\sc xii}
that is not well observable in Fe{\sc ix} (the temperature of 
formation of this spectral line is about 1.3 MK).

The wave is not exactly fully propagating when analyzed using DBDI: it 
produces stationary brightenings (bright points and elongated in altitude 
structures) on its passage then stops (the diffuse arch stops). All these 
produced structures last for tens of minutes in soft-X ray and about one hour in 
195 \AA. The existence of stationary brightenings of the so-called coronal 
waves is also coherent with 6 previous observations (Delann\'ee 
and Aulanier 1999, Delann\'ee 2000, Delann\'ee et al. 2007a, 
Attrill et al. 2007a). This event is the first one obtained close 
to the limb, and in soft X-ray, presenting stationary brightenings.

This is the fourth case studied using DBDI and comparison with 
magnetic field topology (see also Delann\'ee and Aulanier 1999 and 
Delann\'ee et al. 2007, Delann\'ee 2009). For these four studies 
the stationary brightenings produced on the solar disc on the passage of the 
coronal wave are lying in jumps of connectivity of the magnetic 
field lines.

The wave is produced conjointly with a CME. This is coherent 
with the conclusion of Biesecker et al. (2002): a wave is 
always accompany by a CME. This observation also shows that 
the final last produced stationary brightening is the footprint 
on the sun of one leg of the CME appearing above the coronagraph occulter. 
The dimmings close to the solar limb are located inside the CME footprint. 

The wave front consists of elongated in altitude structures that draw the 
lower parts of magnetic field lines perturbed or opening during the CME.

A streamer is also observed disturbed, taking a small curve as if 
there was a wave propagating along it. The brightness of the streamer 
increases as well as its thickness. This perturbation appears during 
the evolution of the CME, just 12 minutes after the first appearance 
of the CME. The magnetic field extrapolation shows that the streamer 
is rooted at the border of a coronal hole where the wave also stops. 
All these facts 
show that the coronal wave, the CME and the perturbation of the 
streamer are connected through perturbation of the large scale
magnetic field lines. 

There exist several models of those three phenomenons. The models
of the coronal waves can be divided in two types: magnetosonic wave
or signature of the magnetic field restructuring during the opening 
of magnetic field lines. Studying only one event cannot 
permit to distinguish between those two origins. However, 
I exclude the first possible mechanism for the reasons given in the 
introduction section. 
I believe that the coronal waves are the edges of the same opening 
magnetic flux tube that also produce the observed CME (Delann\'ee et al. 2007, 2008). 
This last model is supported by the observations and the analysis performed in this 
article.

The type II burst observed during this coronal wave could be 
interpreted as the presence of a shock. This fact led 
Khan and Aurass (2002) to reject the idea that coronal waves 
are a signature of opening of the magnetic field lines, and to 
reassure that the coronal waves are magnetosonic shocks. However, 
it seems to me that the presence of the 
type II conjointly with a coronal wave has to be more studied from 
the point of view of the opening magnetic field line model.

\begin{acknowledgements}
The author thanks John Leibacher for providing the OSPAN data. 
Douglas Biesecker for providing the SXI data with some corrections
already done (roll, intensity calibration, image cleaning), and his 
advises and comments. Andre\"i Zhukov and Christophe Marqu\'e for 
their deep reading of this manuscript.
\end{acknowledgements}

%\begin{figure*}
%\includegraphics[width=4.0cm]{\Eto\0613.filtre.eps}
%\includegraphics[width=4.0cm]{imagesps.version2/3nov97Ha.eps}
%\includegraphics[width=4.0cm]{imagesps.version2/3nov97sxr.eps}
%\includegraphics[width=4.0cm]{imagesps.version2/3nov97Ha.contour.eps}
%\caption{images from Narukage et al. (2002). Each images is a difference 
%of an image with the previous one. First image is obtained
%in H$\alpha$+0.8 \AA~at 04:38:01 UT at the HIDA observatory. Second image
%is obtained in soft X-ray at 04:37:48 UT using the soft X-ray telescope
%on board the Yohkoh satellite. Third image is the H$\alpha$+0.8 \AA~
%with contour of the soft X-ray wave front. The two wave fronts are co-spatial.}
%\caption{Large view of the sun (right) presented in the image at 06:13 UT of 
%Figure 5 in Eto et al. (2002) and a closer subfield of view of the location of 
%the EIT wave and the oscillating filament obtained by DBDI, intensity hyperbolic smoothing and wavelet filtering. The wave front, maked by the solid line, appears well ahead of the oscillating filament.}
%\label{figeto}
%\end{figure*}

\begin{thebibliography}{}

\bibitem[]{}
Aulanier, G., Pariat, E., D\'emoulin, P. \& DeVore, C.R. 2006, Sol. Phys., 238, 347

\bibitem[2007a]{}
Attrill, G., Harra, L., van Driel-Gesztelyi, L., D\'emoulin, P. 2007, ApJ, 656L, 101

\bibitem[2007]{}
Balasubramaniam, K. S., Pevtsov, A. A., Neidig, D. F. 2007, ApJ, 658, 1372



\bibitem[]{}
Biesecker, D. A., Myers, D. C., Thompson, B. J., Hammer, D. M., Vourlidas, A. 2002, ApJ, 569, 1009

\bibitem[]{}
Brueckner, G. E., Howard, R. A., Koomen, M. J., Korendyke, C. M. 1995, Sol. Phys., 162, 357

\bibitem[2002]{}
Chen, P. F., Wu, S. T., Shibata, K., Fang, C. 2002, ApJ, 572, L99

\bibitem[2005]{}
Chen, P. F., Fang, C., Shibata, K. 2005, ApJ, 622, 1202

\bibitem[2007]{}
Chen, P. F. 2006, ApJ, 641L, 153 

\bibitem[]{}
Cliver, E. W., Laurenza, M., Stolini, M., Thompson, B. J. 2005, ApJ, 631, 604

\bibitem[]{}
Delaboudini\`ere, J.-P., Artzner, G. E., Brunaud, J. et al. 1995, Sol. Phys., 162, 291

\bibitem[1999]{}
Delann\'ee, C. and Aulanier, G. 1999, Sol. Phys., 190, 107

\bibitem[2007]{}
Delann\'ee, C., Hochedez, J. F., Aulanier, G. 2007, A\&A, 465, 603

\bibitem[2009]{}
Delann\'ee C. 2009, A\&A, 495, 571

\bibitem[]{}
Delann\'ee, C., T\"or\"ok, T., Aulanier, G., Hochedez, J.-F. 2008, Sol. Phys., 247, 123

\bibitem[]{}
Dere, K. P., Brueckner, G. E., Howard, R. A. et al. 1997, Sol. Phys., 175, 601

\bibitem[]{}
Eto, S., Isobe, H., Narukage, N. et al. 2002, PASJ, 54, 481

\bibitem[]{}
Gary, G. A. 2001, Sol. Phys., 203, 71

\bibitem[]{}
Gilbert, H. R., Holze, T. E., Thompson, B. J., Burkepile, J. T. 2004, ApJ, 607, 540

\bibitem[]{}
Khan, J. I., Hudson, H. S. 2000, GRL, 27, 1083

\bibitem[]{}
Khan, J. I., Aurass, H. 2002, A\&A, 383, 1018

\bibitem[]{}
Klassen, A., Aurass, H., Mann, G., Thompson, B. J. 2000, A\&AS, 141, 357

\bibitem[2001]{}

Klimchuk, J. A. 2001, Proc. of the Chapman Conference on Space Weather, AGU, Geophysical Monograph Series 125, (eds) Song, P., Singer, H., Siscoe, G., 143

\bibitem[]{}
Lemen, J. R., Duncan, W., Edwards, C. G., Friedlaender F. M., Jurcevich, B. K., et al. 2004, "The solar x-ray imager for GOES", Proc. SPIE 5171, Telescopes and Instrumentation for Solar Astrophysics, 65

\bibitem[]{}
Long D. M., Gallagher, R. T., Mc Ateer, R. T. J., Bloomfield, D. S. 2008, ApJ, 680, 81

\bibitem[]{}
Lynch, B. J., Antiochos, S. K., DeVore, C. R., Luhmann, J. G., 
Zurbuchen, T. H. 2008, ApJ, 683, 1192

\bibitem[]{}
Magara, T., Longcope, D. 2001, ApJ, 559, 55

\bibitem[]{}
Mandrini, C. H., Démoulin, P., Schmieder, B., Deng, Y. Y., Rudawy, P. 
2002, A\&A, 391, 317

\bibitem[]{}
McIntosh, S. W., Leamon, R. J., Davey, A. R., Wills-Davey, M. J. 
2007, ApJ, 660, 1653

\bibitem[]{}
Moreton, G. E. 1960, AJ, 65, 494


\bibitem[]{}
Moreton, G. E., Ramsey, H. E. 1960 PASP, 72, 357 

\bibitem[]{}
Moreton, G. E. 1961, S.\&T., 69, 145

\bibitem[]{}
Narukage, N., Hudson, H. S., Morimoto, T., Akiyama, S., Litai, R., Kurokawa, H., Shibata, K. 2002, ApJ, 572, L109

\bibitem[]{}
Patsourakos, S. and Vourlidas, A. 2009, ApJ, 700, L182

\bibitem[]{}
Pizzo, V. J., Hils, S. M., Balch, C. C. et al. 2005, Sol. Phys., 226, 283

\bibitem[]{}
Scherrer, P. H., Bogart, R. S., Bush, R. I., Hoeksema, J. T., Kosovichev, A. G. 1955, Sol. Phys., 162, 129

\bibitem[]{}
Schrivjer, C., deRosa, M. 2003, Sol. Phys., 212, 165

\bibitem[]{}
Terradas, J., Ofman, L. 2004, ApJ, 610, 523

\bibitem[]{}
Thompson, B. J., Plunket, S. P., Gurman, J. B. et al. 1998, Geophys. Res. Letters, 25, 2461

\bibitem[1999]{}
Thompson, B. J., Gurman, J. B., Neupert, W. M. et al. 1999, ApJ, 517, L151

\bibitem[]{}

Thompson, B. J., Reynolds, B., Aurass, H. et al. 2000, Sol. Phys., 193, 161

\bibitem[]{}
Uchida, Y. 1968, Sol. Phys., 4, 30

\bibitem[2006]{}
Vrsnak, B., Warmuth, A., Temmer, M., et al. 2006, A\&A, 448, 739

\bibitem[]{}
Wang, Y.-M. 2000, ApJ, 543, L89

\bibitem[]{}
Wang, T., Yan, Y., Wang, J., Kurokawa, H., Shibata, K. 2002, ApJ, 572, 598

\bibitem[2001]{}
Warmuth, A., Vrsnak, B., Aurass, H., Hansmeier, A. 2001, ApJ, 560L, 105

\bibitem[2004]{}
Warmuth, A., Vrsnak, B., Magdaleni\'c, J., Hanslmeier, A., Otruba, W. 2004, A\&A, 418, 1101

\bibitem[2005]{}
Warmuth , A., Mann, G., Aurass, H. 2005, ApJ, 626L, 121

\bibitem[]{}
Warmuth, A. 2007, The high Energy Solar Corona Waves, Eruptions, Particules, Lecture Notes in Physics, 725, 107

\bibitem[2000]{}
Wills-Davey, M. J., Thompson, B. J. 1999, Sol. Phys., 190, 467

\bibitem[]{}
Wills-Davey, M. J., DeForest, C. E., Stenflo, J. O. 2007, ApJ, 664, 556

\bibitem[2001]{wu}
Wu, S. T., Zheng, H., Wang, S., et al. 2001, J. Geophys. Res., 106, A11, 25089

\bibitem []{}
Wu, S. T., Li, B., Wang, S., Huinan Zheng 2005, J. Geophys. Res., 110, A11, 102, 

\bibitem[]{}
Zarro, D. M., Sterling, A. C., Thompson, B. J., Hudson, H. S., Nitta, N. 1999, 
 ApJ, 520, L139

\end{thebibliography}
\end{document}